\documentclass[aps,pra,preprint,groupedaddress]{revtex4-2}

\usepackage{xcolor}					

\newcommand{\white}[1]{\textcolor{white}{#1}}
\usepackage{placeins}
\usepackage{float}
\usepackage{graphicx}
\usepackage{gensymb}
\usepackage{amsmath,amssymb}
\usepackage{placeins}
\usepackage{multirow}
\usepackage{booktabs}
\usepackage[english]{babel}
\usepackage{ulem}

\usepackage{subfigure}
\usepackage{overpic}
\usepackage{xfrac}

\usepackage{booktabs}
\usepackage{tikz}
\usetikzlibrary{tikzmark}
\usetikzlibrary{calc}
\usetikzlibrary{arrows.meta}
\usepackage{lineno} 

\usepackage{lipsum}

\definecolor{blue2}{rgb}{0, 0.4470, 0.7410}
\definecolor{red2}{rgb}{0.8500, 0.1250, 0.0480} 
\definecolor{orange2}{rgb}{0.8500, 0.3250, 0.0980} 
\definecolor{yellow2}{rgb}{0.9290, 0.6940, 0.1250}
\definecolor{purple2}{rgb}{0.4940, 0.1840, 0.5560}
\definecolor{green2}{rgb}{0.4660, 0.6740, 0.1880}
\definecolor{ltblue2}{rgb}{0.3010, 0.7450, 0.9330}
\definecolor{dkred2}{rgb}{0.6350, 0.0780, 0.1840}
\definecolor{gray2}{rgb}{0.22, 0.22, 0.3}
\definecolor{ltgray2}{rgb}{0.647, 0.647, 0.647}
\definecolor{blueIV}{rgb}{0, 0, 0.7410}
\definecolor{blueIII}{rgb}{0.2, 0.2, 0.7410}
\definecolor{blueII}{rgb}{0.4, 0.4, 0.7410}
\definecolor{blueI}{rgb}{0.7410, 0.7410, 0.7410}
\definecolor{jetVI}{rgb}{0.9763 0.9831 0.0538}
\definecolor{jetV}{rgb}{0.9264 0.7256 0.2996}
\definecolor{jetIV}{rgb}{0.4783 0.7489 0.4877}
\definecolor{jetIII}{rgb}{0.0282 0.6663 0.7574}
\definecolor{jetII}{rgb}{0.0582 0.4677 0.8589}
\definecolor{jetI}{rgb}{0.2081 0.1663 0.5292}
\definecolor{mode1}{rgb}{0 0.4470 0.7410}
\definecolor{mode2}{rgb}{0.8500 0.3250 0.0980}
\definecolor{mode3}{rgb}{0.9290 0.6940 0.1250}
\definecolor{dkgold}{rgb}{0.5930 0.5150 0.3260}
\definecolor{mode1}{rgb}{0 0.4470 0.7410}
\definecolor{mode2}{rgb}{0.8500 0.3250 0.0980}
\definecolor{mode3}{rgb}{0.9290 0.6940 0.1250}
\definecolor{dkgold}{rgb}{0.5930 0.5150 0.3260}
\definecolor{matblue}{rgb}{0.000 0.447 0.741}
\definecolor{matred}{rgb}{0.850 0.325 0.098}
\definecolor{matyellow}{rgb}{0.9290 0.6940 0.125}
\definecolor{matpurple}{rgb}{0.494 0.184 0.556}
\definecolor{matgreen}{rgb}{0.466 0.674 0.188}
\definecolor{matcyan}{rgb}{0.3010 0.7450 0.9330}

\usepackage{hyperref}
\hypersetup{
	colorlinks = true,
	urlcolor = black,
	citecolor = black,
}

\begin{document}


\title{Interplay between streaks and vortices in shock-boundary layer interactions with conditional bubble events over a turbine airfoil}

\author{Hugo Felippe da Silva Lui}
\email[]{hugolui@unicamp.br}
\affiliation{Universidade Estadual de Campinas, Campinas, S\~ao Paulo, Brazil}

\author{William Roberto Wolf}
\affiliation{Universidade Estadual de Campinas, Campinas, S\~ao Paulo, Brazil}

\date{\today}
\begin{abstract}

The shock-boundary layer interaction (SBLI) over the convex wall of a supersonic turbine vane is studied with a focus on extreme separation bubble events and the interplay between the bubble, streaks, and streamwise vortices. The present analysis is performed on a dataset computed by a wall-resolved large-eddy simulation (LES) of a supersonic turbine at Mach number 2.0 and Reynolds number of $395,000$. Building on findings related to near-wall streaks and vortices that drive the bubble breathing motion, here, we employ conditional analysis to study extreme bubble events. First, mean and turbulence statistics are computed to examine differences between expanding and contracting bubble states. Then, finite-time Lyapunov exponent calculations are used to understand the interactions between streaks and streamwise vortices along large and small separation bubble events. Additionally, a deforming control volume methodology is applied to analyze the mass imbalance within the bubble and to identify regions that are more susceptible to mass inflow and outflow. In events where the recirculation region is small, near-wall high-speed streaks penetrate the bubble and lead to higher tangential Reynolds stresses upstream of the incident shock compared to those with large recirculation. The streaks are accompanied by streamwise vortices that meander and induce intense fluid mixing, leading to higher wall-normal and spanwise Reynolds stresses, and consequently, higher turbulent kinetic energy. This turbulent activity also causes intense fluctuations in wall pressure and skin-friction coefficient along the separation region. In contrast, during events with bubble expansion, high-speed streaks are advected over the separation region, with streamwise vortices appearing only downstream of the shock, resulting in minimal fluid mixing inside the bubble. The analysis of mass flux along the bubble surface reveals that during its contraction phase, mass flux out of the bubble occurs predominantly upstream of the incident shock because of high-speed streaks dragged towards the wall by the streamwise vortices. In the expansion phase, pronounced mass flux into the bubble is observed downstream of the shock, near reattachment. This increased fluid injection is associated with the presence of streamwise vortices and low-momentum flow structures near reattachment, suggesting that fluid entrainment by vortices plays a key role in the mass flux into the bubble.

\end{abstract}

%
%
%



\maketitle
\clearpage

\section{Introduction}
\label{sec:intro}

Shock-wave boundary-layer interaction (SBLI) is a critical phenomenon in high-speed flow applications, including supersonic aircraft engines, overexpanded rocket nozzles, transonic wings of commercial airplanes, turbomachinery cascades, and control surfaces of high-speed vehicles \citep{DELERY1985,GAITONDE2015,Sandberg2022}. In these applications, SBLIs can degrade performance by reducing combustion efficiency, increasing aerodynamic losses, inducing structural fatigue, and diminishing control-surface effectiveness \citep{Gaitonde2023}. These detrimental effects occur due to significant boundary layer separation caused by the intense adverse pressure gradient imposed by the incident shock. 

The separated flow has an unsteady nature, characterized by the breathing motion of the separation bubble, oscillations of the separation and reattachment shocks, as well as the flapping of the shear layer over the bubble. These flow features generate unsteady pressure and friction forces, along with thermal loading on the structure. Therefore, a better understanding of SBLI flow physics can provide insights for the design of high-speed flow devices and guide the development of flow control strategies to reduce the size of the separation bubble and its unsteadiness \citep{clemens2014}.

Several numerical and experimental studies have investigated the unsteadiness of SBLIs in two-dimensional configurations (i.e., spanwise periodic flows), such as flat plates \citep{dupont_haddad_debieve_2006, pirozzoli_bernardini_grasso_2010, Touber2009, morgan2013, Aubard2013, pasquariello_2017, adler_gaitonde_2018, volpiani_2018, bernardini_2023}, compression ramps \citep{ganapathisubramani_clemens_dolling_2009, priebe2012, porter_2019, khobragade_2022}, and steps \citep{hu_2021, hu_2022}. These studies span a variety of flow conditions and geometries, and show that the unsteadiness associated with separation bubble pulsations, shock oscillations, and shear layer flapping typically occurs at low frequencies in the range $0.01 < St_{L} < 0.1$, which is one to two orders of magnitude lower than the turbulent fluctuations in the incoming boundary layer ($St_{L} \geq 1$) \citep{clemens2014}. Here, the Strouhal number $St_{L} = fL_{SB}/u_\infty$ is defined in terms of the frequency $f$, the length of the separation bubble $L_{SB}$, and the freestream velocity $u_\infty$.

The breathing motion of the separation bubble plays a key role in the unsteadiness of the SBLI, as its contraction or expansion induces shock oscillations and shear layer flapping. One proposed mechanism for bubble pulsation suggests that incoming boundary layer fluctuations drive its motion \citep{beresh2002, ganapathisubramani_clemens_dolling_2007, ganapathisubramani_clemens_dolling_2009, porter_2019, baidya_2020}. For example, \citet{beresh2002} suggested that large-scale structures in the incoming boundary layer influence its resistance to separation. In these cases, high-speed streaks are associated with temporarily fuller velocity profiles that are less prone to separation. 
Some studies \citep{baidya_2020,ganapathisubramani_clemens_dolling_2007,ganapathisubramani_clemens_dolling_2009} observed that, as high-speed (low-speed) structures pass through the separated flow, they cause the bubble to contract (expand).

Another proposed mechanism attributes the breathing motion of the bubble to the mass imbalance within it, driven by shear layer entrainment \citep{piponniau2009, samper_2018, jenquin_2023}. It was suggested that the shear layer ejects fluid from the separated flow, while fluid from the surrounding region near reattachment is injected into the recirculation bubble. A greater mass injection leads to bubble expansion, whereas a more pronounced mass ejection causes its contraction. Alternatively, several authors have explored the unsteadiness of the recirculation bubble from the perspective of flow instabilities \citep{Touber2009, priebe2012, priebe_2016, Nichols_2017, adler_gaitonde_2018}, with a particular emphasis on the absolute instability of the bubble. \citet{priebe_2016}, on the other hand, suggested that the curvature of the bubble induces a centrifugal instability, resulting in the formation of streamwise vortices. These vortices generate unsteady, streamwise-elongated regions of low and high momentum within the bubble, thereby driving its unsteadiness.

Regardless of the mechanism driving the bubble breathing motion, recent investigations have reported the occurrence of extreme, intermittent events in SBLIs \citep{jenquin_2023, bernardini_2023, lui_2024}. These extreme events are associated with a sudden contraction when the bubble is large or a rapid expansion when the bubble is small. However, these previous analyses have not thoroughly examined the differences in flow quantities and coherent structures between extreme events involving small and large bubbles. Extensive investigations have been conducted on extreme events in wall-bounded turbulent flow in the absence of shock waves \citep{Hack_Schmidt_2021,Guerrero_Lambert_Chin_2020,Zhang_Wan_Dong_Liu_Sun_Lu_2023,Fan_Kozul_Li_Sandberg_2024,leandro2024}. In such flows, extreme events are characterized by large positive and negative wall shear stress fluctuations and, in compressible cases, by extreme wall heat fluxes. In most of these cases, conditional analysis was applied to understand the influence of extreme wall shear stress events on flow quantities and coherent structures. 

Conditional analysis of SBLIs was performed primarily to investigate the influence of the incoming boundary layer state and separation bubble size on the movement of the separation shock \citep{beresh2002, piponniau2009, Souverein_2010,Agostini_2015}. For instance, \citet{beresh2002} observed that conditionally averaged velocity profiles in the incoming boundary layer are fuller when the shock moves downstream. \citet{piponniau2009} performed a conditional analysis of the velocity fields for small and large bubbles and found that contractions of the bubble are associated with downstream movements of the separation shock. More recently, \citet{jenquin_2023} examined conditionally averaged pressure fluctuation fields in the SBLI region. Their results showed that the expansion of the separation bubble is associated with negative fluctuations in both wall pressure and streamwise velocity in the incoming boundary layer. In general, however, these studies have not fully explored the differences in turbulence quantities and coherent structures between small and large bubbles. 

This work builds on our previous findings \citep{lui2022, lui_2024} obtained by wall-resolved large eddy simulations (LES), where the flow physics and unsteadiness of SBLIs were investigated over the curved surfaces of a supersonic turbine vane. 
Our previous results emphasized the crucial role of near-wall streaks in driving the bubble breathing motion over the blade suction side. The occurrence of extreme flow events associated with strong bubble contractions and expansions was observed and linked to the presence of streamwise vortices. In the present work, the physics of SBLI occurring on the convex wall of the turbine vane is further explored through a conditional analysis of extreme separation bubble events. This study allows for the characterization of the mechanisms governing bubble growth and collapse, as well as their effects on pressure and skin-friction coefficients and other turbulence quantities. Understanding the flow dynamics associated with the bubble’s breathing motion provides valuable insights for the development of flow control strategies aimed at suppressing this unsteadiness, thereby improving aerodynamic performance. 

To elucidate the dynamics of extreme bubble events, the differences in mean flow and turbulence quantities are examined, including the interplay between bubble, streaks, and streamwise vortices. We assess how these interactions influence fluid transport within the separation bubble by computing its surface mass flux during the contraction and expansion motions. This, in turn, allows for the identification of regions that are more susceptible to mass inflow and outflow in the context of SBLI.

The present work is organized as follows: In Section \ref{sec:numerical_methodology}, the numerical methodology and flow configuration are described. Results are then presented in Section \ref{sec:results}, where Section \ref{subsec:motivation} summarizes the main flow characteristics of the SBLI over the convex wall of the turbine vane, providing context for subsequent analyses. Next, Section \ref{subsec:conditional_analysis} presents a conditional analysis of extreme bubble events, examining differences in mean flow and turbulence quantities for instants of contracted and expanded bubbles. 
In Section \ref{subsec:FTLE}, the finite-time Lyapunov exponent is computed to investigate the interplay between bubble, streaks, and streamwise vortices, as well as fluid transport within the separation bubble. Finally, Section \ref{subsec:mass_imbalance} analyzes the mass flux across the bubble surface, being followed by the main conclusions presented in Section \ref{sec:conc}.  

\section{Flow configuration and numerical Methodology} 
\label{sec:numerical_methodology}

A wall-resolved LES is performed as reported by \citet{lui2022} and here an overview of the numerical methodology is provided. The non-dimensional compressible Navier-Stokes equations are solved in general curvilinear coordinates. Length, velocity components, density, pressure, temperature, and time are nondimensionalized using the axial airfoil chord $c_{x}$, the inlet speed of sound $a_{\infty}$, the inlet density $\rho_{\infty}$, $\rho_{\infty} a_{\infty}^2$, $(\gamma - 1)T_{\infty}$, and $c_{x}/a_{\infty}$, respectively. The Reynolds and Mach numbers are defined as $Re = \rho_{\infty} u_{\infty} c_{x}/\mu_{\infty}$ and $M_{\infty} = u_{\infty}/a_{\infty}$, while the Prandtl number is given by $Pr = \mu_{\infty} c_{p}/\kappa_{\infty}$. Here, $u_{\infty}$, $T_{\infty}$, $\mu_{\infty}$, $c_p$, and $\kappa_{\infty}$ represent the inlet flow velocity, temperature, dynamic viscosity, specific heat at constant pressure, and thermal conductivity, respectively. Further details of the governing equations and nondimensionalization procedure can be found in Refs. \citep{Nagarajan2003, bhaskaran_thesis, lui2022}.

\begin{figure}[H]
\begin{overpic}[trim = 1mm 1mm 1mm 1mm, clip,width=0.99\textwidth]{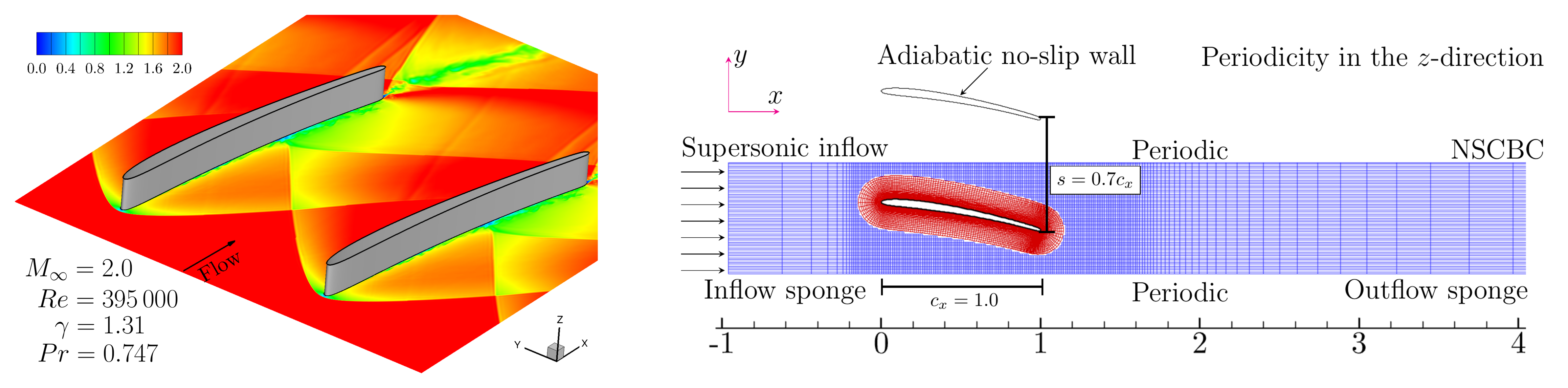}
		\put(0,18){(a)}
		\put(40,18){(b)}		
	\end{overpic} 
	\caption{Schematics of (a) three-dimensional geometry and inlet conditions, with the background showing the Mach number contours, and (b) computational domain and boundary conditions, with every $5$th grid point shown.}
	\label{fig:schematic}
\end{figure}

The airfoil geometry, along with the fluid properties and inflow conditions, is the same as that used by~\citet{LIU2019} and \citet{lui2022}, and this information is shown in Fig. \ref{fig:schematic}.
The vane cascade is exposed to a uniform flow, with an inlet Mach number of $M_{\infty} = 2.0$ and a Reynolds number of $Re = 395\,000$, based on the inlet velocity $u_{\infty}$ and the airfoil axial chord $c_x$. The fluid is modeled as a calorically perfect gas, with a ratio of specific heats of $\gamma = 1.31$, a Prandtl number of $Pr = 0.747$, with a ratio of the Sutherland constant to the inlet temperature given by $S_{\mu}/T_{\infty} = 0.07182$. The spacing between the vanes, $s$, corresponds to $0.7c_x$, and the spanwise size is equal to $0.1c_x$.

An overset mesh procedure is used in the present simulation with two overlapping grids: one is a body-fitted O-grid block, which accurately resolves the turbulent boundary layers around the vane, and the other is an H-grid block, used to facilitate the application of cascade pitchwise periodicity. The O-grid is discretized with $1280 \times 300 \times 144$ points in the streamwise, wall-normal, and spanwise directions, respectively, while the background Cartesian grid contains $960 \times 280 \times 72$ points. In total, the computational grid consists of approximately $75 \times 10^6$ points. 

In both grid blocks, the governing equations are discretized using a sixth-order finite-difference compact scheme, implemented on a staggered grid \citep{Nagarajan2003}. A sixth-order compact interpolation scheme is employed to calculate flow quantities at the different nodes of the staggered grid. The node arrangement is provided by \citet{Nagarajan2003}. To overcome the stiffness issue caused by the fine boundary layer grid in the O-grid block, the time integration of the governing equations is performed using an implicit, second-order-accurate Beam–Warming scheme. In the H-grid block, a low-storage third-order Runge-Kutta scheme is used for time advancement of the governing equations.

At each time step, a sixth-order compact filter \citep{Lele1992} is applied in regions of the flow away from solid boundaries to control high-wavenumber numerical instabilities resulting from mesh non-uniformities and interpolations between overlapping grids. In the present LES, no explicit subgrid-scale model is used; however, the high-order compact filter approximates the effects of such models \citep{Mathew_etal_2003}. To suppress unresolved high-frequency oscillations near the shock waves without affecting small-scale turbulence, the localized artificial diffusivity (LAD) scheme LAD-D2-0 \citep{Kawai2010} is employed. This shock-capturing scheme computes additional artificial bulk viscosity and thermal conductivity, which are then added to their physical counterparts. Further details on the numerical procedure for the LES can be found in Refs. \citep{Nagarajan2003}, \citep{Bhaskaran}, and \citep{lui2022}.

Although turbine vanes are typically exposed to inflow turbulence, a uniform supersonic inflow condition is applied at the inlet boundary. To induce transition to turbulence in the airfoil boundary layers, a body-forcing term is used as a tripping device \citep{WAINDIM2016, lui2022}. This approach enables a more direct comparison with other SBLI studies in the literature for canonical configurations. The tripping setup consists of unsteady random spanwise disturbances with an amplitude selected to trigger bypass transition. On the suction side, the tripping is applied in the range $0.22 < x < 0.27$ along a wall-normal distance of $0 < \Delta n < 0.001c_{x}$. On the pressure side, it is applied in the range $0.10 < x < 0.15$ with the same wall-normal distance. It is worth noting that the current airfoil configuration has limited space in the streamwise direction for flow development, further constrained by the SBLI. As a result, there is insufficient space to establish a fully developed turbulent boundary layer.

Regarding the other boundary conditions, the Navier-Stokes characteristic boundary condition (NSCBC) \citep{Poinsot1992} is used at the outlet. To absorb convected numerical disturbances, damping sponges are applied on both the inflow and outflow boundaries (first and last 15 points of the domain) \citep{nag_thesis, bhaskaran_thesis}. The ramping function and its parameters are the same as those used in Refs. \citep{bhaskaran_thesis, wolf2012}. A no-slip adiabatic condition is enforced on the airfoil surface. To model a linear cascade of vanes, periodic boundary conditions are applied in the $y$-direction of the background grid. Additionally, to avoid modeling complexities near the tip and end-wall and to facilitate comparison with other two-dimensional SBLI results in the literature, periodic boundary conditions are also applied in the spanwise $z$-direction.

The grid resolution in the simulation follows the recommendations for a wall-resolved LES \citep{Georgiadis2010}. Previous studies \citep{lui2022, lui_2024} have verified this by assessing the near-wall grid spacing in terms of wall units and the ratio of the grid spacing to the calculated Kolmogorov length scale. Additionally, based on two-point spanwise correlations of velocity fluctuations at several chord locations, \citet{lui_2024} confirmed that the spanwise domain is sufficiently large to not influence the turbulence dynamics. Throughout this work, flow statistics are computed using both instantaneous three-dimensional snapshots and spanwise-averaged flow fields that are recorded every 0.006 and 0.0012 nondimensional time units, respectively, over a total duration of 45 nondimensional time units, based on the inlet velocity and axial chord.

\section{Results}
\label{sec:results}

\subsection{General flow features}
\label{subsec:motivation}

\begin{figure}
	\centering	
	\begin{overpic}[trim = 1mm 1mm 1mm 1mm, clip,width=0.48\textwidth]{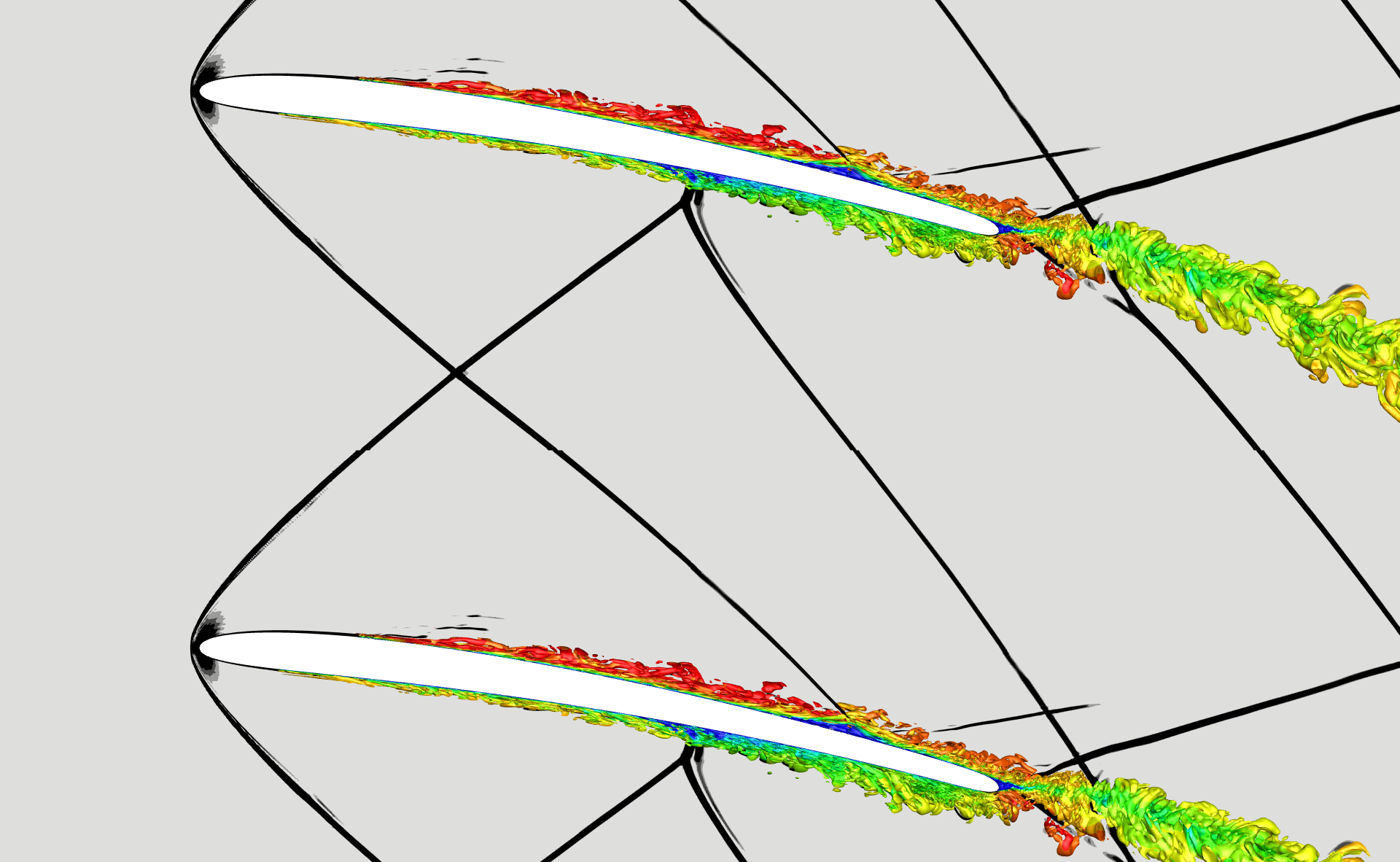}
	\put(1,56){(a)}
	\end{overpic} 
	\begin{overpic}[trim = 1mm 1mm 1mm 1mm, clip,width=0.48\textwidth]{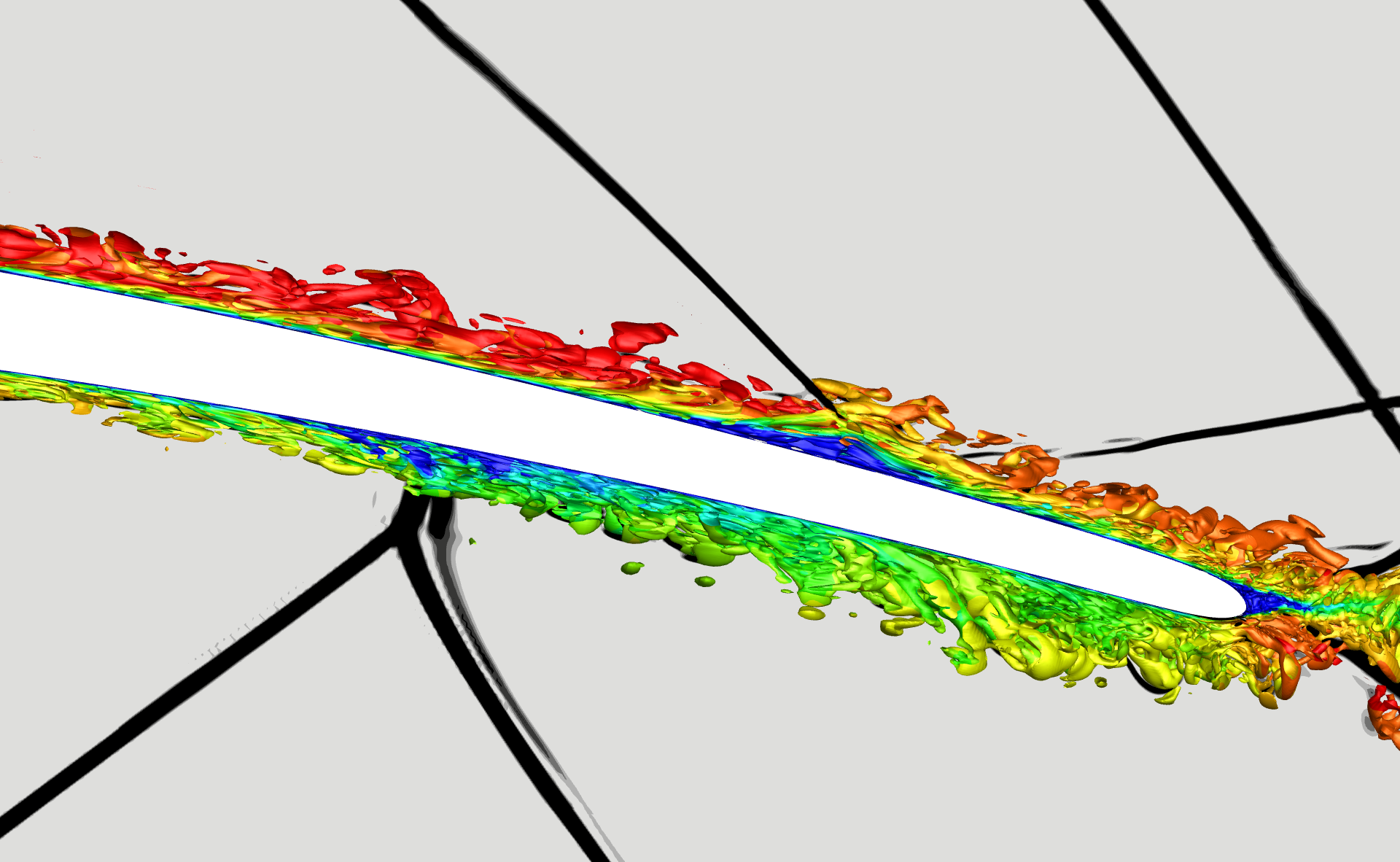}
	\put(1,56){(b)}
	\end{overpic} 
	\caption{Overview of the turbine cascade (left), with a detailed view of the SBLIs (right). The boundary layers are shown by an isosurface of $Q$-criterion colored by streamwise velocity and the background plane highlights the shock waves displayed by the magnitude of density gradient $\| \nabla \rho \|$.}
	\label{fig:iso_surfaces}
\end{figure}

The flow physics of the present supersonic cascade was investigated by \citet{lui2022, lui_2024}, and the main flow features are summarized here to contextualize the conditional analyses of extreme bubble events that are presented in this work. First, instantaneous flow field planes are shown in Fig. \ref{fig:iso_surfaces}. The left column displays the flow through the turbine cascade, while the right column presents detailed views of the SBLIs on both sides of the airfoil. The turbulent boundary layers on the suction and pressure sides are visualized by an isosurface of \( Q \)-criterion colored by streamwise velocity. Shock waves are illustrated by a background plane that shows the magnitude of density gradient \( \| \nabla \rho \| \).

Detached shock waves are generated at the leading edges of the airfoils, becoming oblique shocks that interact with the turbulent boundary layers on the adjacent airfoils. The SBLIs induce the formation of separation bubbles (shown in blue) because of the imposed adverse pressure gradients. On the suction side, an oblique shock impinges on the turbulent boundary layer. However, unlike configurations where the incoming boundary layer develops on flat surfaces (compression ramps and flat plates), the compression waves upstream of the recirculation bubble do not coalesce into a separation shock. On the pressure side, a Mach reflection takes place and forms a thin bubble, despite a strong pressure jump imposed by the normal shock. The difference in the sizes of the separation regions is related to the combined effects of pressure gradients, dilatation, and turbulent kinetic energy (TKE) levels near the wall in the incoming boundary layer, as discussed by \citet{lui_2024}.

\begin{figure}
	\centering	
	\begin{overpic}[trim = 1mm 1mm 1mm 1mm, clip,width=0.48\textwidth]{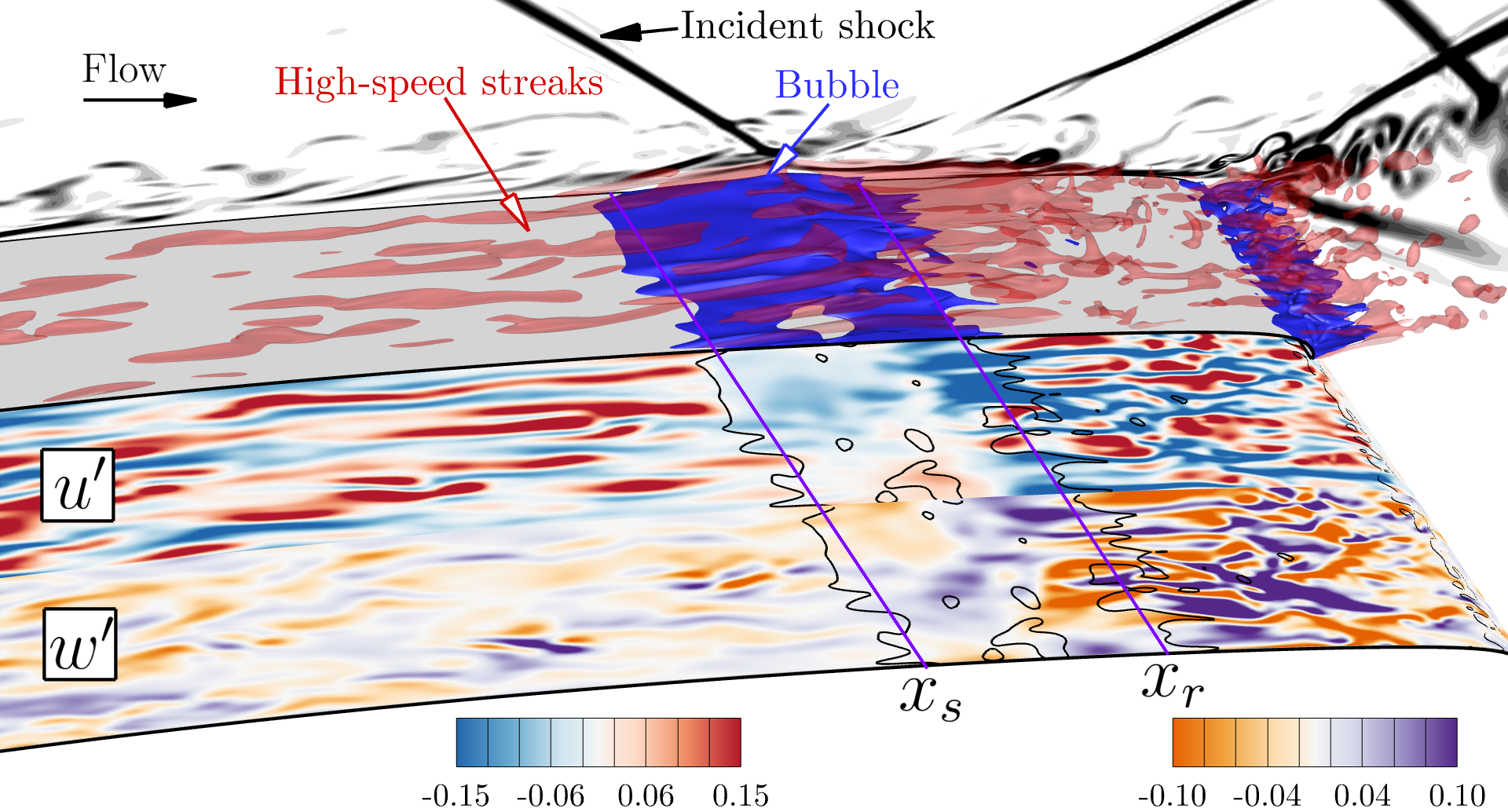}
	\put(1,53){(a)}
	\end{overpic} 
	\begin{overpic}[trim = 1mm 1mm 1mm 1mm, clip,width=0.48\textwidth]{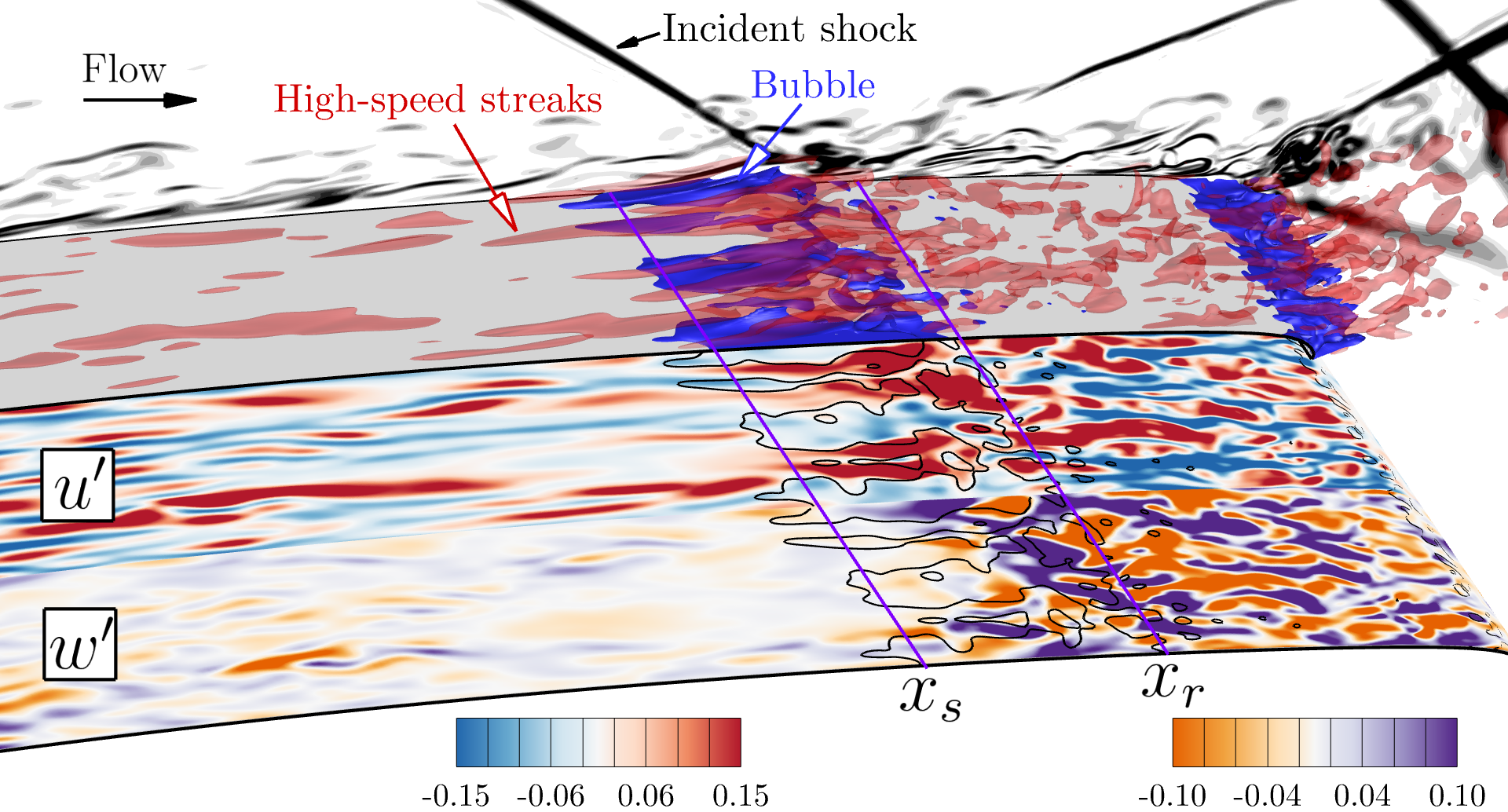}
	\put(1,54){(b)}
	\end{overpic} 
	\caption{Instantaneous flow topology on the airfoil suction side for (a) large and (b) small recirculation bubbles. The topmost strip shows isosurfaces of $u$-velocity fluctuations with $u' = 0.15$. The center and bottommost strips display 
    near-wall planes of $u$- and $w$-velocity fluctuations, respectively. The velocity fluctuations are normalized by the inlet velocity. The separation bubble is visualized by blue isosurfaces corresponding to $u=0$ in the three-dimensional (topmost strip) plots, while in the contour plots (center and bottommost strips) it can be seen as delimited by black lines that conform to the passage of streaks. The shocks are visualized in the background with grayscale contours of $\| \nabla \rho \|$. The mean separation $x_s$ and reattachment $x_r$ positions, computed using the full dataset, are indicated by magenta lines}.
	\label{fig:bubble_3D}
\end{figure}

The suction side separation bubble is further analyzed including the interplay between large-scale boundary layer structures and flow recirculation.
To find a connection between these structures and the state of the bubble, instantaneous flow visualizations are shown in Fig. \ref{fig:bubble_3D} when the bubble is (a) large and (b) small. The top strip displays red isosurfaces of $u' = 0.15$, with blue isosurfaces of $u = 0$ depicting the recirculation bubble. Here, $u$ denotes the streamwise velocity component, and $u'$ corresponds to the streamwise velocity fluctuations. The center and bottom strips show 
near-wall planes on the airfoil surface, displaying $u$- and $w$-velocity fluctuations, respectively. The velocity fluctuations are normalized by the inlet velocity. In the plots, black lines delimit the separated flow region, and magenta lines mark the mean separation $x_s$ and reattachment $x_r$ locations. The shocks are displayed in the background by grayscale contours of $\| \nabla \rho \|$.

The top strips reveal the interaction between high-speed streaks and the separated flow. When the bubble is large, the streaks are transported over the separation region. On the other hand, when high-speed streaks penetrate the bubble, they cause local flow reattachment, resulting in a smaller recirculation region. The presence of low- and high-speed large-scale structures in the incoming boundary layer is displayed in the center strips. The separation (black) line aligns with these structures, where positive (negative) velocity fluctuations cause the separation point to move downstream (upstream). These findings are in agreement with previous studies \citep{beresh2002,ganapathisubramani_clemens_dolling_2009,porter_2019,baidya_2020,lui2022,lui_2024}. Moreover, it is interesting to note that when the bubble is small, high $u$- and $w$-velocity fluctuations occur between the mean separation and reattachment locations. In contrast, when the bubble is large, only small $u$- and $w$-velocity fluctuations are present.


\begin{figure}
\centering	
\begin{overpic}[trim = 1mm 1mm 1mm 1mm, clip,width=0.99\textwidth]{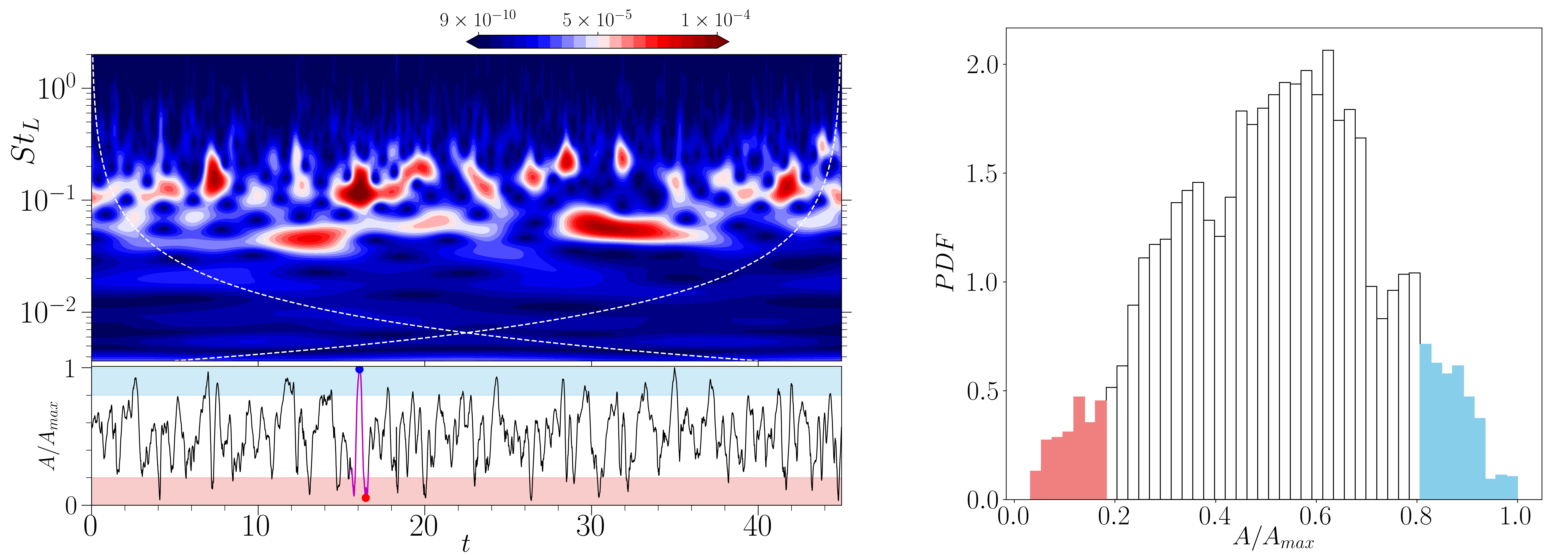}
    \put(1,33){(a)}
    \put(0,11){(b)}
    \put(58.3,34){(c)}
	\end{overpic} 
	\caption{Analysis of intermittent events using (a) the scalogram of the normalized bubble area $A/A_{max}$, (b) its corresponding signal, and (c) the histogram of $A/A_{max}$}.
	\label{fig:wavelet}
\end{figure}

\begin{figure}
\centering	
\begin{overpic}[trim = 1mm 1mm 1mm 1mm, clip,width=0.49\textwidth]{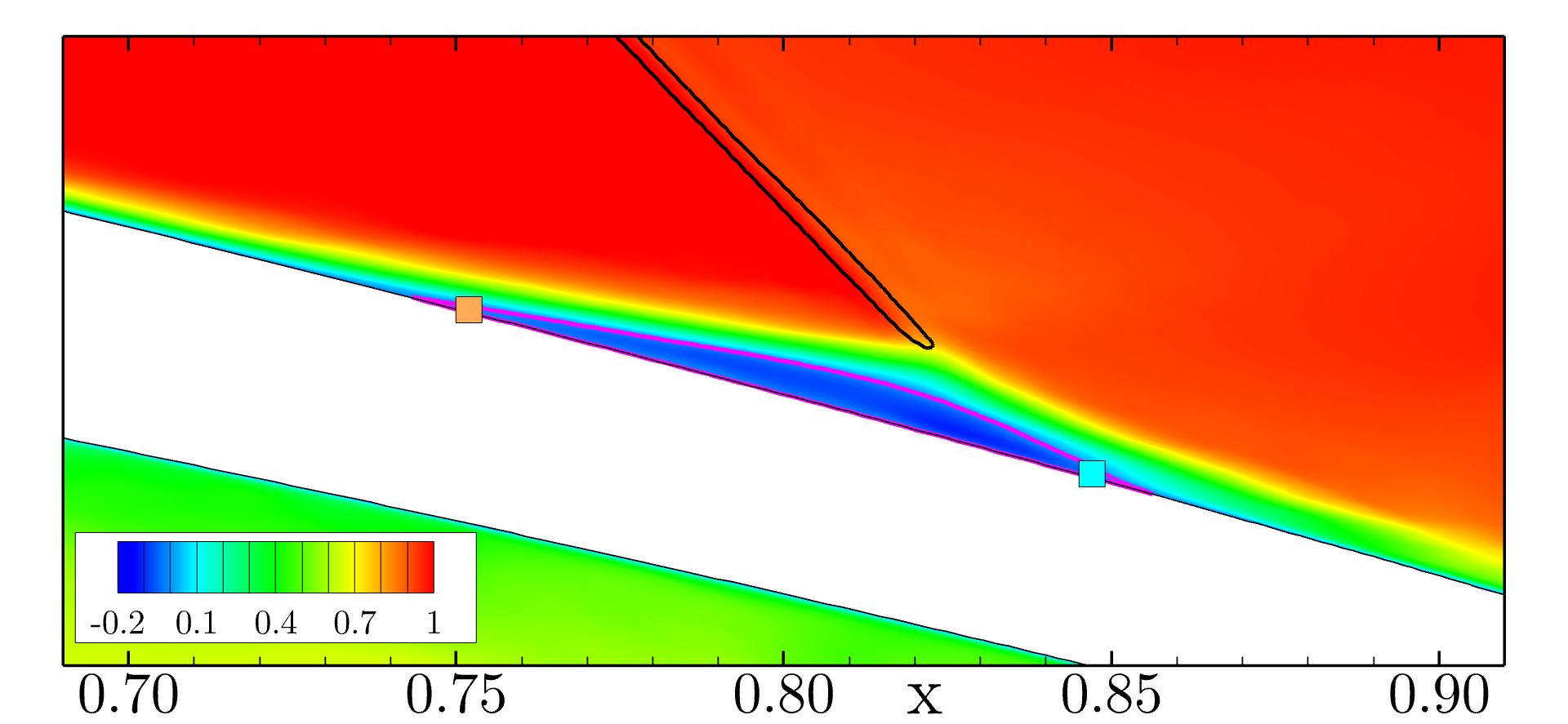}
    \put(5,38){(a)}
	\end{overpic} 
\begin{overpic}[trim = 1mm 1mm 1mm 1mm, clip,width=0.49\textwidth]{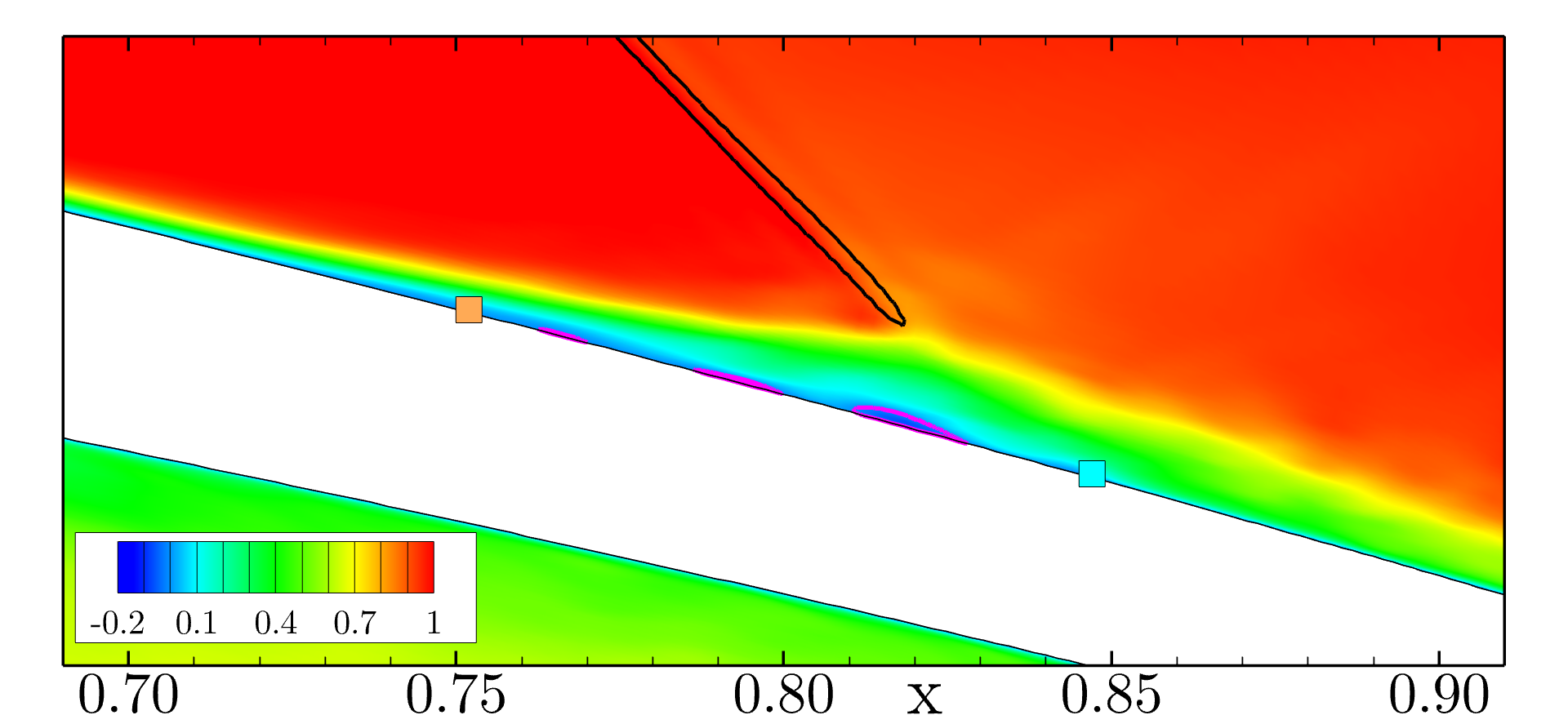}
    \put(5,38){(b)}
	\end{overpic} 
	\caption{Instantaneous spanwise-averaged contours of streamwise velocity for (a) a large bubble and (b) a small bubble. The time instants correspond to the blue and red circles in Fig. \ref{fig:wavelet}(b). The black lines display the incident shock through the pressure gradient magnitude $\| \nabla p \|$, while the purple lines delimit the separated flow regions. The orange and cyan squares represent the mean separation and reattachment positions, respectively.}
	\label{fig:bubble_2d}
\end{figure}

To highlight the intermittent nature of the suction side bubble breathing motion, a time-frequency analysis is performed using a continuous wavelet transform (CWT) with a complex Morlet wavelet \cite{Farge_1992}. The intermittent characteristics of SBLIs have also been observed in previous studies \citep{poggie_1997,bernardini_2023,jenquin_2023}. Figure \ref{fig:wavelet}(a) displays the scalogram of the normalized bubble area $A/A_{max}$. The white dashed lines delimit the cone of influence, where distortions in the scalogram may occur due to the signal endpoints. Figure \ref{fig:wavelet}(b) shows the time evolution of $A/A_{max}$, with blue and red circles marking two specific extreme events when the bubble is large and small, respectively. These events are illustrated by instantaneous spanwise-averaged contours of streamwise velocity in Figs. \ref{fig:bubble_2d}(a) and \ref{fig:bubble_2d}(b), respectively. The black lines represent the incident shock, visualized by the magnitude of the pressure gradient $\| \nabla p \|$. For reference, the mean separation and reattachment locations are indicated in the figures by orange and cyan squares, respectively.

The spectrogram shows several localized bursts (red regions), which indicate intense signal fluctuations at low frequencies ranging from $0.04 \le St_L \le 0.2$, occurring at specific time periods. A closer examination of the temporal signal of $A/A_{max}$ reveals that these intermittent events are associated with bubble contractions and expansions, as illustrated in Figs. \ref{fig:bubble_2d}(a) and \ref{fig:bubble_2d}(b). These events occur on various time scales, including sudden bubble size variations and sequences of consecutive contractions and expansions. As shown by \citet{lui2022} and \citet{lui_2024}, and illustrated in Fig. \ref{fig:bubble_3D}, the bubble contractions in the present configuration are related to the passage of high-speed streaks through the bubble. 

\subsection{Conditional analysis of extreme bubble events}
\label{subsec:conditional_analysis}

The results of the previous section show that the bubble undergoes significant contractions and expansions. Figure \ref{fig:bubble_3D} suggests that strong velocity fluctuations occur in the SBLI region when the bubble is small, while the opposite occurs when the bubble is large. To analyze the flow statistics of these extreme events separately, conditional analyses are performed. Throughout this study, the term `extreme events' refers to both small and large bubble events, with the normalized bubble area $A/A_{max}$ used as a metric to identify the extreme events. Here, a small bubble is defined as when $A/A_{max} \le 0.2$, while a large bubble is defined by $A/A_{max} \ge 0.8$. The time instants associated with these extreme events are highlighted by the light red and blue shaded areas in Fig. \ref{fig:wavelet}(b). 

The present LES dataset comprises 7502 snapshots, with 1074 extreme events ($14.3\%$ of the total dataset), including 431 occurrences of small bubble events ($5.7\%$ of the total dataset) and 643 occurrences of large bubble events ($8.6\%$ of the total dataset). The histogram of normalized bubble area $A/A_{max}$ shown in Fig. \ref{fig:wavelet}(c) indicates that extreme events occur in the tails of the distribution, highlighted by the red and blue regions that represent small and large bubble events, respectively. A convergence study of the conditional averaging procedure is presented in the Appendix. Here, the conditionally averaged quantities are denoted by $\overline{\cdot}$ and $\langle \cdot \rangle$ , where these symbols represent the Reynolds-averaged and Favre-averaged quantities, respectively. The averaging operation is performed in time and the spanwise direction. The superscripts $'$ and $''$ represent the fluctuations of Reynolds-averaged and Favre-averaged variables, respectively.

\subsubsection{Conditionally averaged flow fields}

\begin{figure}[H]
	\centering	
	\begin{overpic}[trim = 1mm 1mm 15mm 15mm, clip,width=0.49\textwidth]{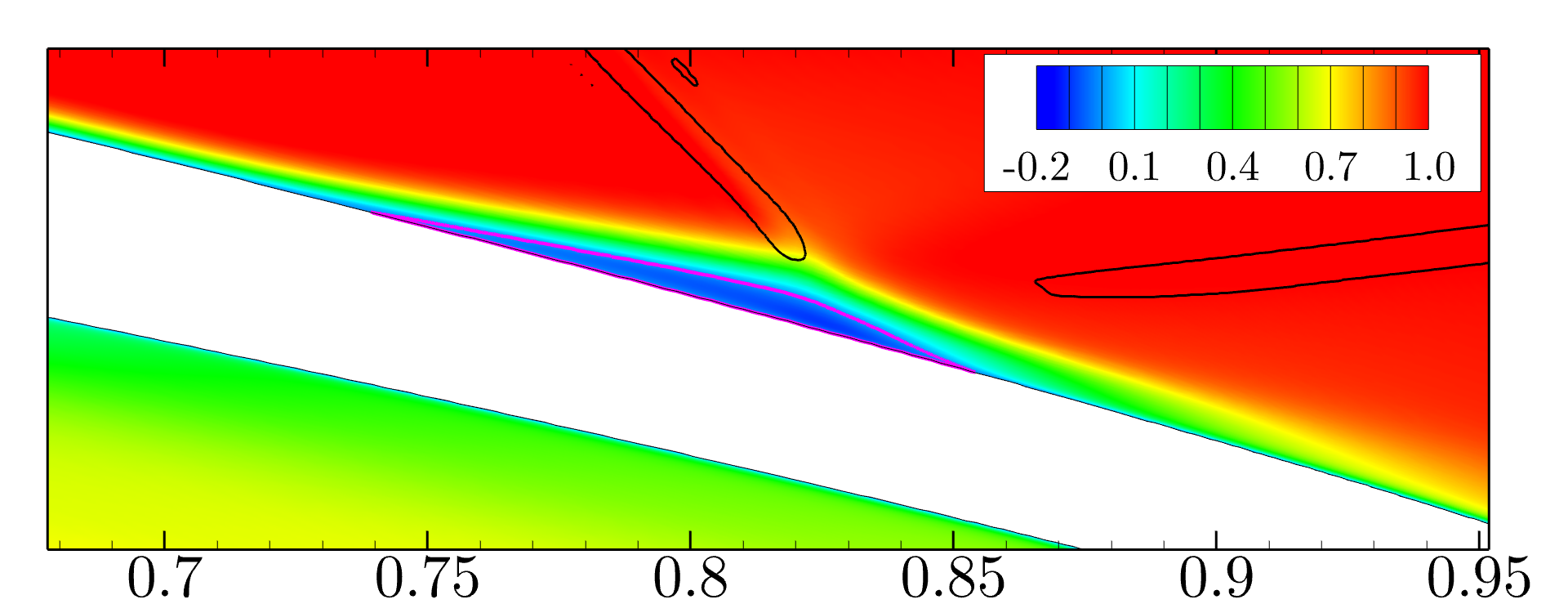}
	\put(3,8){(a)}
	\end{overpic} 
	\begin{overpic}[trim = 1mm 1mm 15mm 15mm, clip,width=0.49\textwidth]{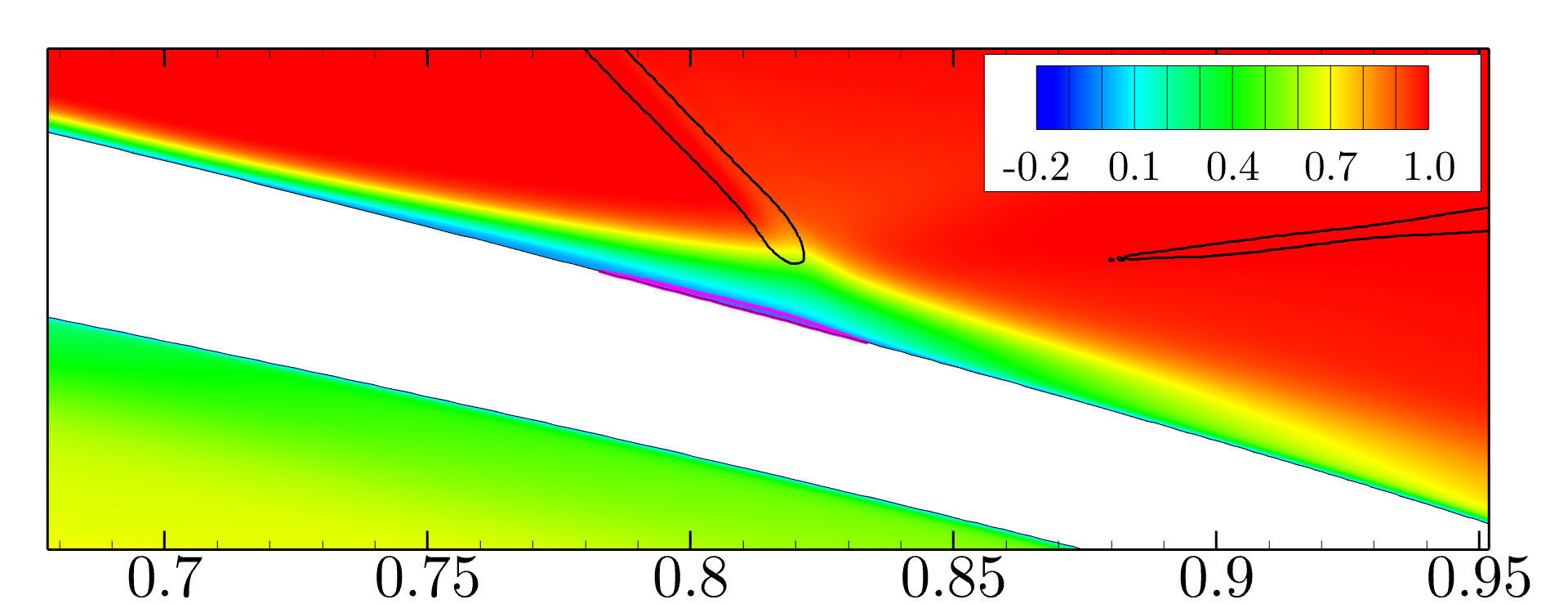}
	\put(3,8){(b)}
	\end{overpic} 
    
	\caption{Contours of conditionally averaged wall-tangential velocity $\overline{u_t}$ for (a) large bubble events and (b) small bubble events. Results are normalized by the inlet velocity. The magenta lines delimit the separation region, while the black lines display the shocks through the quantity $\vec{V} \cdot \nabla p$.}
	\label{fig:Ut_mean}
\end{figure}

\begin{figure}
\centering	
\begin{overpic}[trim = 1mm 1mm 1mm 1mm, clip,width=0.99\textwidth]{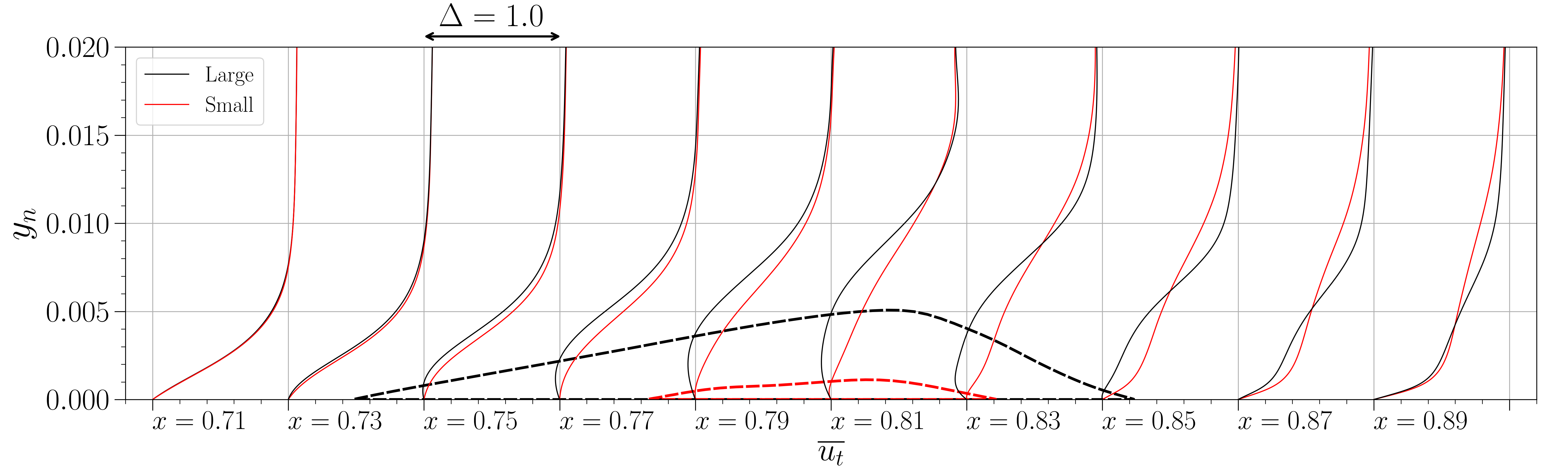}
	\end{overpic} 
	\caption{Wall-normal profiles of conditionally averaged $\overline{u_t}$ for large bubble events (solid black) and small bubble events (solid red), normalized by the inlet velocity. The black and red dashed lines delimit the separation bubbles for the large and small bubble events, respectively.}
	\label{fig:ut_profiles}
\end{figure}

The conditionally averaged wall-tangential velocity $\overline{u_t}$ contours are presented in Fig. \ref{fig:Ut_mean}(a) for large bubble events and Fig. \ref{fig:Ut_mean}(b) for small bubble events. The values are normalized by the inlet velocity $U_{\infty}$. The separation region is delimited by magenta lines, and the incident and reattachment shocks are represented by black lines, visualized using $\vec{V} \cdot \nabla p$, where $\vec{V}$ is the velocity vector. Substantial differences are observed between the flow fields. As expected, the recirculation region is much smaller for small bubble events, resulting in a thicker shear layer over the bubble, which in turn affects the tangential velocity values in the SBLI region. This effect is highlighted in the wall-normal profiles of conditionally averaged $\overline{u_t}$, shown at several chordwise locations in Fig. \ref{fig:ut_profiles}.


In the incoming boundary layer, at $x = 0.71$, the velocity profiles for large and small bubble events appear similar, but differences become apparent downstream. Along the recirculation region, the velocity profiles show a greater deceleration of the fluid near the wall for the large bubble events, while flow reversal is barely noticeable in small bubble events, suggesting a weak flow separation. This may be explained by the fuller velocity profiles upstream of the small bubble events, particularly in the near-wall region. A fuller velocity profile increases resistance to boundary layer separation \citep{DELERY1985,beresh2002}. Downstream of flow reattachment, the velocity profiles resemble those typically observed in a mixing layer \citep{priebe2012,fang2020}. The velocity difference between the low-speed flow near the wall and the high-speed external flow is more pronounced in large bubble events compared to small bubble events. For the former, lower values of $\overline{u_t}$ are observed near the wall, while higher values are seen after the inflection point.

\begin{figure}
\centering	
\begin{overpic}[trim = 1mm 1mm 1mm 1mm, clip,width=0.95\textwidth]{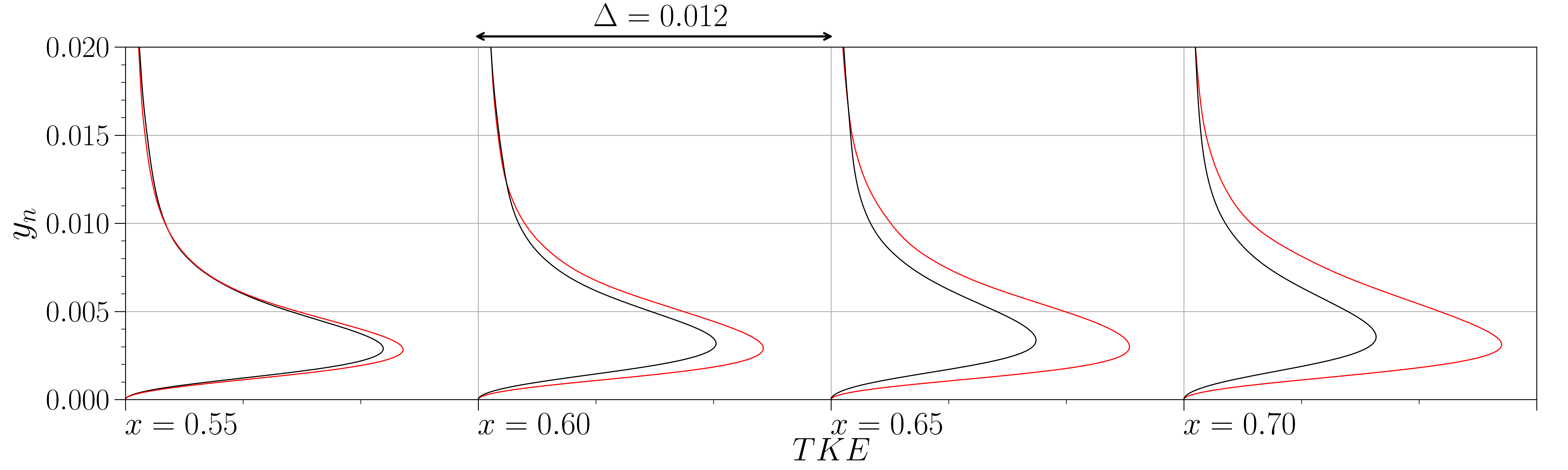}
	\put(0,30){(a)}
    \end{overpic} 

\vspace{5.1pt}

\begin{overpic}[trim = 1mm 1mm 1mm 1mm, clip,width=0.95\textwidth]{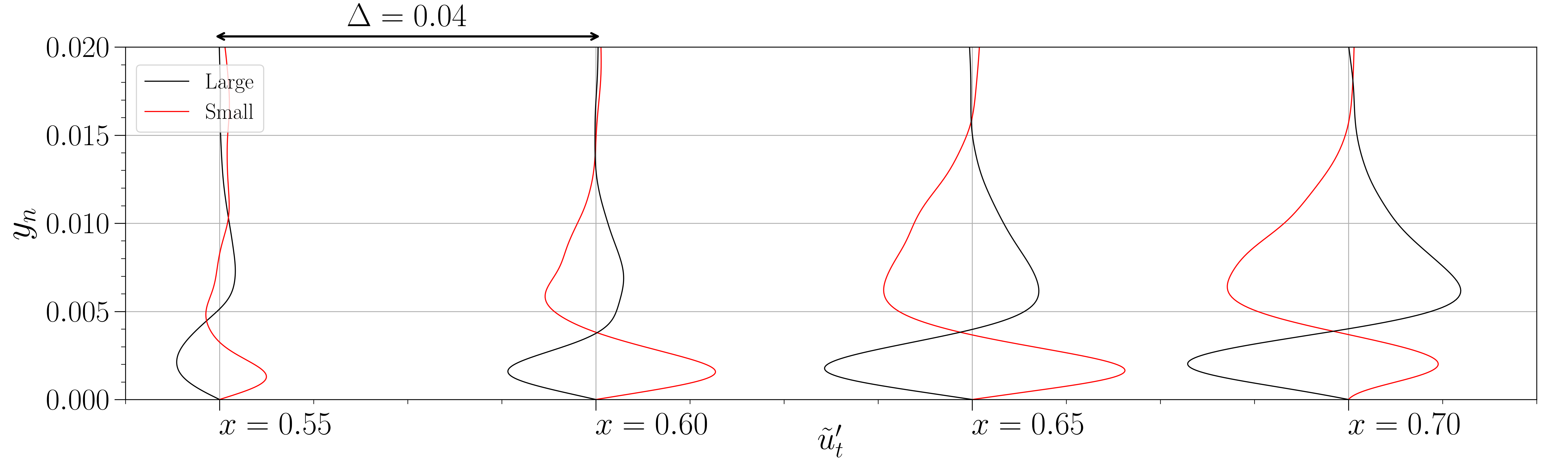}
	\put(0,30){(b)}
    \end{overpic} 

	\caption{Conditionally averaged profiles of (a) turbulence kinetic energy and (b) tangential velocity fluctuations in the incoming boundary layer for large bubble (black) and small bubble (red) events.}
	\label{fig:turbulece_statistics_profiles_upstream}
\end{figure}

To elucidate differences in the incoming boundary layer between large and small bubble events, Fig. \ref{fig:turbulece_statistics_profiles_upstream} presents conditionally averaged TKE profiles in the wall-normal direction, and  profiles of tangential velocity fluctuations for both large and small bubble events. Velocity fluctuations of the conditional events are computed as $\tilde{u}_t' = \overline{u_t - \hat{u}_t}$, where $\hat{u}_t$ represents the mean flow field computed with the entire dataset, and $\overline{(\cdot)}$ represents the conditional Reynolds averaging applied for large or small bubble events. To characterize the state of the incoming boundary layer associated with extreme bubble events, the conditional data account for the convective time delay of the fluctuation field, representing the time required for boundary layer structures to reach the bubble separation point. This approach is similar to that used by \citet{beresh2002} and \citet{porter_2019}, who studied the relationship between boundary layer velocity fluctuations and the motion of separation shock. Here, the time delay is determined between $x = 0.55$ (within the incoming boundary layer) and the respective separation points ($x = 0.739$ for large bubbles and $x = 0.782$ for small bubbles), using a convection velocity of 0.67$u_{\infty}$, estimated from the space–time correlation of our previous work \citep{lui2022}. The computed time delays are $t - 0.282 c_x/u_{\infty}$ for large-bubble events, and $t - 0.346 c_x/u_{\infty}$ for small-bubble events.

Figure \ref{fig:turbulece_statistics_profiles_upstream}(a) shows that the TKE is higher along the boundary layer during small bubble events, particularly near the wall. Moreover, conditional profiles of the tangential velocity fluctuations with time delay, $\tilde{u}_t'$, shown in Fig. \ref{fig:turbulece_statistics_profiles_upstream}(b), also exhibit substantial differences. For small bubble events, $\tilde{u}_t'$ tends to be positive near the wall, whereas it is negative for large bubble events; this behavior reverses away from the wall. These observations are consistent with previous studies \citep{ganapathisubramani_clemens_dolling_2007,baidya_2020,lui2022, lui_2024}, which indicate that the passage of near-wall high-speed streaks through the separated flow induces bubble contraction. Meanwhile, large bubble events occur when high-speed streaks are transported farther from the wall, passing over the bubble.

\begin{figure}
	\centering	
	\begin{overpic}[trim = 1mm 1mm 15mm 15mm, clip,width=0.49\textwidth]{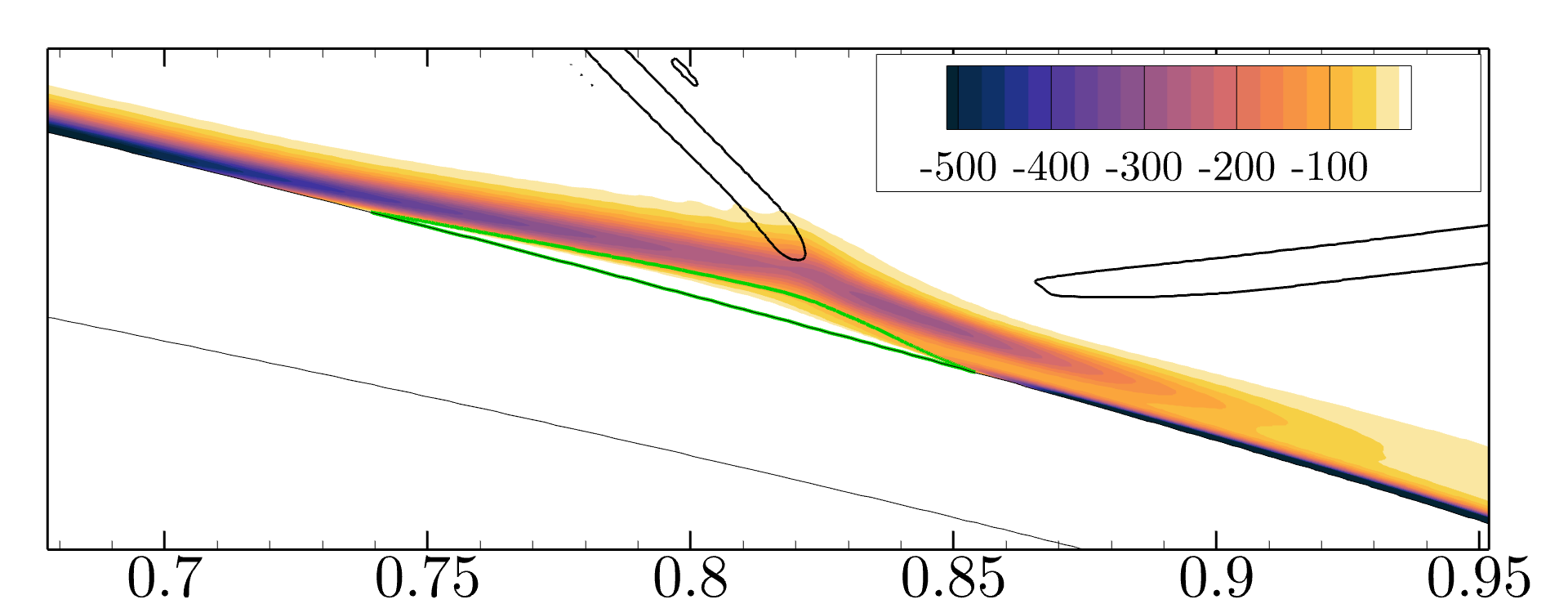}
	\put(3,8){(a)}
	\end{overpic} 
	\begin{overpic}[trim = 1mm 1mm 15mm 15mm, clip,width=0.49\textwidth]{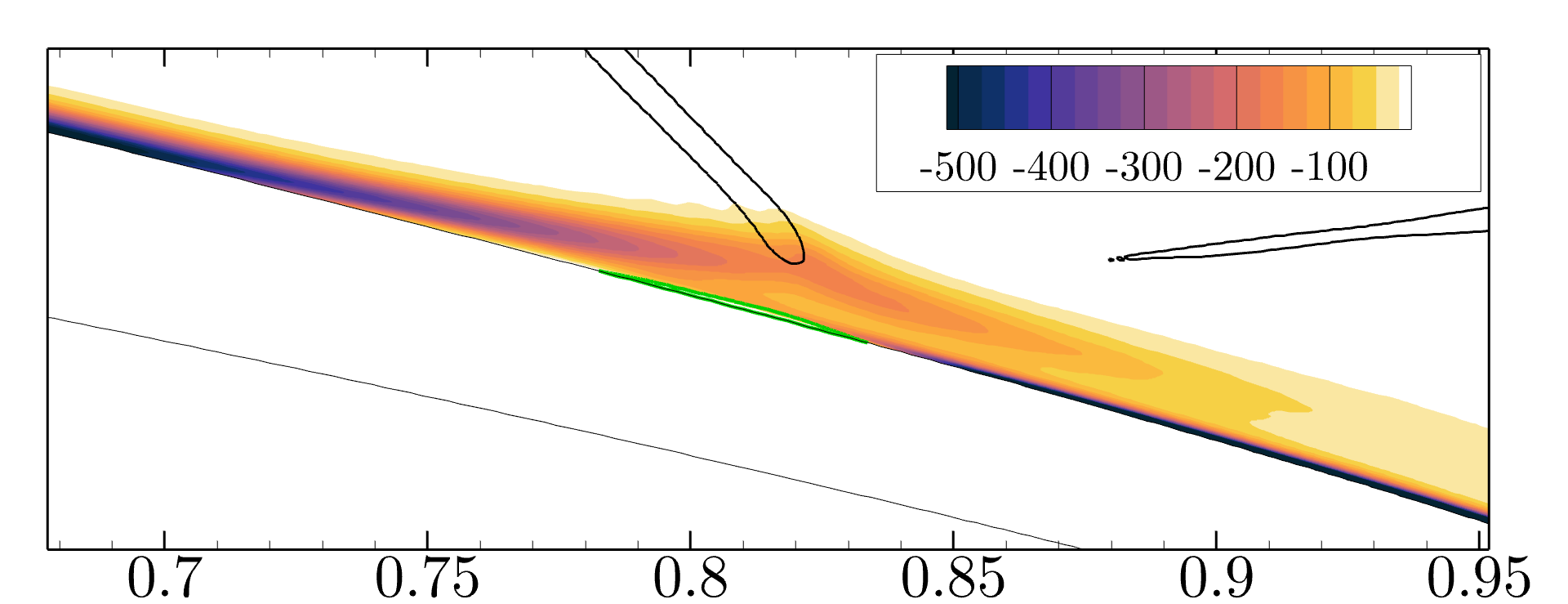}
	\put(3,8){(b)}
	\end{overpic} 
    
	\caption{Contours of conditionally averaged $z$-vorticity $\overline{\omega_z}$ for (a) large bubble events and (b) small bubble events. Results are normalized by $c_{x}/u_{\infty}$. The green lines delimit the recirculation bubble, and the black lines display the shocks through the quantity $\vec{V} \cdot \nabla p$.}
	\label{fig:z_vorticity_fields}
\end{figure}

\begin{figure}
\centering	
\begin{overpic}[trim = 1mm 1mm 1mm 1mm, clip,width=0.99\textwidth]{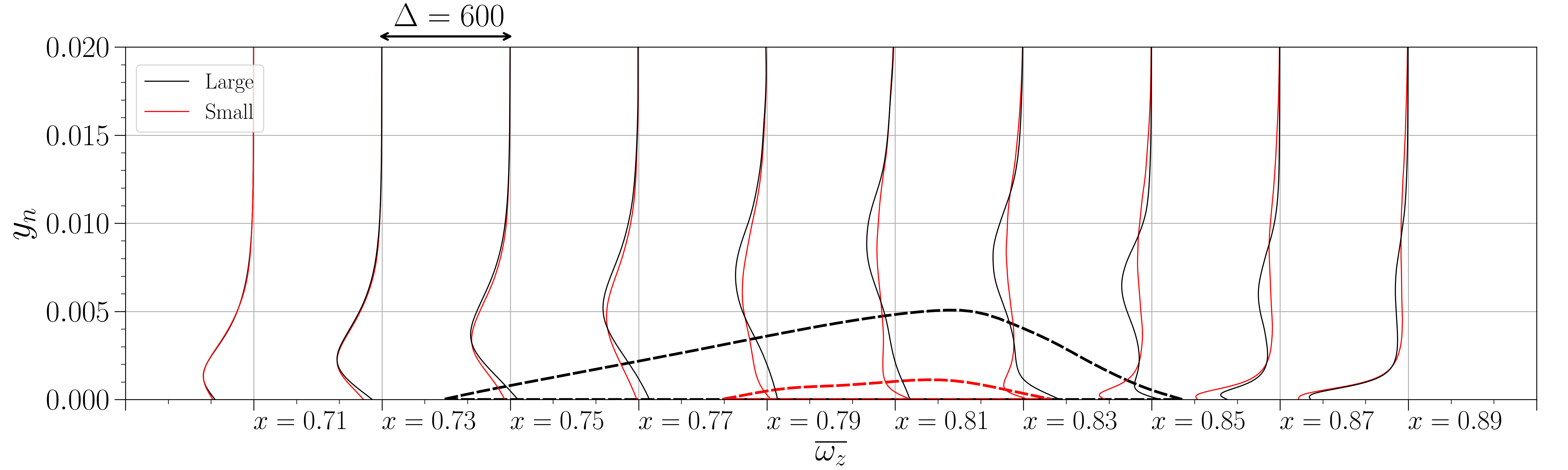}
	\end{overpic} 
	\caption{Wall-normal profiles of conditionally averaged $\overline{\omega_z}$ for large bubble events (solid black) and small bubble events (solid red). Results are normalized by $c_{x}/u_{\infty}$. The black and red dashed lines delimit the separation bubbles for the large and small bubble events, respectively.}
	\label{fig:z_vorticity_profiles}
\end{figure}

To examine the shear layer structure in large and small bubble events, contours and profiles of conditionally averaged $z$-vorticity $\overline{\omega_z}$ are shown in Figs. \ref{fig:z_vorticity_fields} and \ref{fig:z_vorticity_profiles}, respectively. The results are normalized by $c_{x}/u_{\infty}$. The green lines define the separation bubble, while the black lines represent the shock waves, identified by the quantity $\vec{V} \cdot \nabla p$. The contour plots show a region with high values of $\overline{\omega_z}$ extending above the separation bubbles because of their respective shear layers. Higher values of negative vorticity are observed for large bubble events, which can be attributed to the stronger shear between the reversed flow and the high-speed external flow. However, near the wall, larger negative vorticity values are observed for small bubble events downstream of reattachment. The higher values arise from the fuller velocity profiles observed at $x=0.85$ and $0.87$ in Fig. \ref{fig:ut_profiles}. In such cases, the near-wall vorticity is mainly based on the wall normal derivative of the streamwise velocity and, as will be shown later in Fig. \ref{fig:gradient_profiles}, the values of $\partial \langle u_t \rangle / \partial y_n$ are considerably higher for small bubble events. These observations are also supported by wall normal profiles of conditionally averaged $\overline{\omega_x}$ along the recirculation region.

Downstream of the incident shock, two high $\overline{\omega_z}$ regions are evident: the first is located away from the wall and corresponds to the free shear layer, while the second forms along the wall due to flow reattachment. As expected from the more widespread tangential velocity profiles in Fig. \ref{fig:ut_profiles}, the contours and profiles of $\overline{\omega_z}$ display lower vorticity values along the free shear layer in small bubble events, indicating a lower shear between the low-speed flow near the wall and the high-speed external flow. Instantaneous visualizations by \citet{lui_2024} showed that the free shear layer exhibits less coherence and is more diffuse when the bubble is small.

\begin{figure}[H]
	\centering	
	\begin{overpic}[trim = 1mm 1mm 15mm 15mm, clip,width=0.49\textwidth]{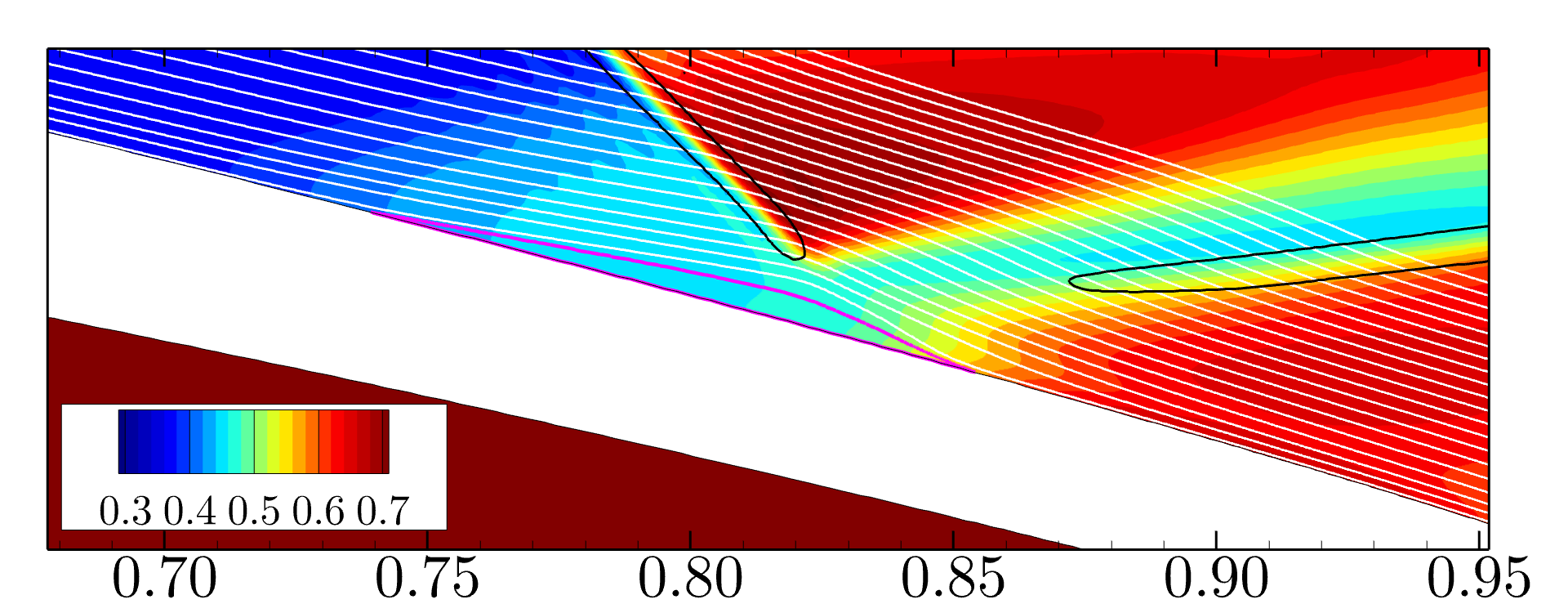}
	\put(0,38){(a)}
	\end{overpic} 
	\begin{overpic}[trim = 1mm 1mm 15mm 15mm, clip,width=0.49\textwidth]{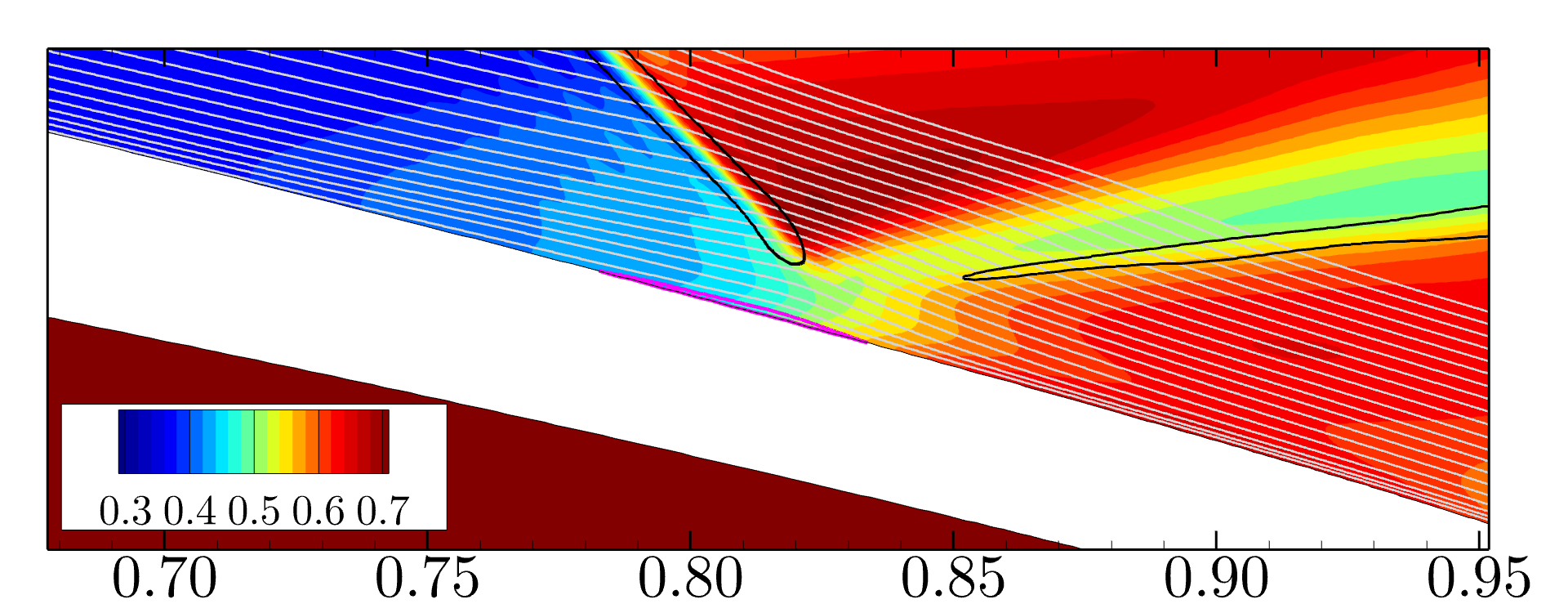}
	\put(0,38){(b)}
	\end{overpic} 
    
	\caption{Contours of conditionally averaged pressure $\overline{p}$ for (a) large bubble events and (b) small bubble events. Results are normalized by the inlet pressure. The purple lines delimit the separation bubble, and the black lines display the shocks through the quantity $\vec{V} \cdot \nabla p$. The white lines indicate the streamlines.}
	\label{fig:pressure_fields}
\end{figure}

The contours of the conditionally averaged pressure $\overline{p}$, superimposed with white streamlines, are plotted in Figs. \ref{fig:pressure_fields} (a) and (b) for large and small bubble events, respectively. The values are normalized by the inlet pressure $p_{\infty}$. The purple lines delimit the recirculation region, while the shocks are illustrated by black lines using $\vec{V} \cdot \nabla p$. The streamlines emphasize the change in flow direction caused by the curvature of the separation bubble, which is more pronounced for large bubble events. After the incident shock, the convex shape of the larger bubble induces a more pronounced flow expansion compared to that of the small bubble (wider cyan contours in Fig. \ref{fig:pressure_fields}(a)). Following this expansion, compression waves coalesce into a reattachment shock, realigning the flow tangent to the wall and causing a pressure increase.

\subsubsection{Conditionally averaged wall pressure and skin-friction coefficients}

\begin{figure}
\centering	
\begin{overpic}[trim = 1mm 1mm 1mm 1mm, clip,width=0.49\textwidth]{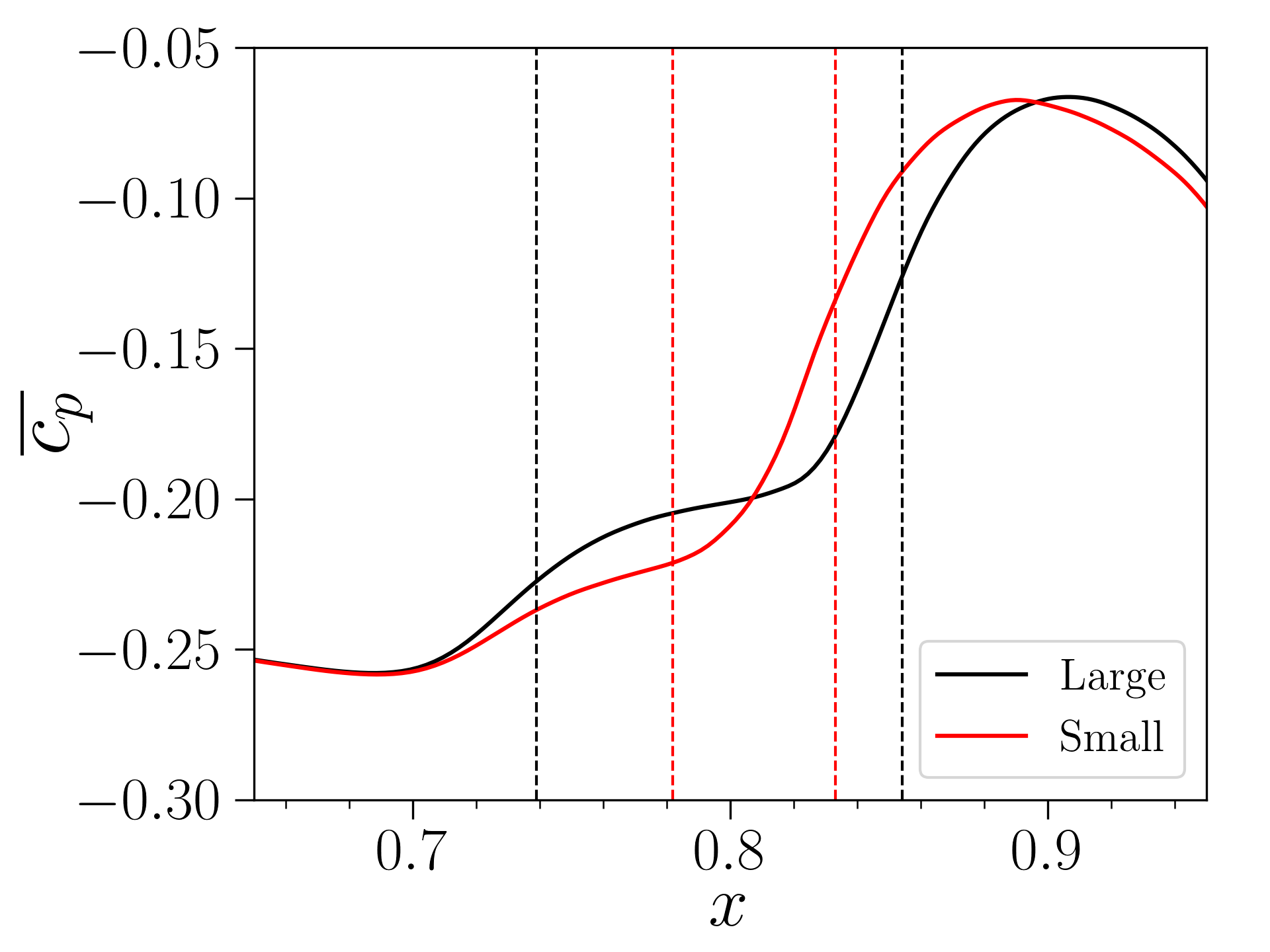}
	\put(-1,70){(a)}
    \end{overpic} 
    \begin{overpic}[trim = 1mm 1mm 1mm 1mm, clip,width=0.49\textwidth]{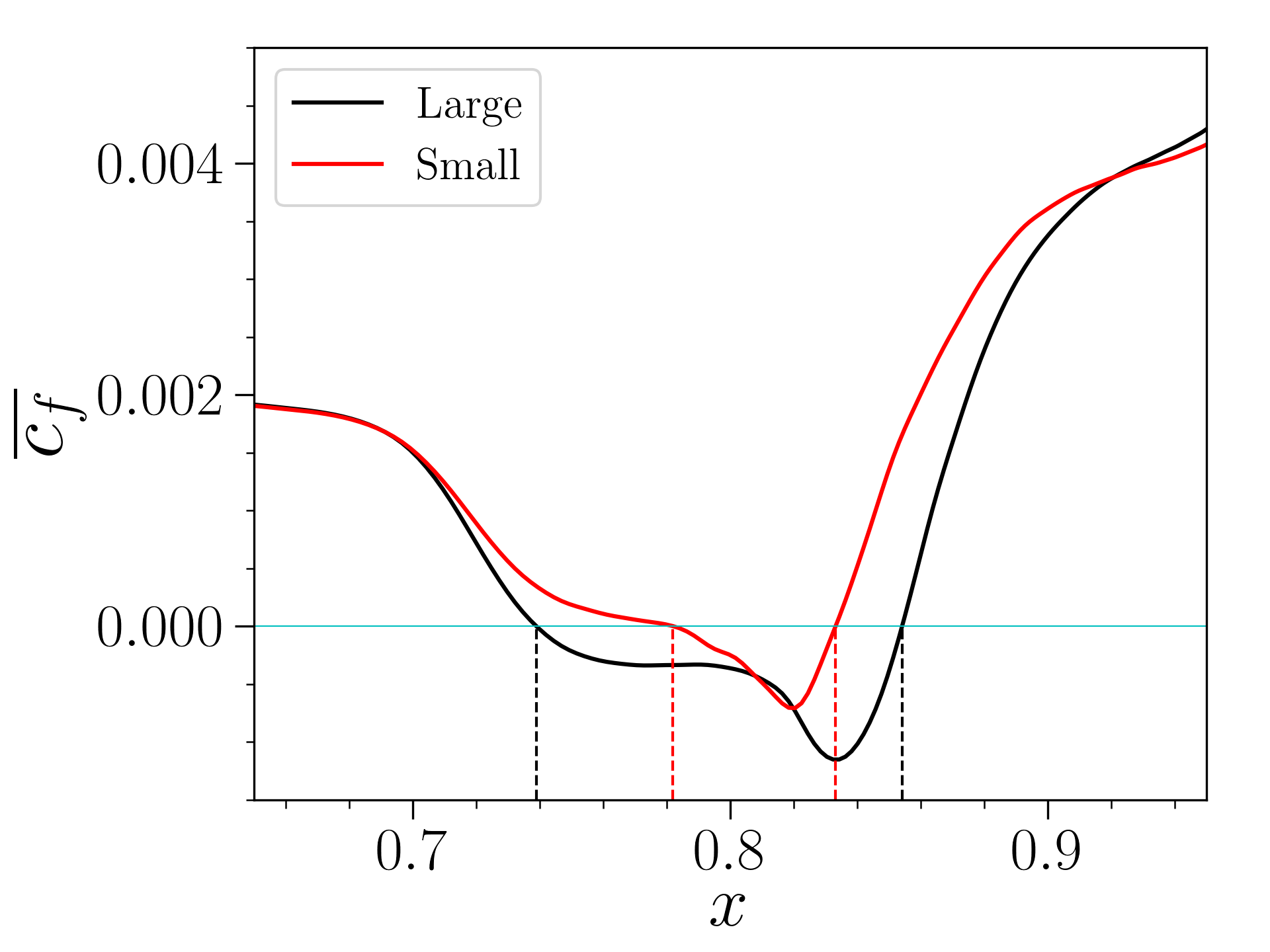}
	\put(0,70){(b)}
    \end{overpic} 
    
	\caption{Distribution of conditionally averaged (a) pressure coefficient $\overline{c_p}$ and (b) skin-friction coefficient $\overline{c_f}$ for large bubble events (solid black) and small bubble events (solid red). The vertical black and red dashed lines delimit the mean separation regions for the large and small bubble events, respectively.}
	\label{fig:wall_quantities}
\end{figure}

The key differences in wall-measured quantities between large and small bubble events are analyzed through the conditionally averaged pressure coefficient $\overline{c_p} = \frac{\overline{p_{w}} - p_{\infty}}{0.5 \rho_{\infty} u_{\infty}^2}$ and skin-friction coefficient $\overline{c_f} = \frac{\overline{\tau_{w}}}{0.5 \rho_{\infty} u_{\infty}^2}$. Figure \ref{fig:wall_quantities}(a) shows the $\overline{c_p}$ distribution along a portion of the suction side, including the SBLI region. The incoming boundary layer develops under a favorable pressure gradient, with similar values of $\overline{c_p}$ observed for both large and small bubble events. For both extreme events, two regions of pressure rise are identified. The first occurs upstream of the bubble due to compression waves, with higher wall pressure values observed for the large bubble events caused by the stronger flow compression associated with the larger separation bubble. The second pressure rise results from the incident shock and occurs earlier (shifted upstream) in small bubble events, as the flow reattaches sooner (see Fig. \ref{fig:pressure_fields}). After flow reattachment, the wall pressure continues to increase, associated with the reattachment shock. Both extreme events reach similar maximum $\overline{c_p}$ values, though at different locations, with the peak occurring farther downstream in large bubble events. After the peak, the wall pressure begins to decrease again in both extreme events as the flow expands toward the trailing edge.

Figure \ref{fig:wall_quantities}(b) shows the distribution of the conditionally averaged skin-friction coefficient $\overline{c_f}$, where $\overline{\tau_{w}}$ represents the wall shear stress. Similar values of $\overline{c_f}$ are observed in the incoming boundary layer, but after entering the compression zone, there is a significant drop in $\overline{c_f}$ accompanied by an increase in $\overline{c_p}$, which is associated with boundary layer deceleration. The decrease in $\overline{c_f}$ is more pronounced for large bubble events. Due to flow reattachment, there is a significant increase in $\overline{c_f}$, reaching values higher than those of the incoming boundary layer. The $\overline{c_f}$ distribution also allows for the quantification of the conditionally averaged separation bubble length $\overline{L_{SB}}$, characterized by locations where $\overline{c_f} < 0$.  For small bubble events, $\overline{L_{SB}} = 0.051$, and for large bubble events, $\overline{L_{SB}} = 0.115$. 

\begin{figure}
	\centering	
	\begin{overpic}[trim = 1mm 1mm 1mm 1mm, clip,width=0.49\textwidth]{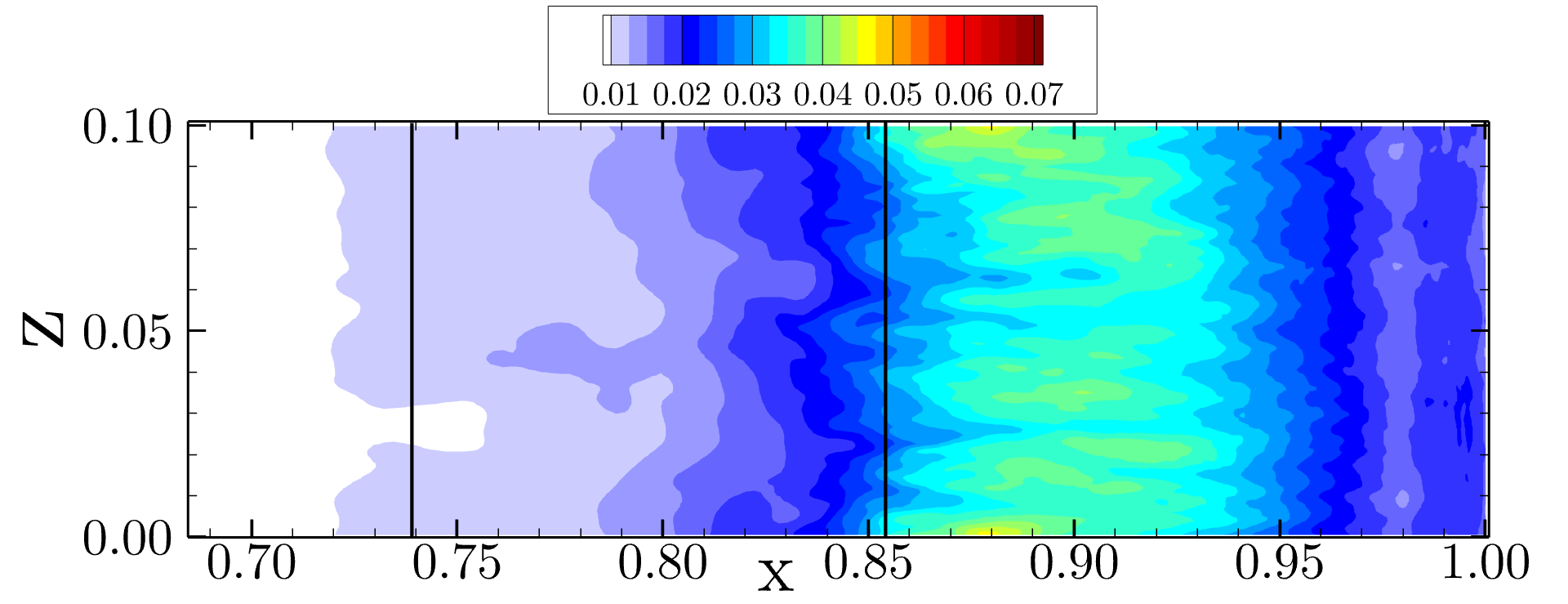}
	\put(1,34){(a)}
	\end{overpic} 
	\begin{overpic}[trim = 1mm 1mm 1mm 1mm, clip,width=0.49\textwidth]{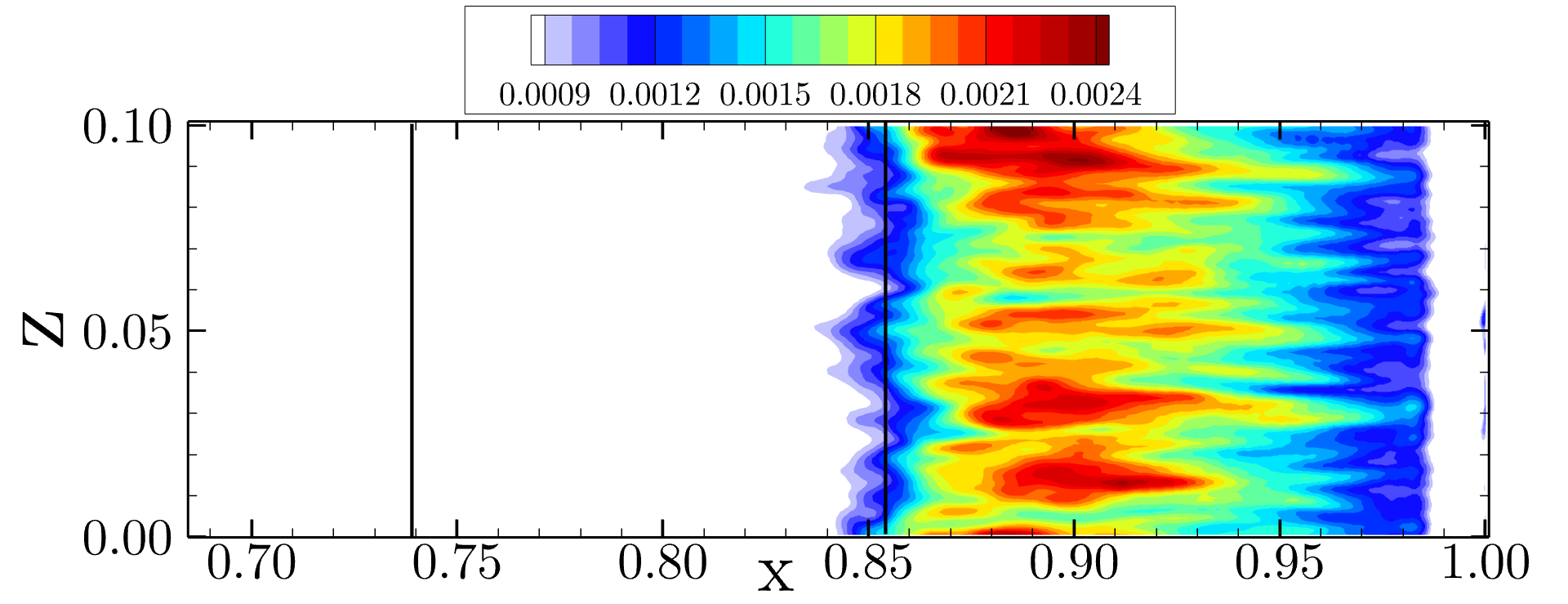}
	\put(1,34){(b)}
	\end{overpic} 

    \begin{overpic}[trim = 1mm 1mm 1mm 1mm, clip,width=0.49\textwidth]{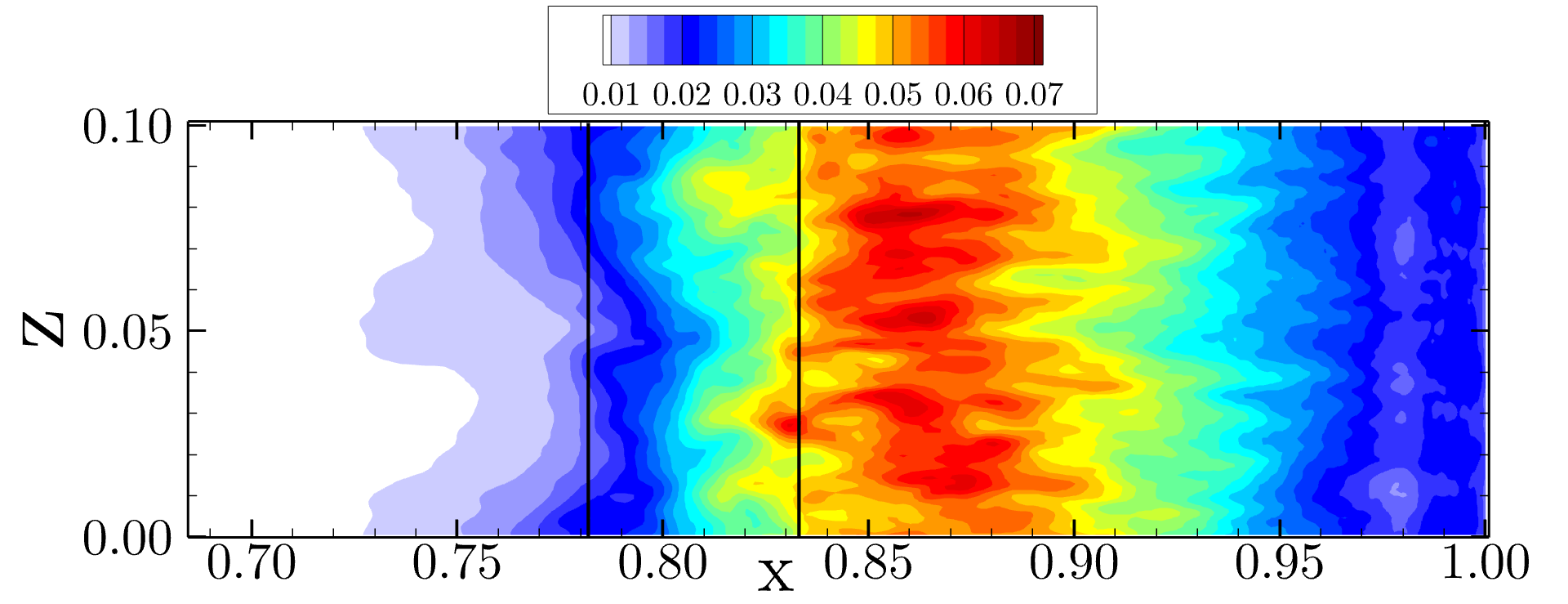}
	\put(1,34){(c)}
	\end{overpic} 
	\begin{overpic}[trim = 1mm 1mm 1mm 1mm, clip,width=0.49\textwidth]{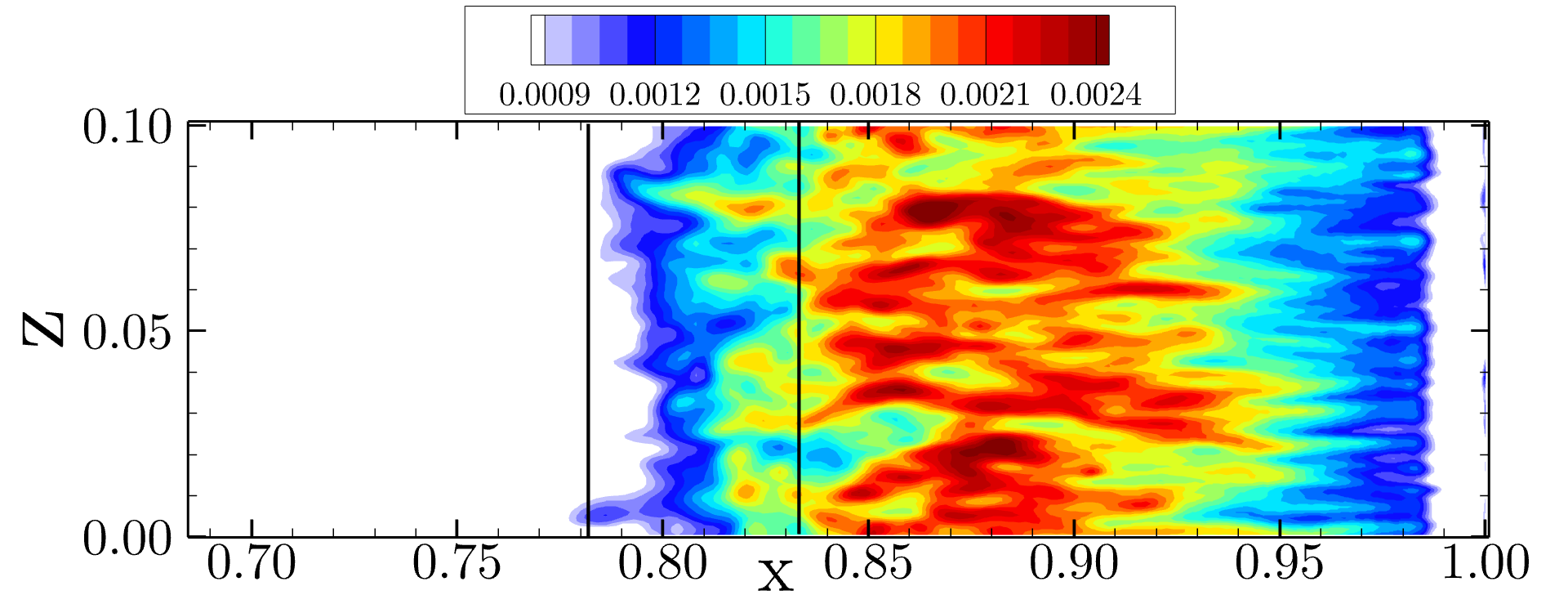}
	\put(1,34){(d)}
	\end{overpic} 
    
	\caption{Contours of conditionally averaged root-mean-square (a,c) normalized wall pressure and (b,d) skin-friction coefficient for large bubble events (top) and small bubble events (bottom). The vertical black lines represent the conditionally averaged separation and reattachment locations.}
	\label{fig:rms}
\end{figure}

The effect of bubble size on the root-mean-square (RMS) values of normalized wall pressure $p_{RMS} = \sqrt{\overline{p'p'}}/p_{\infty}^2$ and skin-friction coefficient $c_{f,RMS} = \sqrt{\overline{c_f' c_f'}}$ is analyzed in Fig. \ref{fig:rms}. In this analysis, the averaging operation is performed only over time, with the mean field obtained through conditional time averaging at each spatial location. For reference, the vertical black lines indicate the conditionally averaged separation and reattachment locations. The plots on the left, Figs. \ref{fig:rms}(a) and \ref{fig:rms}(c), show the conditionally averaged RMS values of normalized wall pressure for the large and small bubble events, respectively. A key observation is that significantly higher values of $p_{RMS}$ are found along the separation region and downstream of reattachment for small bubble events. In both extreme events, maximum values of $p_{RMS}$ are observed downstream of the flow reattachment, with strong fluctuations also occurring upstream in small bubble events as the flow reattaches earlier. The region of maximum $p_{RMS}$ shows traces of longitudinal turbulent structures (streaks), as observed in previous studies \citep{trofimov_2011,GRILLI2013,priebe_2016,pasquariello_2017,jenquin_2023}. 
Streaky structures are also evident in the contours of the conditionally averaged RMS values of skin-friction coefficient for both large and small bubble events, as shown in Figs. \ref{fig:rms}(b) and \ref{fig:rms}(d), respectively. These structures appear more streamwise-oriented in the large bubble events. Similar maximum values of $c_{f,RMS}$ are found downstream of reattachment for both extreme events. Another key observation from these figures is that significant values of $c_{f,RMS}$ are observed along the separation region for small bubble events, whereas no significant values are seen in this region for large bubble events. This suggests that wall shear stress varies more intensely during small bubble events, while it remains more steady for large bubbles.

\subsubsection{Conditionally averaged turbulence quantities}

\begin{figure}
	\centering	
	\begin{overpic}[trim = 1mm 1mm 15mm 15mm, clip,width=0.49\textwidth]{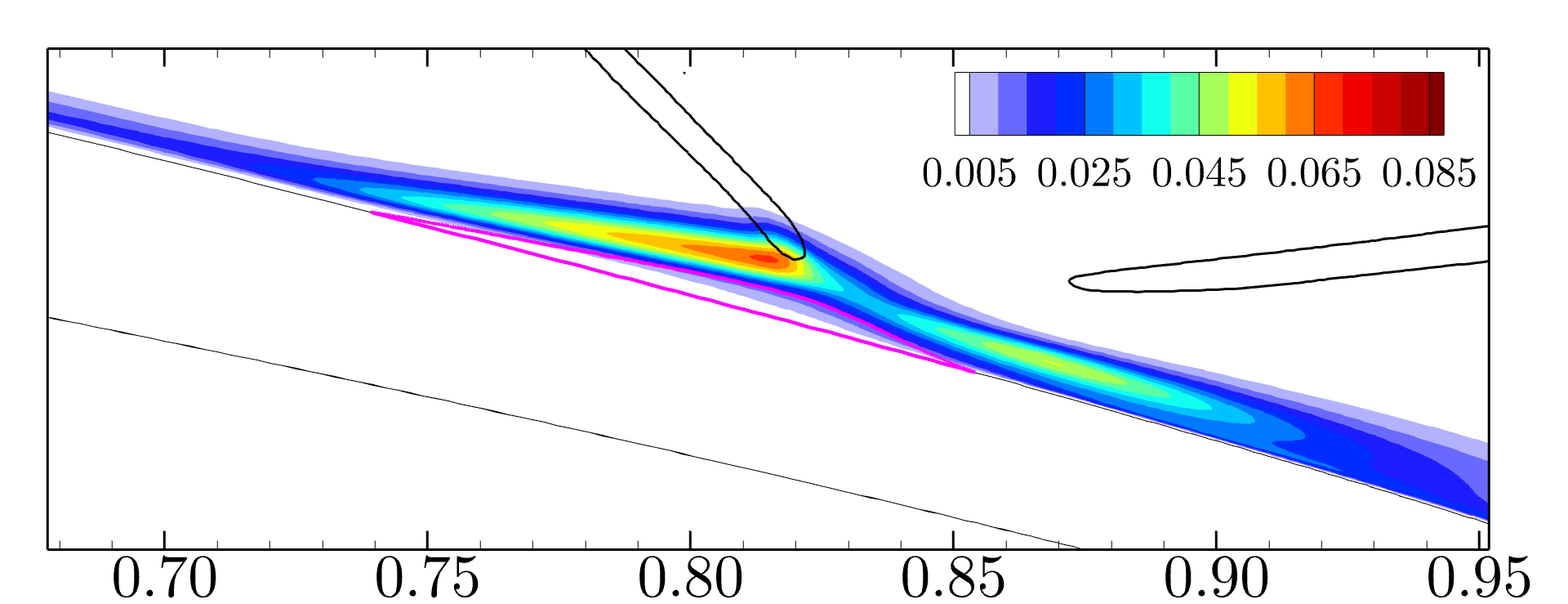}
	\put(5,8){(a)}
	\end{overpic} 
	\begin{overpic}[trim = 1mm 1mm 15mm 15mm, clip,width=0.49\textwidth]{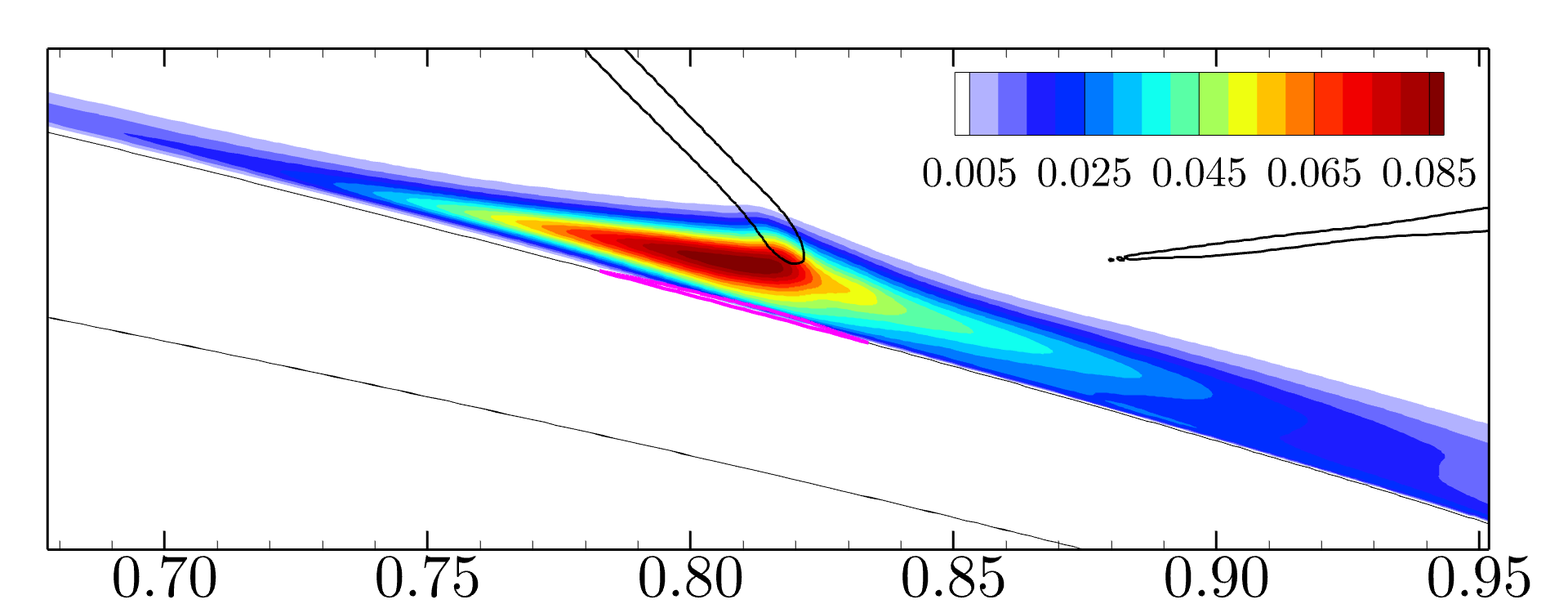}
	\put(5,8){(b)}
	\end{overpic} 

    \begin{overpic}[trim = 1mm 1mm 15mm 15mm, clip,width=0.49\textwidth]{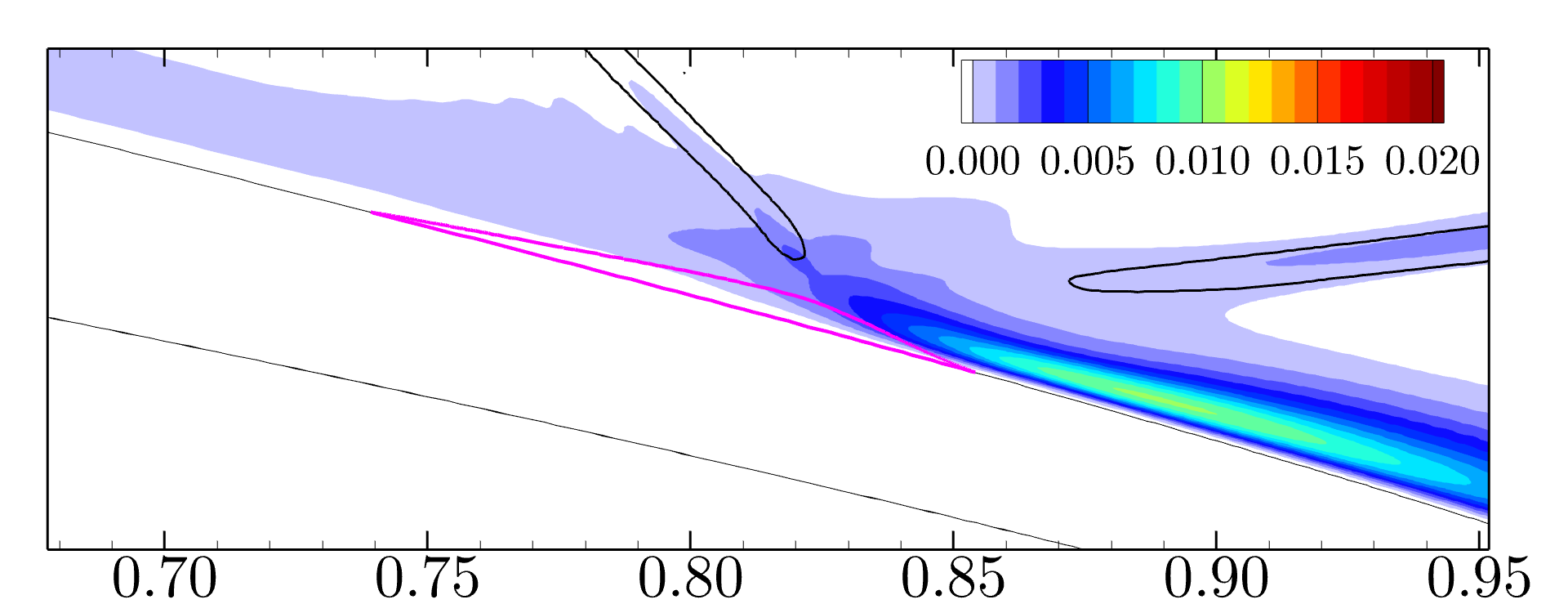}
	\put(5,8){(c)}
	\end{overpic} 
\begin{overpic}[trim = 1mm 1mm 15mm 15mm, clip,width=0.49\textwidth]{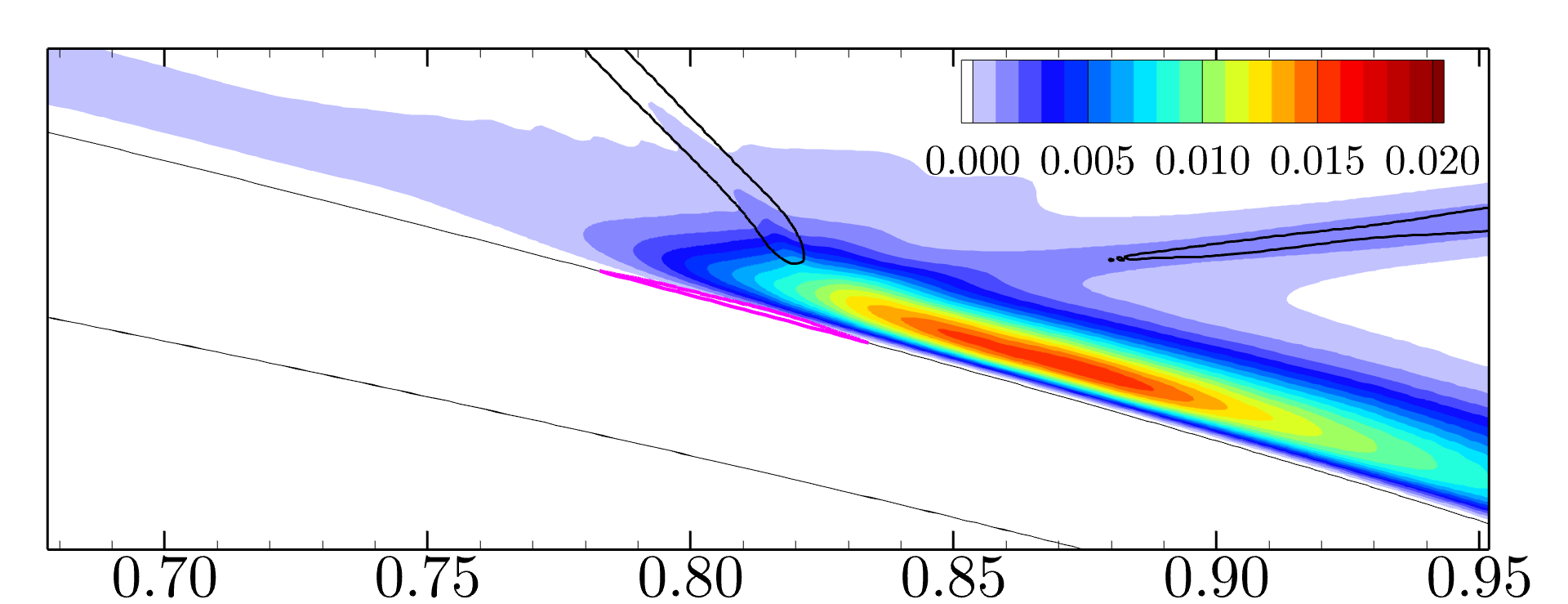}
	\put(5,8){(d)}
	\end{overpic} 

    \begin{overpic}[trim = 1mm 1mm 15mm 15mm, clip,width=0.49\textwidth]{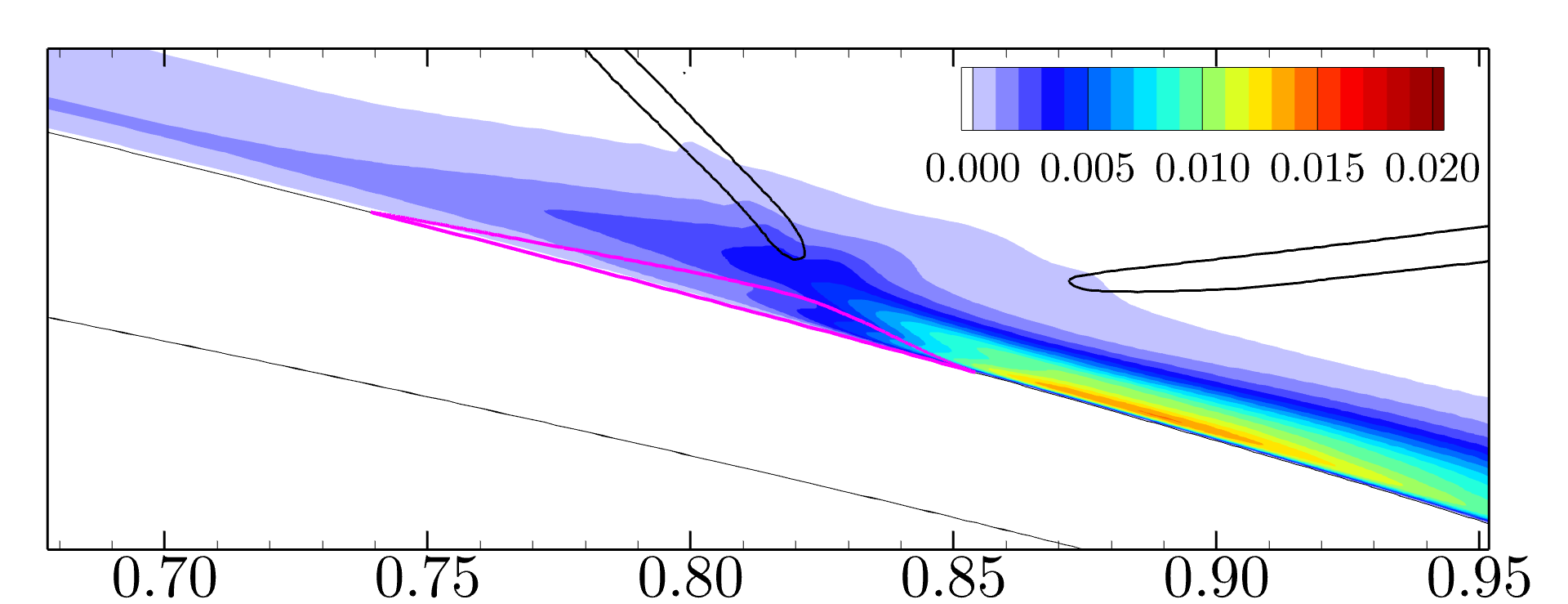}
	\put(5,8){(e)}
	\end{overpic} 
\begin{overpic}[trim = 1mm 1mm 15mm 15mm, clip,width=0.49\textwidth]{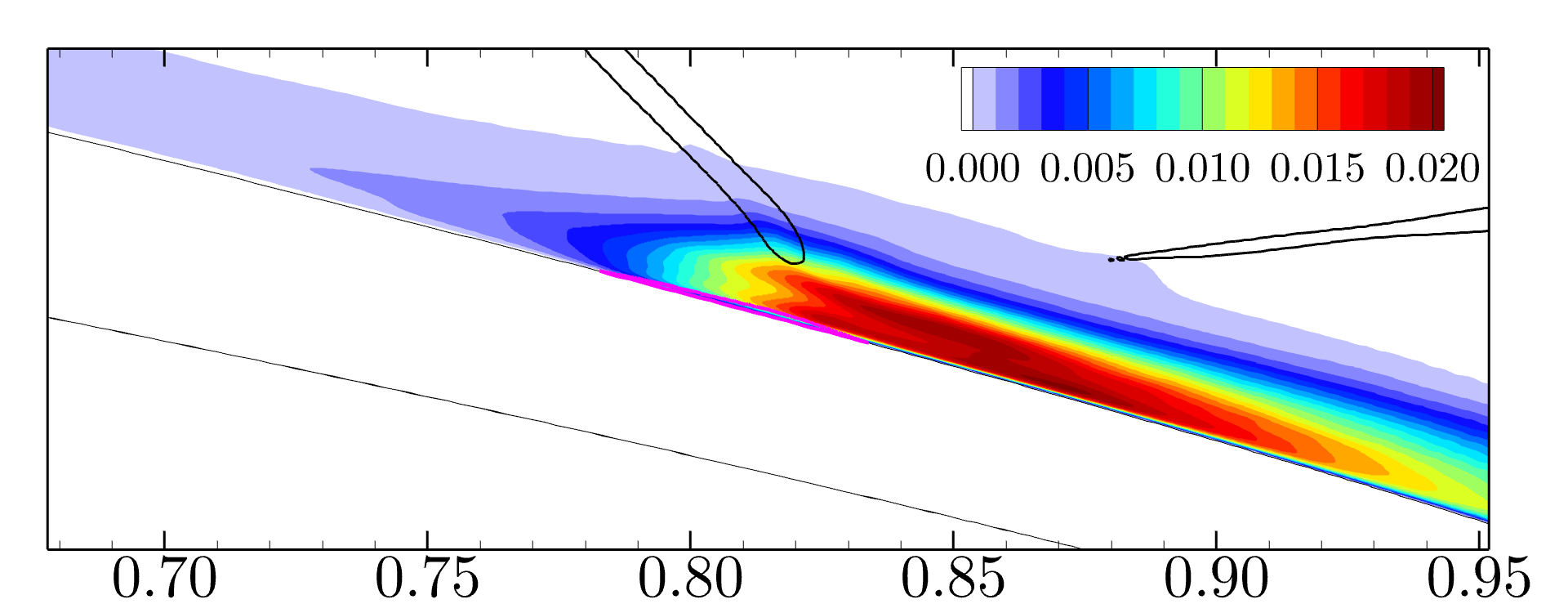}
	\put(5,8){(f)}
	\end{overpic} 

        \begin{overpic}[trim = 1mm 1mm 15mm 15mm, clip,width=0.49\textwidth]{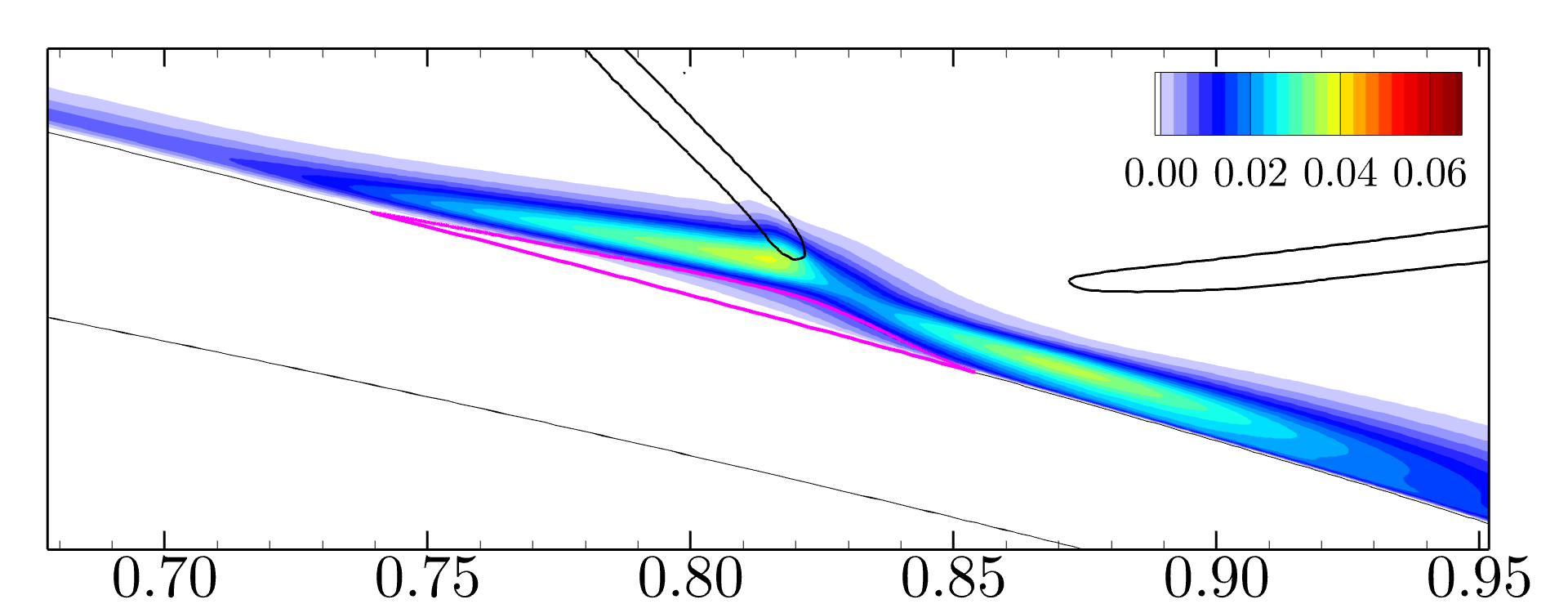}
	\put(5,8){(g)}
	\end{overpic} 
\begin{overpic}[trim = 1mm 1mm 15mm 20mm, clip,width=0.49\textwidth]{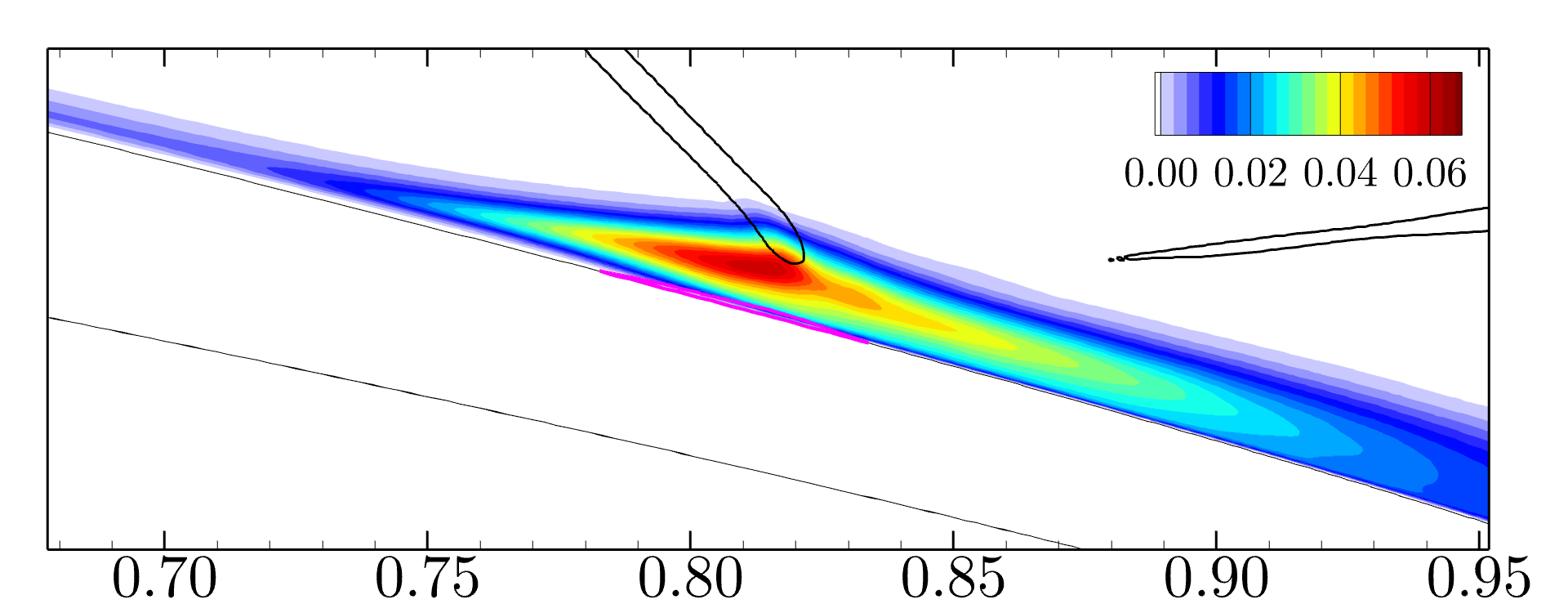}
	\put(5,8){(h)}
	\end{overpic} 

        \begin{overpic}[trim = 1mm 1mm 15mm 15mm, clip,width=0.49\textwidth]{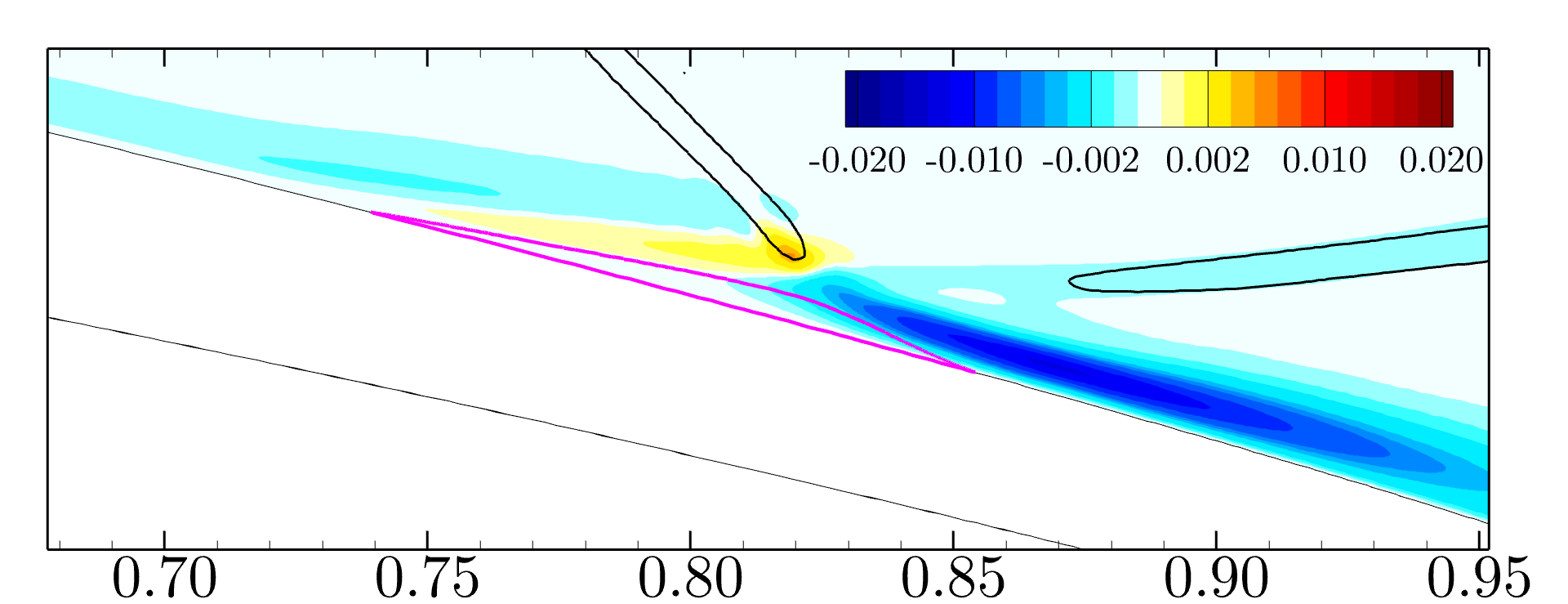}
	\put(5,8){(i)}
	\end{overpic} 
\begin{overpic}[trim = 1mm 1mm 15mm 15mm, clip,width=0.49\textwidth]{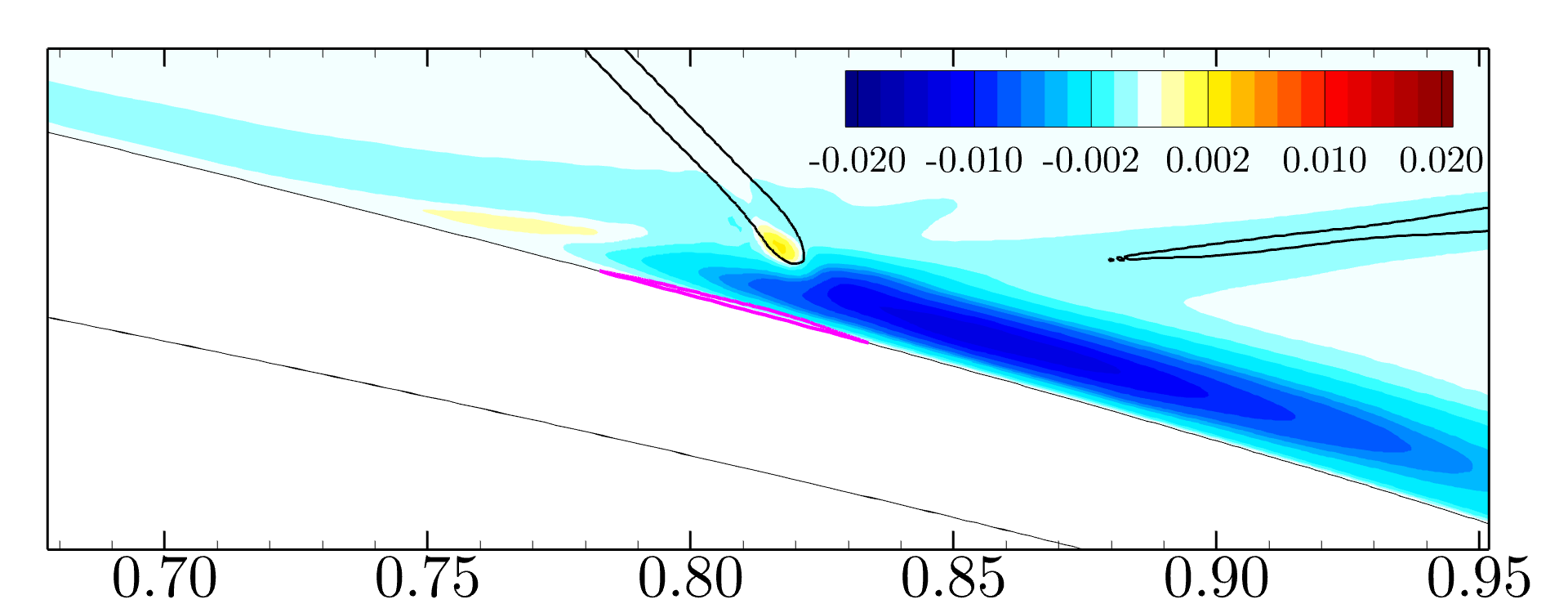}
	\put(5,8){(j)}
	\end{overpic} 
    
	\caption{Contours of conditionally averaged turbulence quantities for (left) large bubble events and (right) small bubble events: (a,b) $\langle u_t''u_t'' \rangle$, (c,d) $\langle u_n''u_n''\rangle$, (e,f) $\langle w''w'' \rangle$, (g,h) TKE, and (i,j) $\langle u_t''u_n'' \rangle$. All variables are normalized by $u_{\infty}^2$. The purple lines delimit the separation bubbles, and the black lines display the shock waves through the quantity $\vec{V} \cdot \nabla P$.}
	\label{fig:turbulece_statistics}
\end{figure}

\begin{figure}
\centering	
\begin{overpic}[trim = 1mm 1mm 1mm 1mm, clip,width=0.79\textwidth]{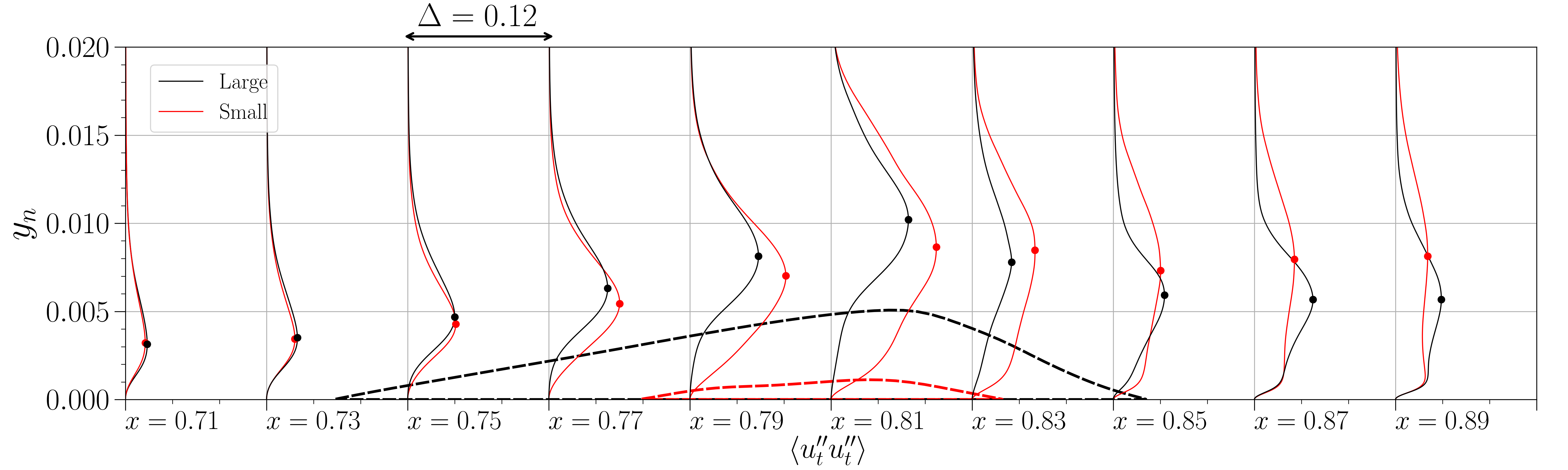}
	\put(0,30){(a)}
    \end{overpic} 

\vspace{5.1pt}

\begin{overpic}[trim = 1mm 1mm 1mm 1mm, clip,width=0.79\textwidth]{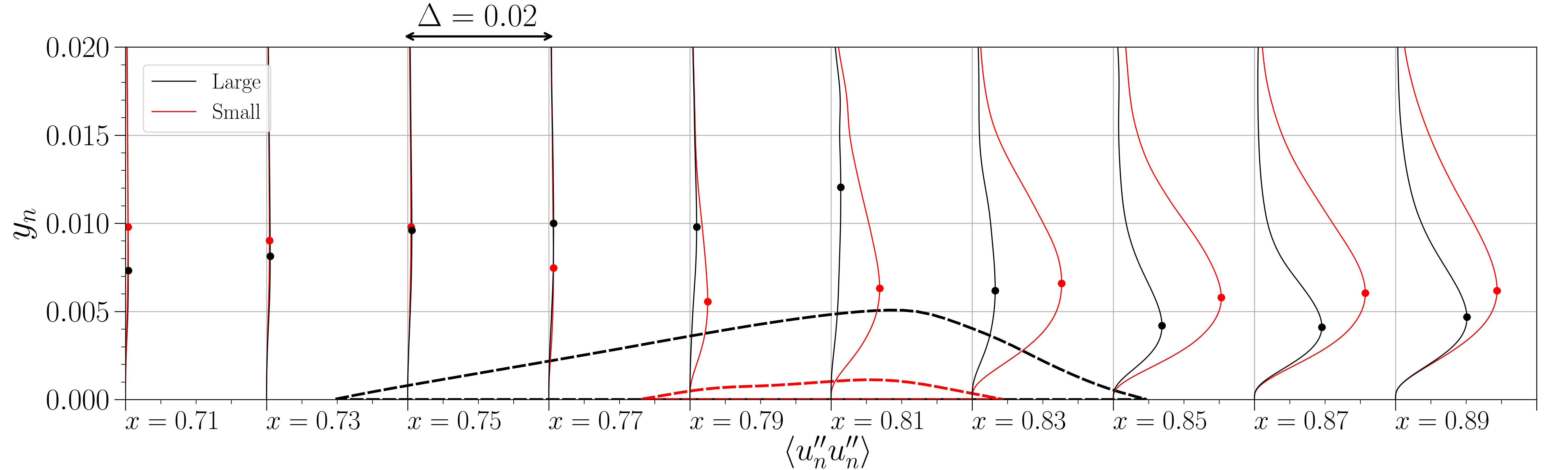}
	\put(0,30){(b)}
    \end{overpic} 

\vspace{5.1pt}

\begin{overpic}[trim = 1mm 1mm 1mm 1mm, clip,width=0.79\textwidth]{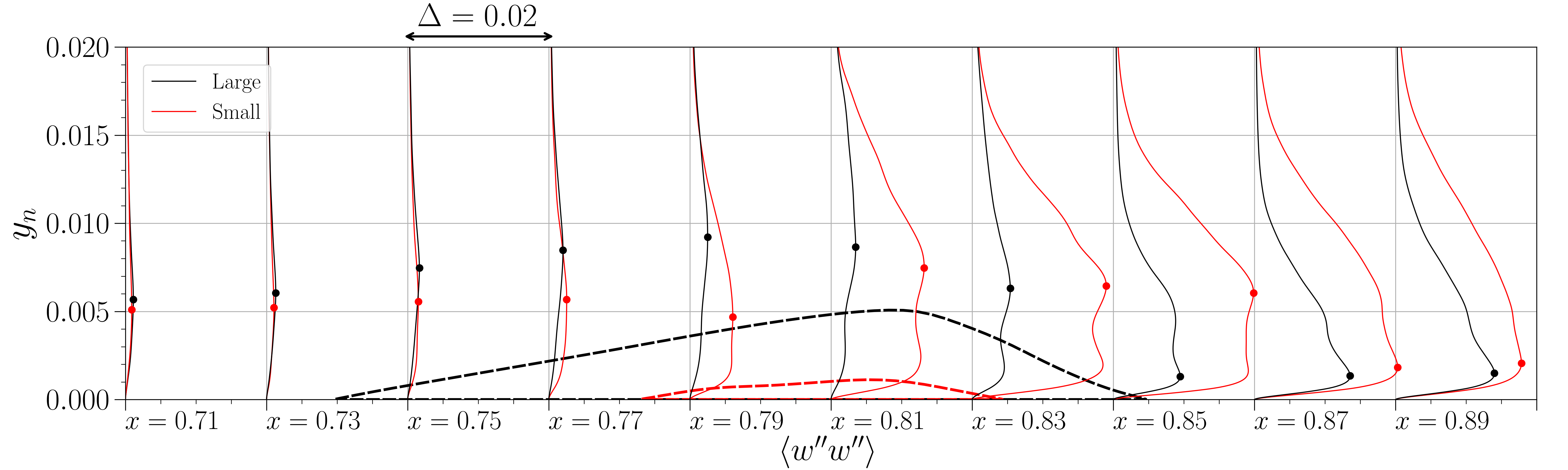}
	\put(0,30){(c)}
    \end{overpic} 

\vspace{5.1pt}

\begin{overpic}[trim = 1mm 1mm 1mm 1mm, clip,width=0.79\textwidth]{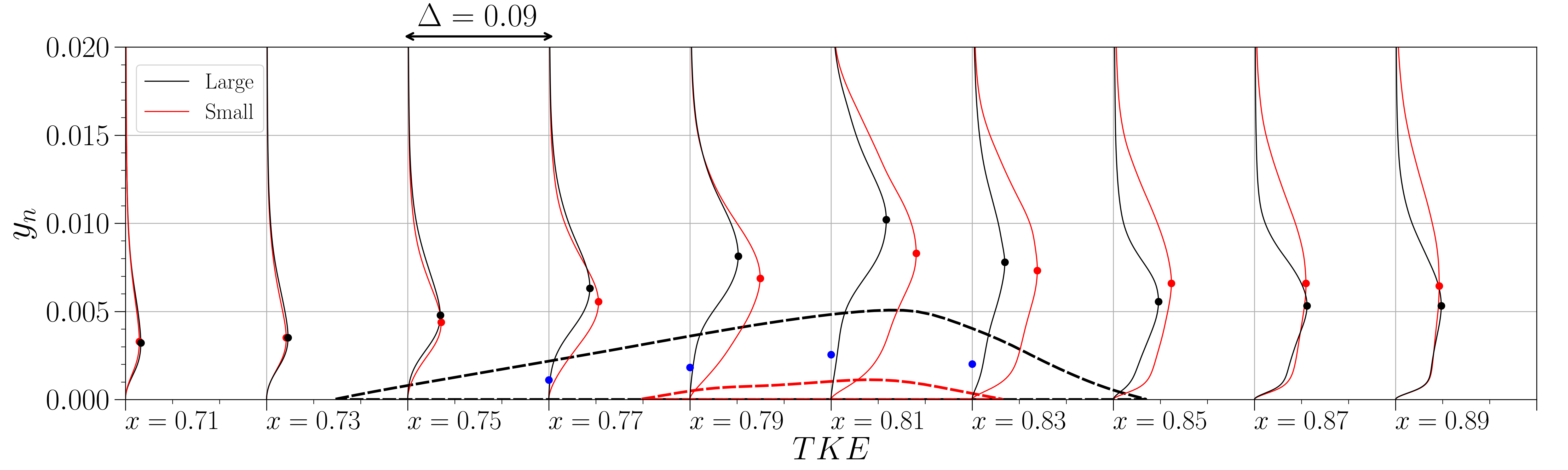}
	\put(0,30){(d)}
    \end{overpic} 

\vspace{5.1pt}

\begin{overpic}[trim = 1mm 1mm 1mm 1mm, clip,width=0.79\textwidth]{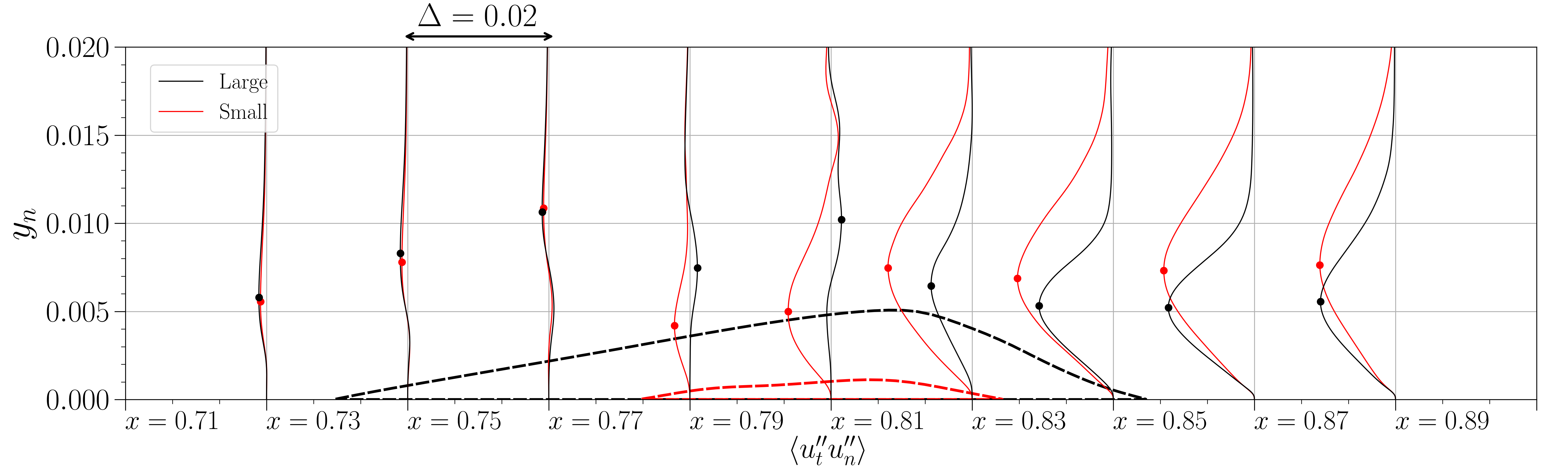}
	\put(0,30){(e)}
\end{overpic} 

	\caption{Profiles of conditionally averaged turbulence quantities for large bubble events (solid black) and small bubble events (solid red): (a) $\langle u_t''u_t'' \rangle$, (b) $\langle u_n''u_n'' \rangle$,  (c)$\langle w''w'' \rangle$, (d) $TKE$, and (e) $\langle u_t''u_n'' \rangle$. All quantities are normalized by $u_{\infty}^2$. The black and red dashed lines delimit the separation bubble for large and small bubble events, respectively. The circles mark the peak values of the corresponding quantities.}
	\label{fig:turbulece_statistics_profiles}
\end{figure}

The impact of extreme events on the Reynolds stress components and TKE is analyzed in Figs. \ref{fig:turbulece_statistics} and \ref{fig:turbulece_statistics_profiles}. These turbulence quantities are normalized by $u_{\infty}^2$. In the contour plots shown in Fig. \ref{fig:turbulece_statistics}, the left column displays the conditionally averaged quantities for large bubble events, while the right column presents the results for small bubble events. The black lines represent the shock waves computed using the quantity $\vec{V} \cdot \nabla P$, and the purple lines delimit the recirculation bubbles. The profiles of conditionally averaged turbulence quantities shown in Fig. \ref{fig:turbulece_statistics_profiles} highlight the spatial evolution of these quantities. For reference, the black and red
dashed lines delimit the recirculation bubbles for large and small bubble events, respectively. The colored circles mark the locations of the maximum values for the corresponding quantities.

In Figs. \ref{fig:turbulece_statistics}(a) and (b), the tangential Reynolds stress component $\langle u_t''u_t'' \rangle$ exhibits high values along the shear layer, extending over the bubble and upstream of the incident shock in both extreme events, with maximum values occurring near the foot of the incident shock. Significantly higher values of $\langle u_t''u_t'' \rangle$ are observed in small bubble events. Large values are also present along the free shear layer in both extreme events, most likely because of the shedding of energetic structures. 

In the profiles of $\langle u_t''u_t'' \rangle$, shown in Fig. \ref{fig:turbulece_statistics_profiles}(a), very small values of $\langle u_t''u_t'' \rangle$ are observed within the separation region for large bubble events. In contrast, for small bubble events, higher values of $\langle u_t''u_t'' \rangle$ appear near the wall along the SBLI region, possibly due to the passage of near-wall streaks, as illustrated in Fig. \ref{fig:bubble_3D}. In both extreme events, the peaks of $\langle u_t''u_t'' \rangle$ in the SBLI region occur along the shear layer. Downstream of reattachment, these peaks also follow the core of the free shear layer, with slightly higher peak values observed in large bubble events.

For the contours of the wall-normal Reynolds stress component $\langle u_n''u_n''\rangle$, shown in Figs. \ref{fig:turbulece_statistics}(c) and (d), the highest fluctuations in both extreme events occur along the free shear layer, indicating strong turbulent motion in the wall-normal direction, likely caused by vortical structures, with higher values observed in small bubble events. Significant values of $\langle u_n''u_n'' \rangle$ are also present along the reattachment shock, associated with its unsteady motion. 
The profiles of $\langle u_n''u_n''\rangle$, shown in Fig. \ref{fig:turbulece_statistics_profiles}(b), indicate that $\langle u_n''u_n''\rangle$ begins to amplify earlier in small bubble events, occurring at the bubble leading edge. Significant values are observed within the shear layer over the bubble, which indicates strong turbulent mixing in the wall-normal direction. In contrast, in large bubble events, $\langle u_n''u_n''\rangle$ starts to amplify only after flow reattachment, closer to the wall. Similar to the $\langle u_t''u_t'' \rangle$ results, the peaks follow the core of the free shear layer; however, higher peak values are observed in small bubble events.

The contours of the spanwise Reynolds stress component $\langle w''w'' \rangle$, presented in Figs. \ref{fig:turbulece_statistics}(e) and \ref{fig:turbulece_statistics}(f), show more pronounced values of this quantity in small bubble events, particularly along the free shear layer and downstream of reattachment, near the wall. Similar to the behavior of $\langle u_n''u_n''\rangle$, Fig. \ref{fig:turbulece_statistics_profiles}(c) indicates that in small bubble events, $\langle w''w'' \rangle$ amplifies earlier, beginning at the bubble leading edge and reaching significant levels within the shear layer over the bubble. This suggests an intense mixing in the spanwise direction. Conversely, in large bubble events, $\langle w''w'' \rangle$ begins to amplify only after flow reattachment. Further downstream, intense values of $\langle w''w'' \rangle$ are observed near the wall, with higher magnitudes in small bubble events. \citet{pasquariello_2017} suggested that the amplification of $\langle w''w'' \rangle$ downstream of the reattachment, near the wall, is associated with the presence of streamwise vortices.

The turbulent kinetic energy, given by TKE $= 0.5 \left( \langle u_t''u_t'' \rangle + \langle u_n''u_n'' \rangle + \langle w''w'' \rangle \right)$, is presented as contour plots in Figs. \ref{fig:turbulece_statistics}(g) and (h). Turbulence amplification in the SBLI region is evident in both extreme events. Upstream of the incident shock, the $\langle u_t''u_t'' \rangle$ component dominates, with maximum TKE values observed along the shear layer. Downstream of the incident shock, all Reynolds stress components contribute to the TKE, but with $\langle u_t''u_t'' \rangle$ still providing the primary contribution. Significant TKE values are found along the free shear layer and downstream of reattachment near the wall. The profiles of TKE, shown in Fig. \ref{fig:turbulece_statistics_profiles}(d), highlight higher TKE values along the bubble and within the shear layer in small bubble events, especially near the wall. The intense turbulence activity in this region will be associated in a later section with the presence of streaks and streamwise vortices.

The contours of the shear Reynolds stress component $\langle u_t''u_n'' \rangle$ are plotted in Figs. \ref{fig:turbulece_statistics}(i) and (j). For large bubble events, positive values of $\langle u_t''u_n'' \rangle$ appear along the shear layer before the incident shock, while negative values are observed after the shock, over the bubble. In contrast, for small bubble events, only negative values of $\langle u_t''u_n'' \rangle$ are present over the bubble. These observations are further illustrated in the profiles of $\langle u_t''u_n'' \rangle$, shown in Fig. \ref{fig:turbulece_statistics_profiles}(e). The highest negative fluctuations are observed along the free shear layer in both extreme events.
Significant negative values of $\langle u_t''u_n'' \rangle$ suggest that low-momentum fluid is transported upward (ejections), while high-momentum fluid is transported downward (sweeps). This fluid transport could be associated with streamwise vortices, which facilitate the movement of low-momentum flow away from the wall and high-momentum flow toward it. Therefore, the results suggest the presence of vortices over the entire bubble in small bubble events, whereas in large bubble events, these vortical structures are more likely to appear near reattachment. Along the free shear layer, this fluid transport is more intense in both extreme events.

To further investigate the differences in the TKE values between extreme events, the production of turbulent kinetic energy $P$ is analyzed. Here, we split the contributions of its different terms as follows
\begin{equation}
P = \underbrace{-\overline{\rho} \langle u_t''u_n'' \rangle \frac{\partial \langle u_t \rangle}{\partial y_n}}_{P_{st}} \underbrace{- \,\overline{\rho} \langle u_t''u_n'' \rangle \frac{\partial \langle u_n \rangle}{\partial x_t}}_{P_{sn}}  \underbrace{-  \, \overline{\rho} \langle u_t''u_t'' \rangle \frac{\partial \langle u_t \rangle}{\partial x_t}}_{P_{xt}}  \underbrace{- \,\overline{\rho} \langle u_n''u_n'' \rangle \frac{\partial \langle u_n \rangle}{\partial y_n}}_{P_{yn}} \mbox{ ,}
\end{equation}
where $P_{st}$ denotes the wall-tangential shear term, $P_{sn}$ is the wall-normal shear term, $P_{xt}$ is the tangential deceleration term, and $P_{yn}$ the wall-normal deceleration term, as described by \citet{fang2020}. All quantitiers are normalized by $c_{x}/\rho_{\infty} u_{\infty}^3$. Since $P_{sn}$ and $P_{yn}$ are negligible in the present case, only the dominant terms $P_{st}$ and $P_{xt}$, and the total production of TKE are discussed here. 

\begin{figure}
	\centering	
	\begin{overpic}[trim = 1mm 1mm 15mm 15mm, clip,width=0.49\textwidth]{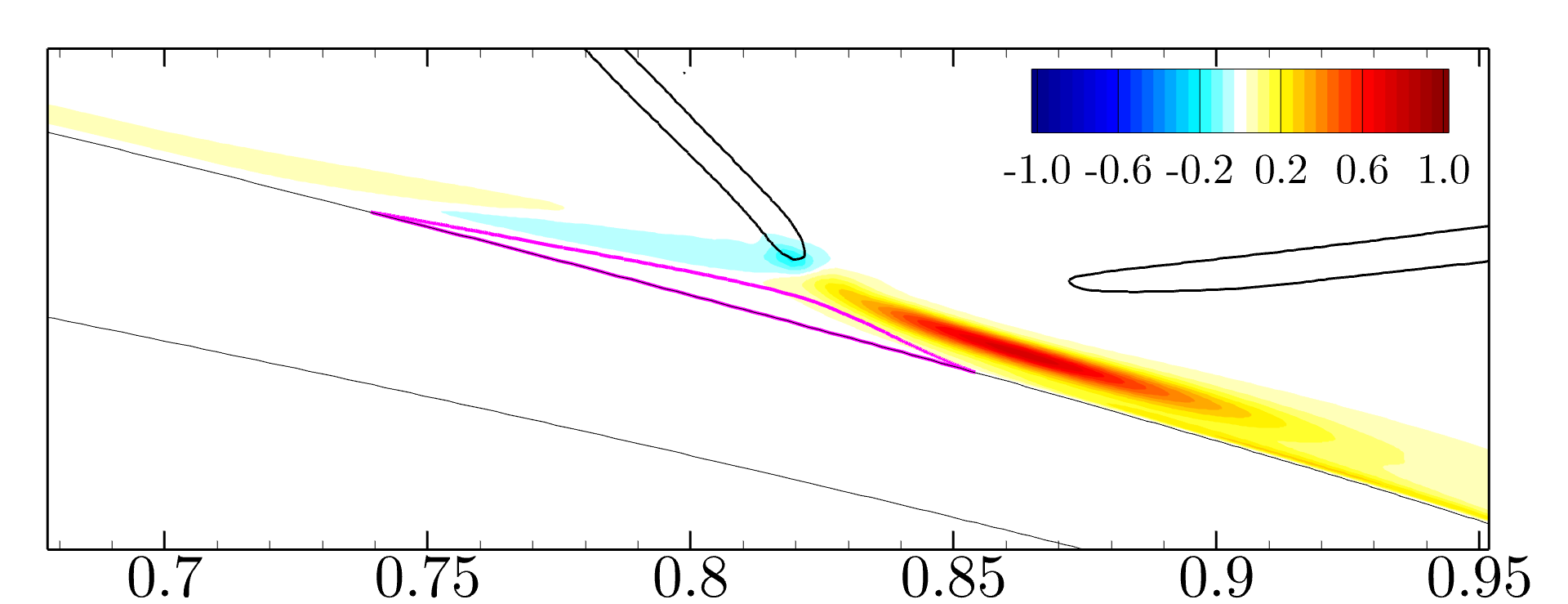}
	\put(5,8){(a)}
	\end{overpic} 
	\begin{overpic}[trim = 1mm 1mm 15mm 15mm, clip,width=0.49\textwidth]{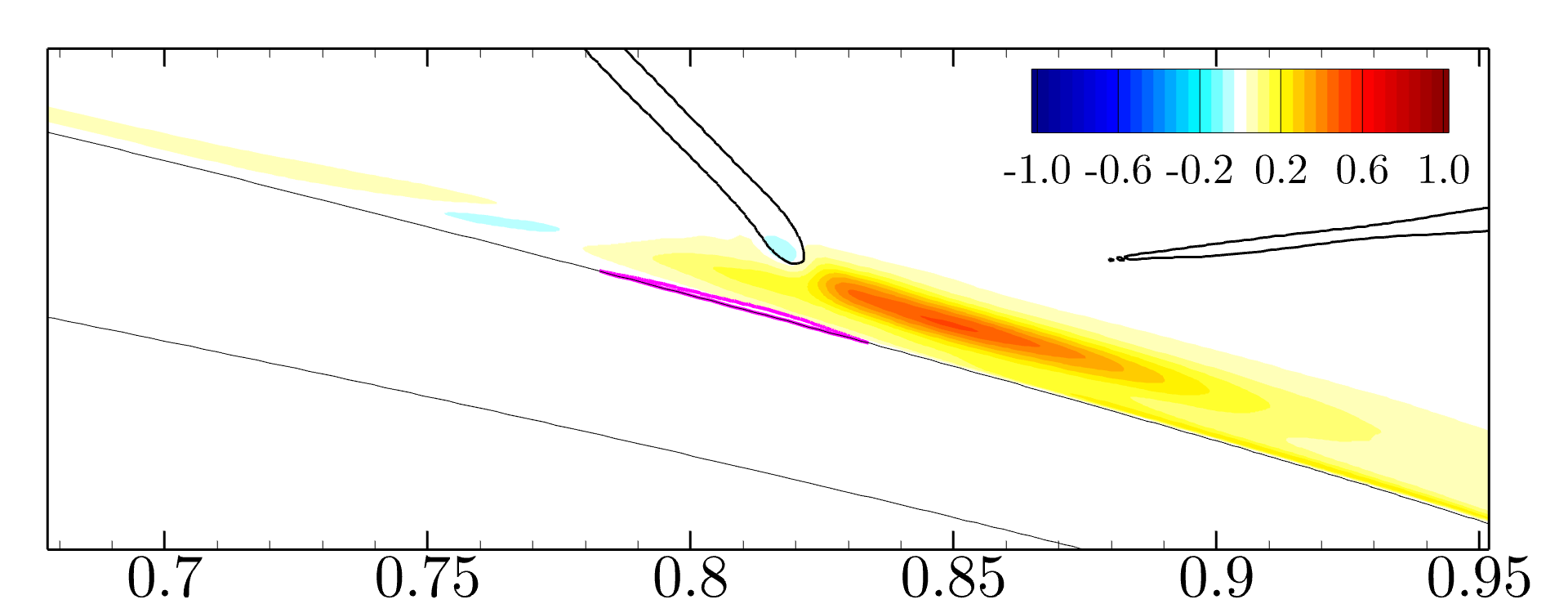}
	\put(5,8){(b)}
	\end{overpic} 

    \begin{overpic}[trim = 1mm 1mm 15mm 15mm, clip,width=0.49\textwidth]{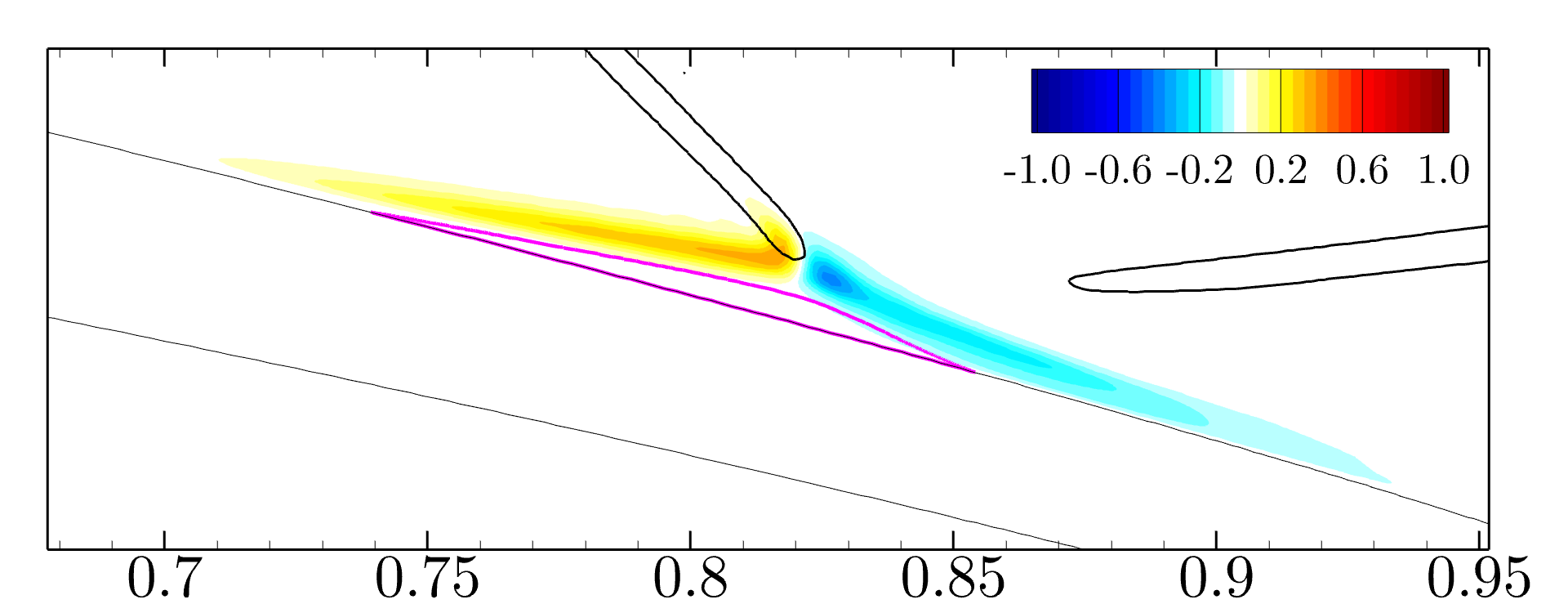}
	\put(5,8){(c)}
	\end{overpic} 
\begin{overpic}[trim = 1mm 1mm 15mm 15mm, clip,width=0.49\textwidth]{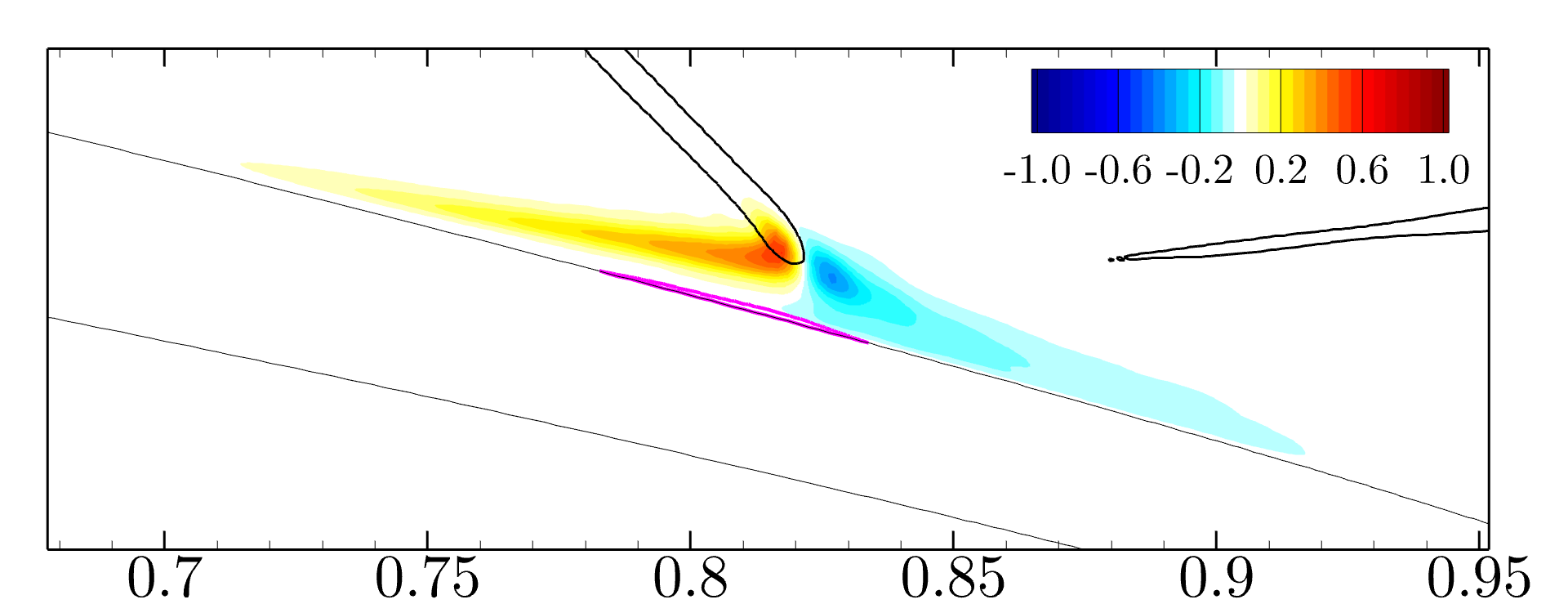}
	\put(5,8){(d)}
	\end{overpic} 

        \begin{overpic}[trim = 1mm 1mm 15mm 15mm, clip,width=0.49\textwidth]{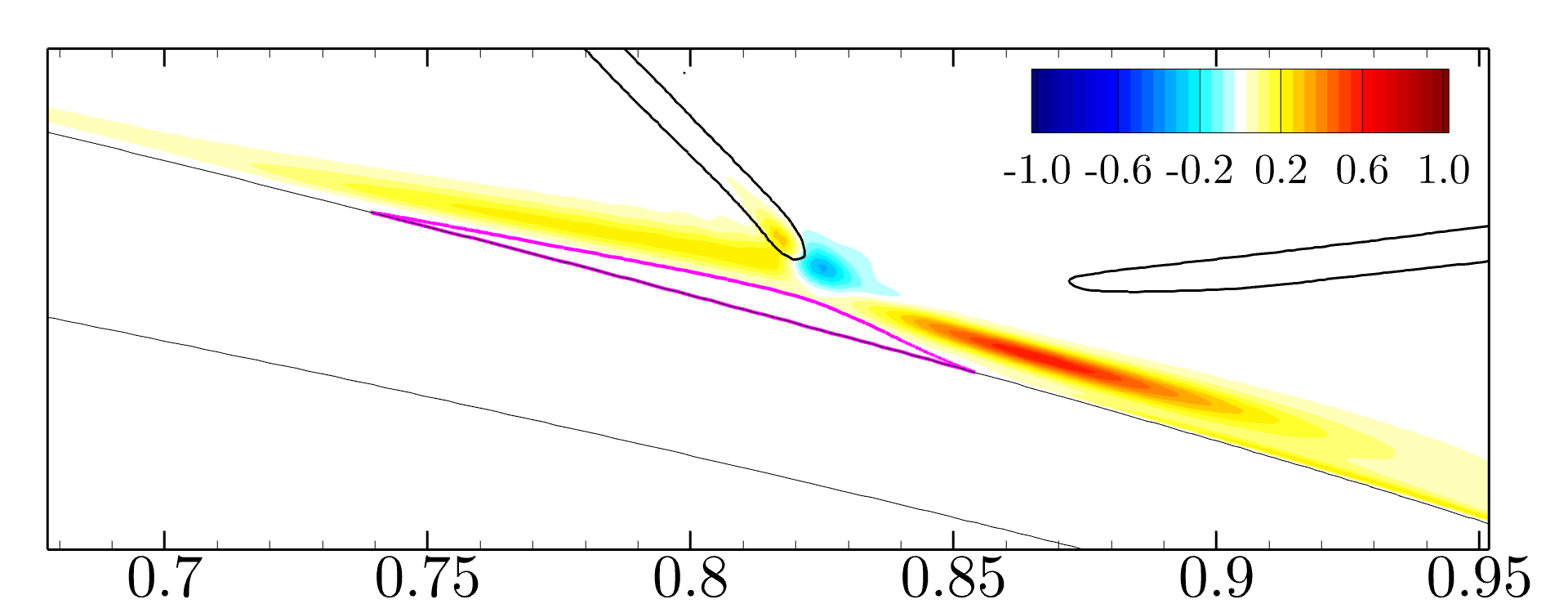}
	\put(5,8){(e)}
	\end{overpic} 
\begin{overpic}[trim = 1mm 1mm 15mm 20mm, clip,width=0.49\textwidth]{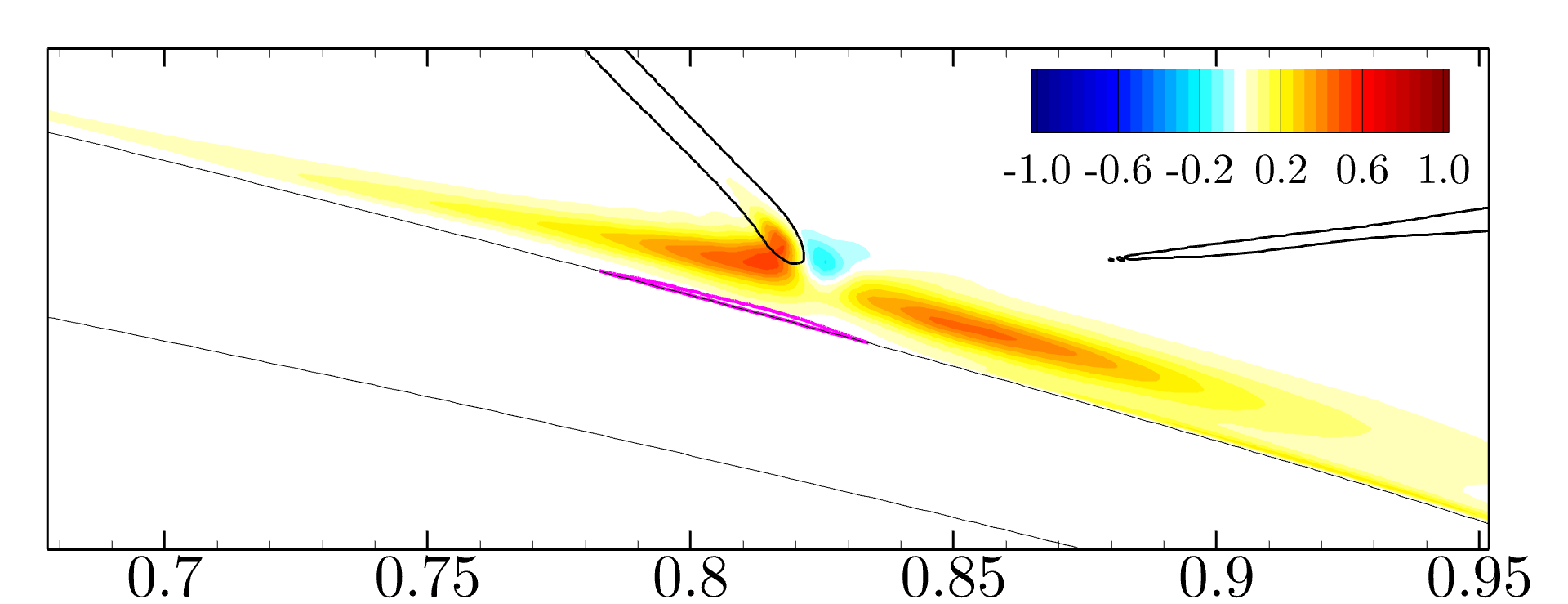}
	\put(5,8){(f)}
	\end{overpic} 
    
	\caption{Contours of conditionally averaged TKE production terms for (left) large bubble events and (right) small bubble events: (a,b) wall-tangential shear production $P_{s_{t}}$, (c,d) wall-tangential deceleration $P_{xt}$, (e,f) and total production $P$. All terms are normalized by $c_{x}/\rho_{\infty} u_{\infty}^3$. The purple lines delimit the mean separation bubble.} 
	\label{fig:tke_production}
\end{figure}

\begin{figure}
\centering	
\begin{overpic}[trim = 1mm 1mm 1mm 1mm, clip,width=0.99\textwidth]{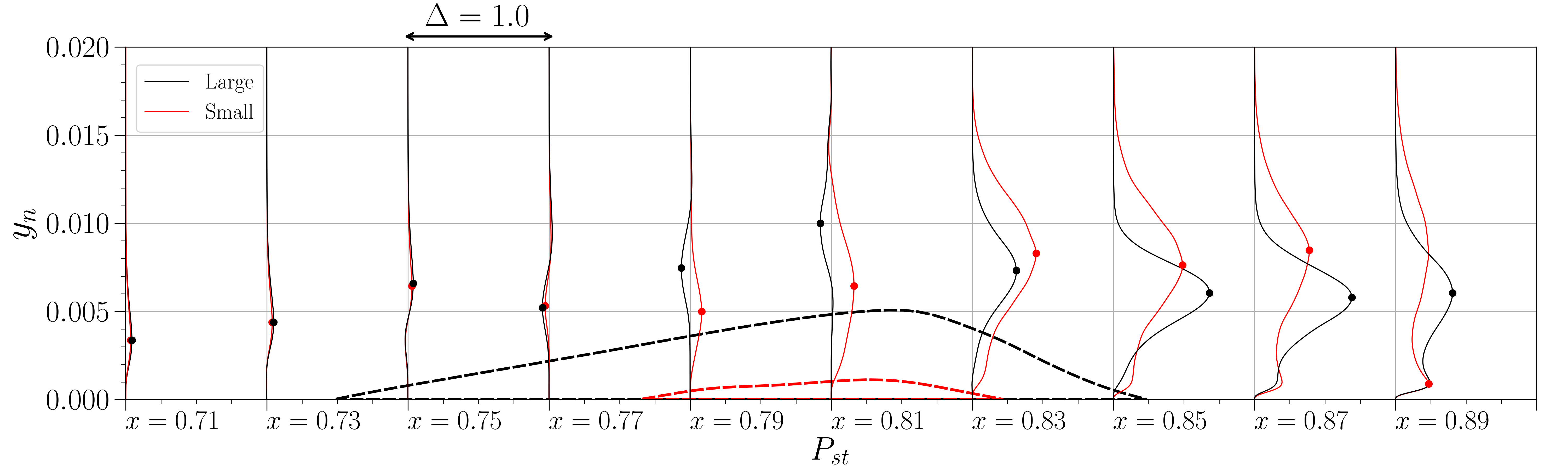}
	\put(0,30){(a)}
    \end{overpic} 

\vspace{5.1pt}

\begin{overpic}[trim = 1mm 1mm 1mm 1mm, clip,width=0.99\textwidth]{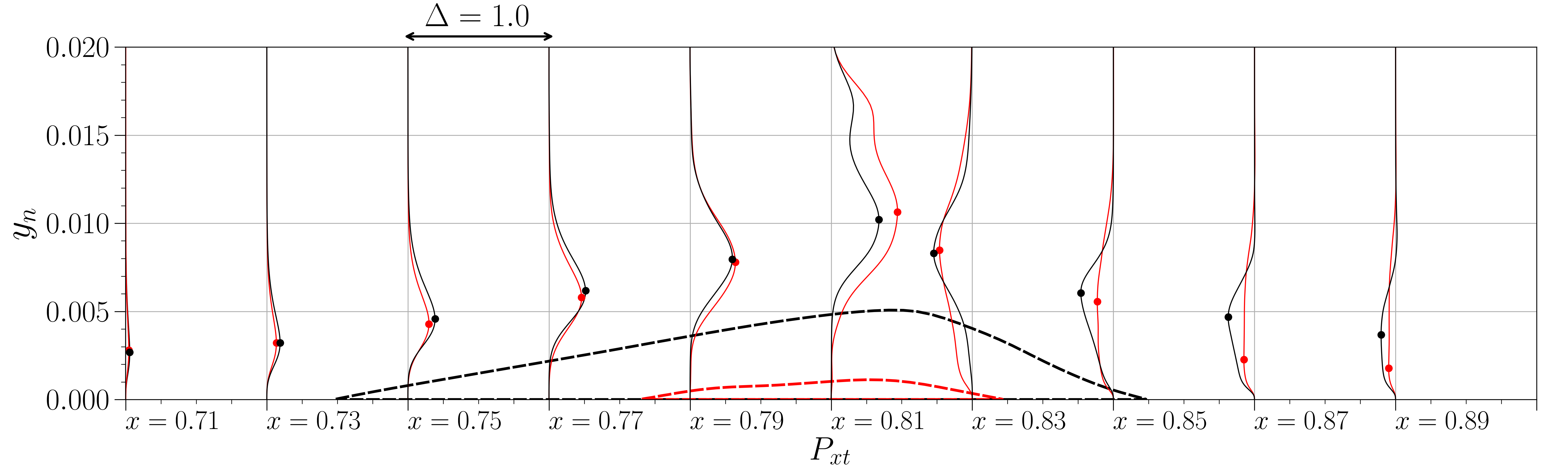}
	\put(0,30){(b)}
    \end{overpic} 

\vspace{5.1pt}

\begin{overpic}[trim = 1mm 1mm 1mm 1mm, clip,width=0.99\textwidth]{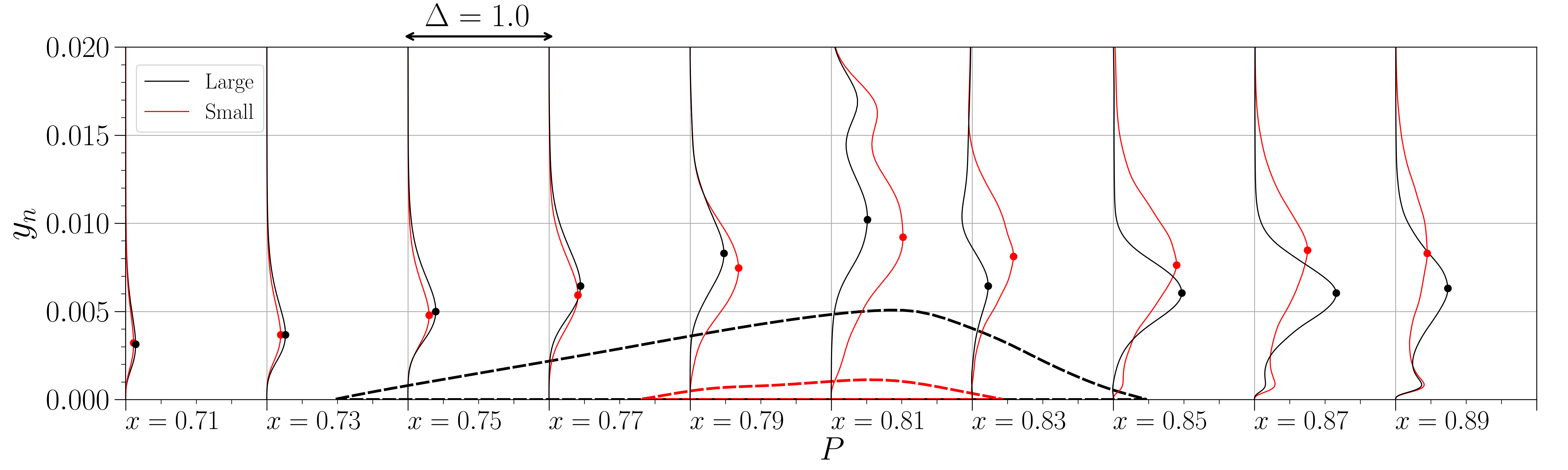}
	\put(0,30){(c)}
    \end{overpic}

	\caption{Profiles of conditionally averaged TKE production for large bubble events (black) and small bubble events (red): (a) wall-tangential shear production $P_{s_{t}}$, (b) wall-tangential deceleration $P_{xt}$, and (c) total production $P$. All terms are normalized by $c_{x}/\rho_{\infty} u_{\infty}^3$. The black and red dashed lines outline the recirculation bubbles for large and small bubble events, respectively, while the circles indicate the peak values of the corresponding quantities.}
	\label{fig:tke_production_profiles}
\end{figure}

\begin{figure}
\centering	
\begin{overpic}[trim = 1mm 1mm 1mm 1mm, clip,width=0.99\textwidth]{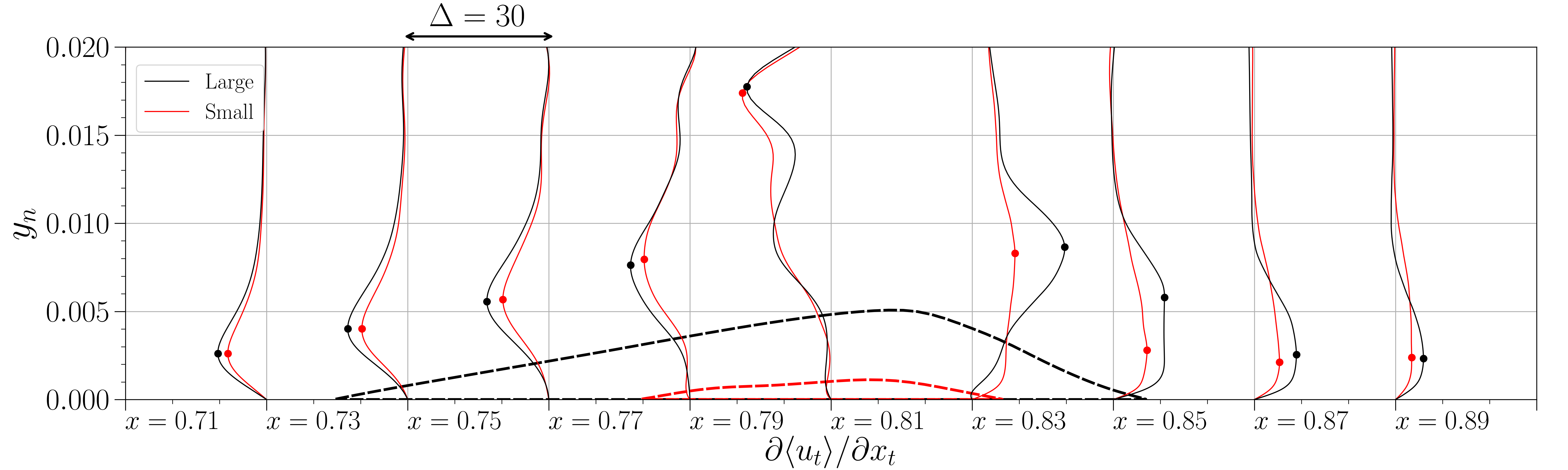}
	\put(0,30){(a)}
    \end{overpic} 

\vspace{5.1pt}

\begin{overpic}[trim = 1mm 1mm 1mm 1mm, clip,width=0.99\textwidth]{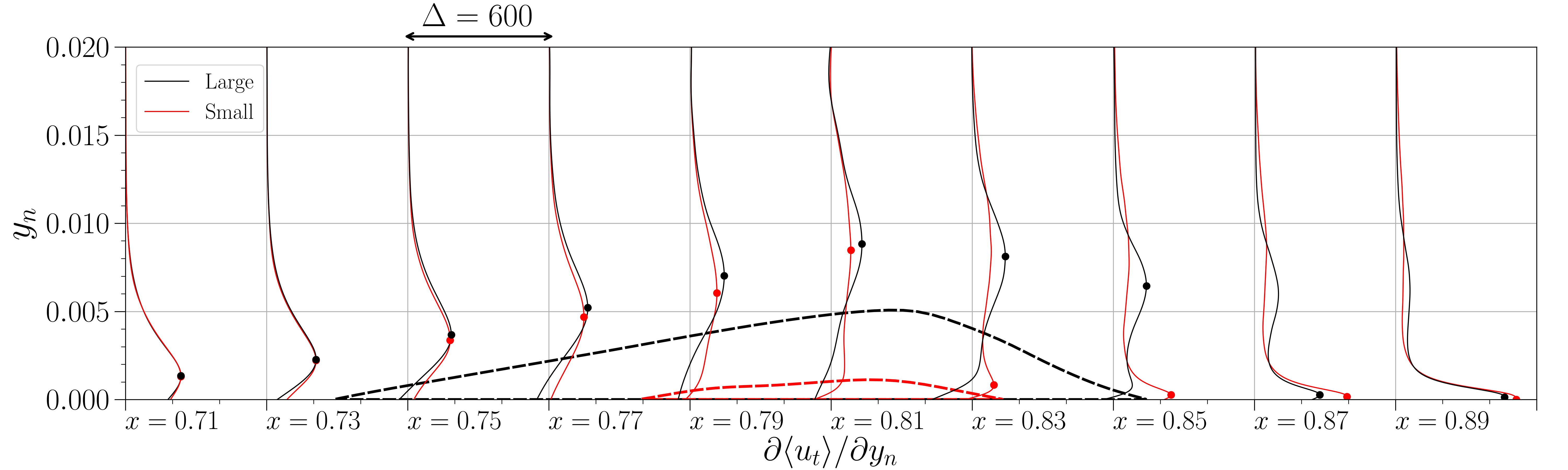}
	\put(0,30){(b)}
    \end{overpic} 

	\caption{Profiles of conditionally averaged velocity gradients for large bubble events (solid black) and small bubble events (solid red): (a) $\partial \langle u_t \rangle / \partial x_t$, and (b) $\partial \langle u_t \rangle / \partial y_n$. The velocity gradients are normalized by $c_{x}/u_{\infty}$. The black and red dashed lines delimit the separation bubbles for large and small bubble events, respectively. The circles mark the peak values of the corresponding quantities.}
	\label{fig:gradient_profiles}
\end{figure}

Figure \ref{fig:tke_production} presents the contours of the conditionally averaged TKE production term. The left column displays results for large bubble events, while the right column corresponds to small bubble events. The separation bubbles are delimited by purple lines, while black lines represent the shock waves. The profiles of the conditionally averaged TKE production term are shown in Fig. \ref{fig:tke_production_profiles} to highlight their spatial evolution and behavior along the SBLI region. For reference, the black and red dashed lines outline the recirculation bubbles for large and small bubble events, respectively. Colored circles indicate the locations of the maximum values for each corresponding quantity. To support the discussion, the profiles of conditionally averaged velocity gradients for both extreme bubble events are displayed in Fig. \ref{fig:gradient_profiles}.

The contours of the wall-tangential shear production term $P_{st}$ are shown in Figs. \ref{fig:tke_production}(a) and (b). Before the incident shock and along the bubble, small negative values of $P_{xt}$ appear along the shear layer in large bubble events, while positive values are observed in small bubble events. This difference in sign is associated with the values of $\langle u_t''u_n'' \rangle$, as shown in Fig. \ref{fig:turbulece_statistics_profiles}(e). While the Reynolds shear stress switches from positive to negative between large and small bubble events, the mean shear is always positive according to Fig. \ref{fig:gradient_profiles}(b). After the shock, high positive $P_{st}$ values are present along the free shear layer in both extreme events, with higher values appearing for large bubble events. These observations are highlighted in the conditionally averaged profiles of $P_{st}$ presented in Fig. \ref{fig:tke_production_profiles}(a). Since similar peak values of $\langle u_t''u_n'' \rangle$ are observed in both extreme events, the higher $P_{st}$ values seen in large bubble events can be attributed to the higher mean shear values along the free shear layer, as seen in Fig. \ref{fig:gradient_profiles}(b). 

Figures \ref{fig:tke_production}(c) and \ref{fig:tke_production}(d) display the contours of the wall-tangential deceleration production term $P_{xt}$. Upstream of the incident shock, positive values of $P_{xt}$ appear along the bubble shear layer in both extreme events due to the mean flow deceleration $\partial \langle u_t \rangle / \partial x_t$, as illustrated in Fig. \ref{fig:gradient_profiles}(a). Although flow deceleration is more pronounced in large bubble events, the increase in $\langle u_t''u_t'' \rangle$ is more pronounced in small bubble events, resulting in higher $P_{xt}$ values compared to large bubbles, especially near the incident shock. Downstream of the shock, the flow accelerates due to the curvature of the bubble, as shown in Fig. \ref{fig:gradient_profiles}(a), resulting in negative values of $P_{xt}$ along both the bubble shear layer and the free shear layer. Higher negative values of $P_{xt}$ are observed downstream of the separation for large bubble events, as highlighted in Fig. \ref{fig:tke_production_profiles}(b). This behavior can be attributed to the higher peak values of both $\langle u_t''u_t'' \rangle$ and $\partial \langle u_t \rangle / \partial x_t$ in these events.

Based on the previous results, the total production of TKE is further examined in Figs. \ref{fig:tke_production}(e) and (f), which show contours of the total production of TKE for both extreme events. In addition, the profiles of $P$ in specific streamwise locations are shown in Fig. \ref{fig:tke_production_profiles}(c). A key observation from these plots is that, before the incident shock, the wall-tangential deceleration $P_{xt}$ is the main contributor to the total production of TKE. Interestingly, in large bubble events, the shear term $P_{st}$ slightly suppresses TKE production over the bubble, upstream of the shock. On the other hand, downstream of the incident shock, TKE production is primarily associated with the shear term $P_{st}$, with $P_{xt}$ suppressing the production of turbulence. These results are in agreement with those reported by \citet{fang2020}.

In terms of magnitude of $P$ between the extreme events, before the incident shock, TKE production is higher for small bubble events due to the higher values of $\langle u_t''u_t'' \rangle$. This occurs despite the lower values of $\partial \langle u_t \rangle / \partial x_t$ observed for these events. Downstream of reattachment, TKE production is higher for large bubble events, mainly due to the higher mean shear along the free shear layer, since similar values of $\langle u_t''u_n'' \rangle$ are observed for both extreme events.

\subsubsection{Probability density functions of velocity fluctuations}
\label{subsec:pdf}

\begin{figure}[H]
\begin{overpic}[trim = 1mm 1mm 1mm 1mm, clip,width=0.99\textwidth]{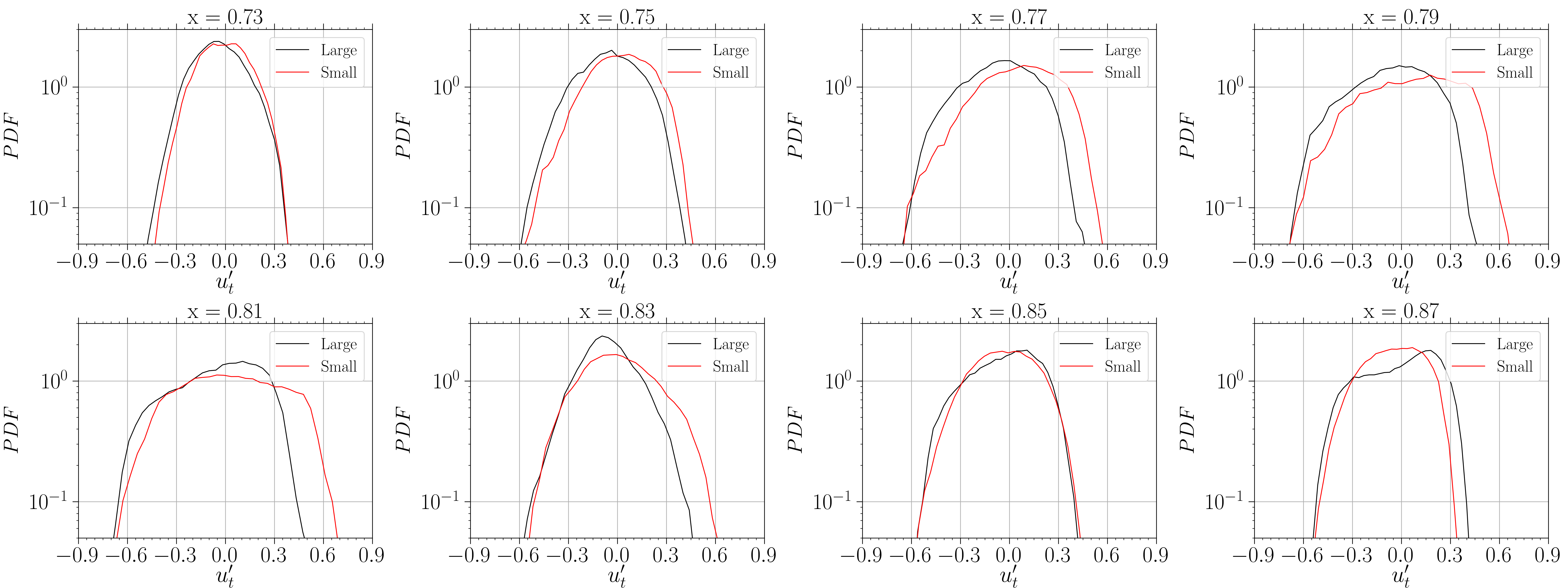}	
	\end{overpic} 
	\caption{Probability density functions of normalized tangential velocity fluctuations $u_t'$ at the locations of maximum TKE for several chordwise positions (red and black circles in Fig. \ref{fig:turbulece_statistics_profiles}(d)). The curves represent the statistics of small (red) and large (black) bubble events.}
	\label{fig:pdf_ut}
\end{figure}

\begin{figure}
\begin{overpic}[trim = 1mm 1mm 1mm 1mm, clip,width=0.99\textwidth]{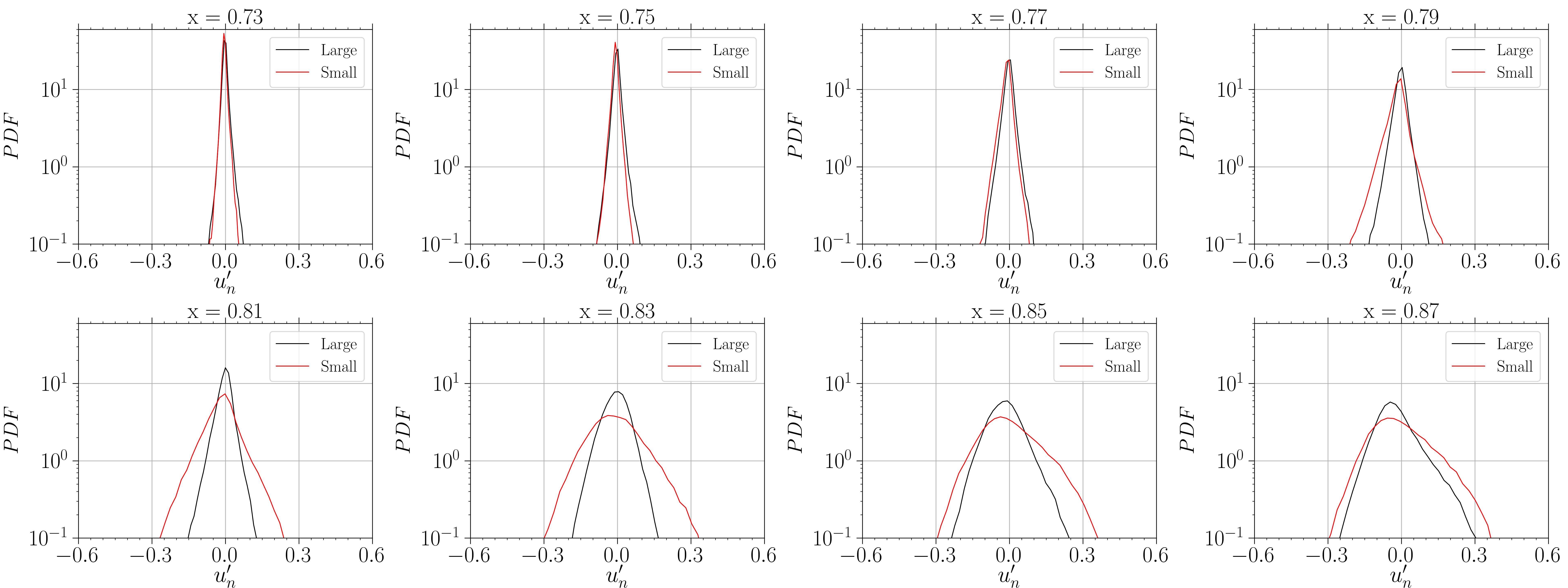}	
	\end{overpic} 
    
\caption{Probability density functions of normalized tangential velocity fluctuations $u_n'$ at the locations of maximum TKE for several chordwise positions (red and black circles in Fig. \ref{fig:turbulece_statistics_profiles}(d)). The curves represent the statistics of small (red) and large (black) bubble events.}
	\label{fig:pdf_un}
\end{figure}

\begin{figure}
\begin{overpic}[trim = 1mm 1mm 1mm 1mm, clip,width=0.99\textwidth]{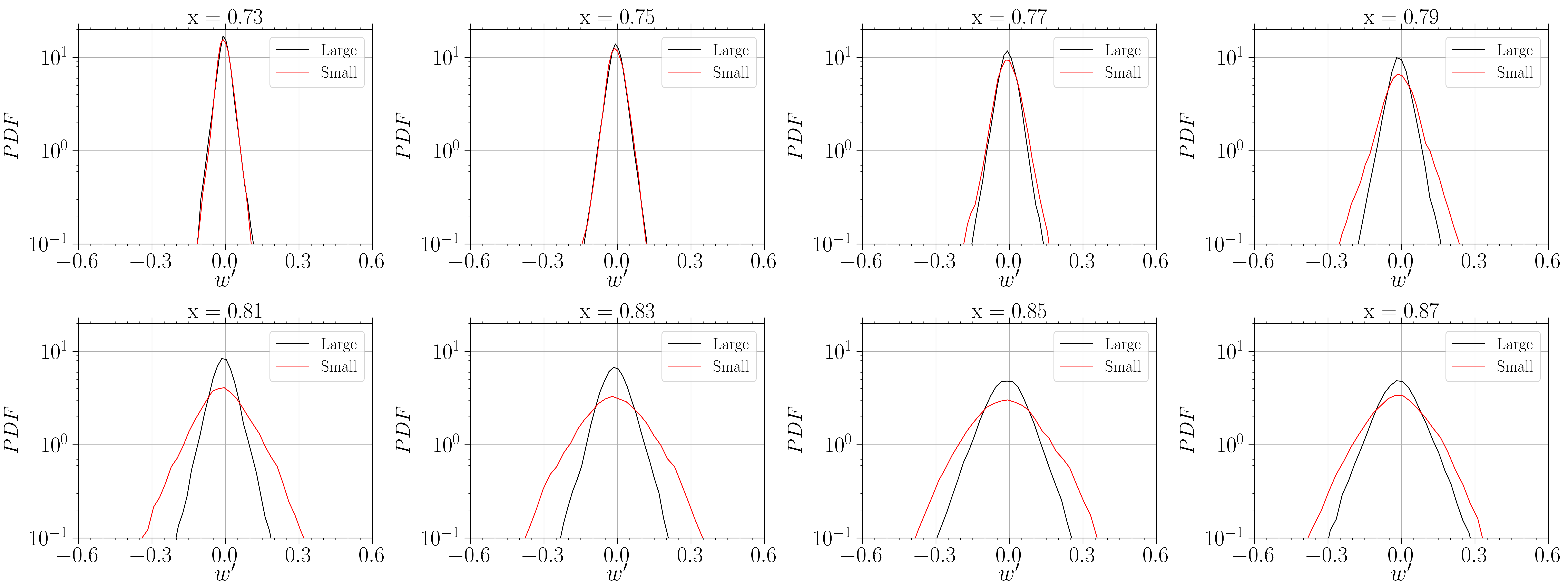}	
	\end{overpic} 
    
\caption{Probability density functions of normalized tangential velocity fluctuations $w'$ at the locations of maximum TKE for several chordwise positions (red and black circles in Fig. \ref{fig:turbulece_statistics_profiles}(d)). The curves represent the statistics of small (red) and large (black) bubble events.}
	\label{fig:pdf_w}
\end{figure}

To statistically compare velocity fluctuations associated with small and large bubble events, probability density functions (PDFs) of tangential, normal, and spanwise velocity fluctuations are presented in Figs. \ref{fig:pdf_ut}, \ref{fig:pdf_un}, and \ref{fig:pdf_w}, respectively. The $x-y$ positions from which the data are extracted are shown as red and black circles in Fig. \ref{fig:turbulece_statistics_profiles}(d), corresponding to small and large bubble events, respectively. These positions mark the regions of maximum TKE, and at each of these coordinates, data are sampled across all points along the homogeneous z-direction. Velocity fluctuations are normalized by $u_{\infty}$ and computed using the mean over the entire dataset. Consequently, the conditional averages of the fluctuations are nonzero. The PDFs therefore represent the distribution of velocity fluctuations conditioned on either small or large bubble events.



The conditional PDFs of tangential velocity fluctuations $u_t'$ are presented in Fig. \ref{fig:pdf_ut}. From upstream of the bubble to flow reattachment ($x = 0.73$ to $x = 0.83$), the PDF distributions reveal a tendency for higher positive $u_t'$ values in small bubble events, while higher negative $u_t'$ values are observed for large bubble events, especially along the shear layer over the bubble. These results suggest a higher probability of occurrence of high-speed streaks in small bubble events and low-speed streaks in large bubble events. Moreover, the tangential velocity fluctuations appear more dispersed at locations along the shear layer and upstream of the shock ($x = 0.77$ to $x = 0.81$) for the small bubble events, indicating more turbulent mixing. Downstream of flow reattachment ($x = 0.85$ to $x = 0.87$), the PDFs for small bubble events are more symmetric, with a small bias towards negative values. In contrast, for large bubble events, PDFs exhibit a higher asymmetry, also being left-skewed, especially at $x = 0.87$. Our previous studies \cite{lui2022, lui_2024} show that the free shear layer is less coherent and more diffuse for small bubble events compared to large bubble events. This may explain the observed differences in PDF results downstream of flow reattachment.


Conditional PDFs of normal velocity fluctuations $u_n'$ are shown in Fig. \ref{fig:pdf_un}. Between $x = 0.73$ and $x = 0.77$, the distributions are highly concentrated around the mean and the PDFs are very similar between small and large bubble events. At downstream locations, the distributions show a tendency toward higher positive and negative $u_n'$ values in small bubble events, along with more dispersed normal velocity fluctuations, suggesting a more intense fluid motion in the wall-normal direction, as indicated by the results in Figs. \ref{fig:turbulece_statistics}(c,d) and Fig. \ref{fig:turbulece_statistics_profiles}(b).

Figure \ref{fig:pdf_w} displays the conditional PDFs of spanwise velocity fluctuations $w'$. Again, the PDF distributions become wider along the chordwise direction, with this effect being more pronounced for small bubble events, especially at locations along the shear layer above the bubble and the free shear layer. As mentioned above, the conditional averages of the velocity fluctuations are nonzero; therefore, the PDFs are not centered around zero. The second observation is a tendency toward higher positive and negative $w'$ values, especially in small bubble events. These results indicate more intense fluid motion in the spanwise direction, most likely associated with streamwise vortices that will be discussed in more detail in the next section.

\begin{figure}
\begin{overpic}[trim = 1mm 1mm 1mm 1mm, clip,width=0.99\textwidth]{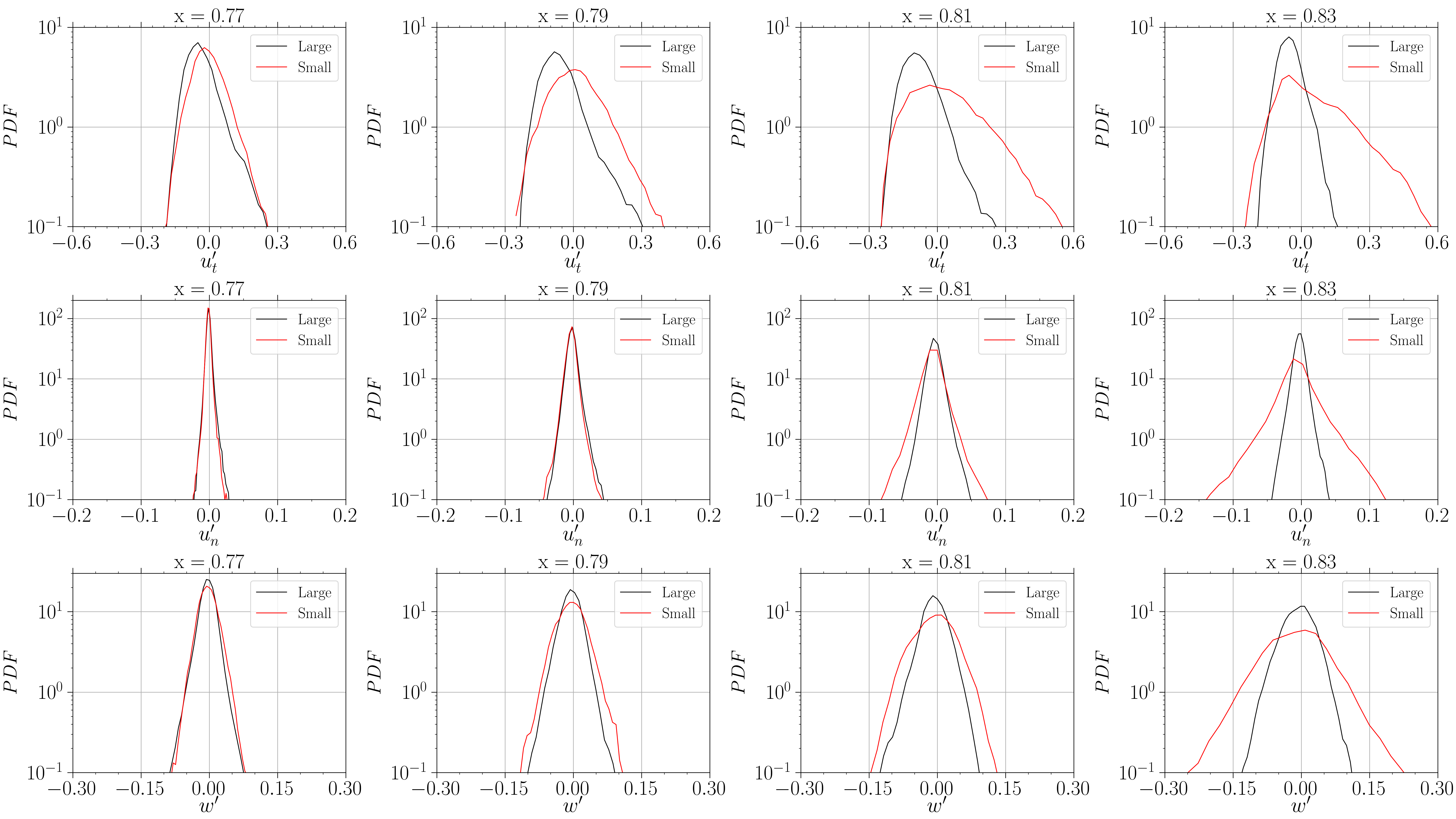}	
	\end{overpic} 
    
	\caption{Probability density functions of normalized velocity fluctuations at near-wall locations between the separated flow and the shear layer (blue circles in Fig. \ref{fig:turbulece_statistics_profiles}(d)). The curves represent the statistics of small (red) and large (black) bubble events.}
	\label{fig:pdf_velocity_fluctuations}
\end{figure}

The conditional PDFs of velocity fluctuations in Fig. \ref{fig:pdf_velocity_fluctuations} are evaluated at wall-normal positions between the separated flow and the shear layer, as indicated by the blue circles in Fig. \ref{fig:turbulece_statistics_profiles}(d). For large bubble events, these locations lie within the separation region, whereas for small bubble events, they coincide with the shear layer. Conditional PDFs of $u_t'$, shown in the first row of Fig. \ref{fig:pdf_velocity_fluctuations}, exhibit a tendency toward higher positive values during small-bubble events. This behavior is likely associated with the passage of near-wall high-speed streaks, as demonstrated in Ref. \citep{lui2022}. Similarly to the results at the highest TKE locations, the conditional PDFs of $u_n'$ and $w'$ exhibit a tendency toward higher positive and negative values during small-bubble events, further indicating enhanced fluid motion in the wall-normal and spanwise directions.

\subsection{Interaction between bubble, streaks, and vortices}
\label{subsec:FTLE}

The results have shown that small bubble events exhibit higher values of Reynolds stress components and turbulent kinetic energy along the bubble and shear layers. Hence, these regions are more likely to experience strong velocity fluctuations. Moreover, previous studies \citep{lui2022,lui_2024} identified the passage of high-speed streaks and streamwise vortices through the bubble in the present configuration. Therefore, the physical mechanism responsible for modifying the turbulence quantities between the extreme events is most likely associated with the interaction between bubble, streaks, and streamwise vortices. To elucidate the interplay between these flow features, the finite-time Lyapunov exponent (FTLE) is computed to investigate fluid transport within the separation region for small and large bubbles.

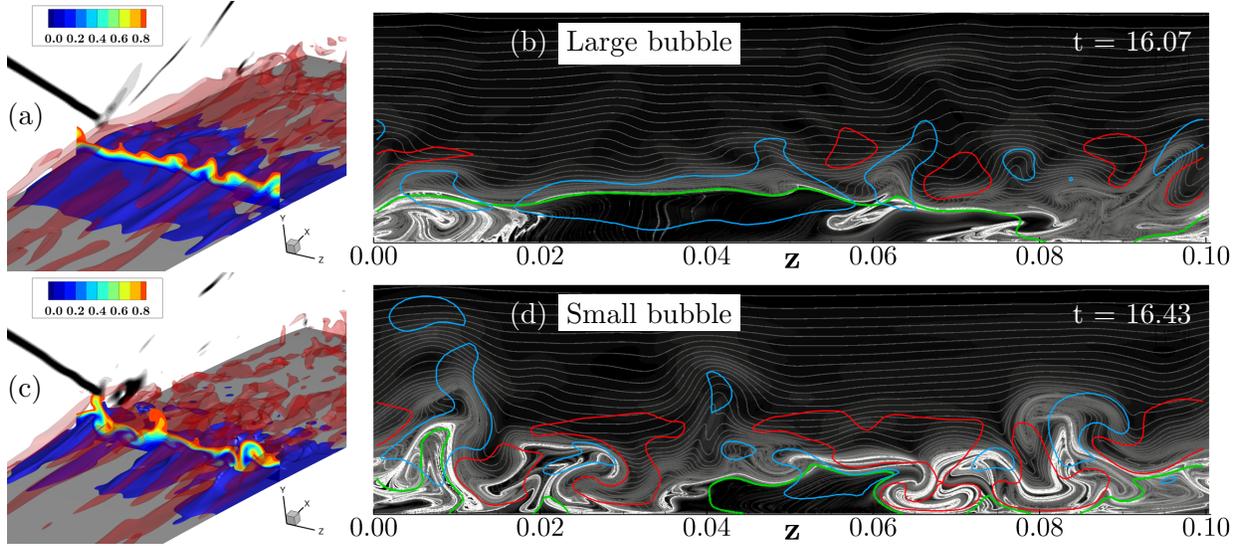
\begin{figure}
\centering	
\begin{overpic}[trim = 1mm 1mm 1mm 1mm, clip,width=0.99\textwidth]{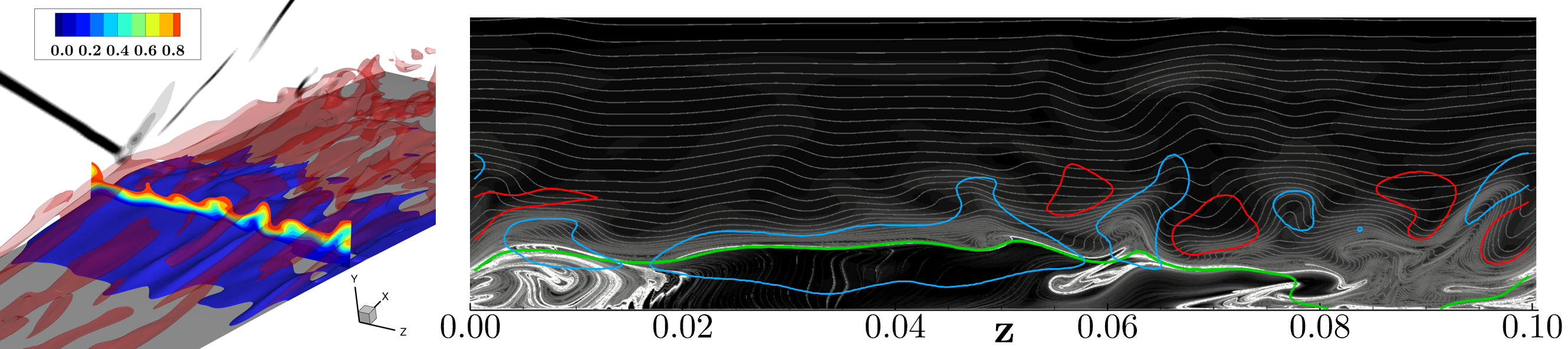}
		\put(0,12){(a)}
            \put(41,18){\white{(b)}}
            \put(45,18){\colorbox{white}{Large bubble}}
            \put(87,18){\white{t = 16.07}}
	\end{overpic} 
 \begin{overpic}[trim = 1mm 1mm 1mm 1mm, clip,width=0.99\textwidth]{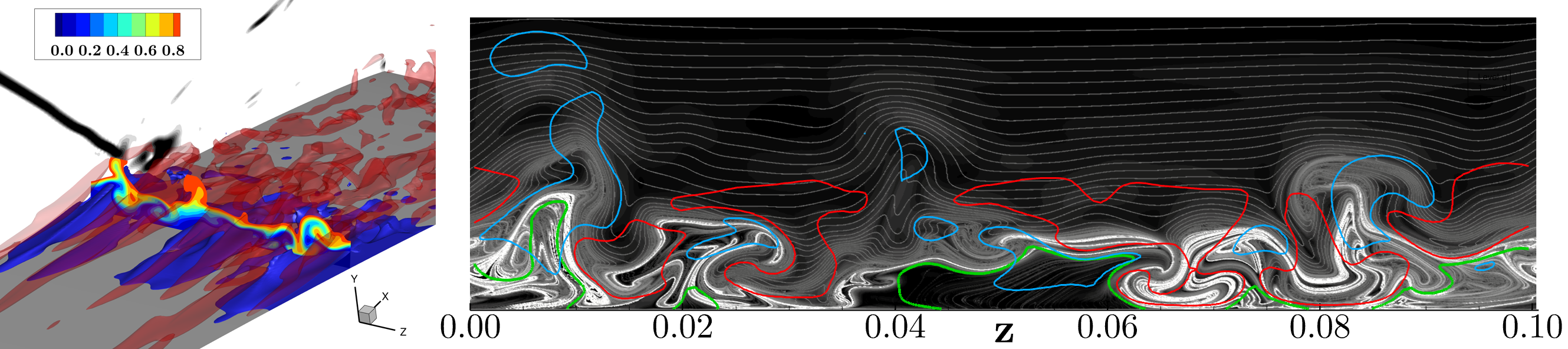}
		\put(0,12){(c)}
            \put(41,18){\white{(d)}}
             \put(45,18){\colorbox{white}{Small bubble}}
             \put(87,18){\white{t = 16.43}}
	\end{overpic} 
	\caption{Interaction between bubble, streaks and vortices upstream of the incident shock for (a,b) large bubble event and (c,d) small bubble event. (a,c) show isosurfaces of instantaneous $u = 0$ velocity (blue) and velocity fluctuations $u' = 0.15$ (red). Shocks are visualized in the background with grayscale contours of $| \nabla \rho |$. (b,d) show instantaneous $z$-$y_n$ planes of the finite-time Lyapunov exponent (white). The green lines delimit the separation regions, and the red and blue lines show streamwise velocity fluctuations of $u' = 0.15$ and $u' = -0.15$, respectively.}
	\label{fig:lcs_LE}
\end{figure}

In this work, the FTLE approach described by \citet{lucas2025} is employed. Here, snapshots of $z$-$y_n$ planes of vertical and spanwise instantaneous velocities are used. These snapshots correspond to a specific moment when the bubble undergoes a strong expansion followed by a contraction, as highlighted by the magenta line in Fig. \ref{fig:wavelet}(b). An initial grid of $1400 \times 1200$ particles is created across the $z$-$y_n$ planes, and these particles are integrated within the velocity fields over a period of $3.0c_x/u_{\infty}$. The integration time was selected based on a sensitivity analysis, which showed that the FTLE fields captured the same flow structures for integration times greater than $1.5c_x/u_{\infty}$. A backward-time integration of the particles is performed using a second-order Adams-Bashforth method, enabling a direct assessment of material transport in forward time, which mimics experimental flow visualizations using tracers. Bright regions in the FTLE fields indicate locations of higher fluid particle attraction.

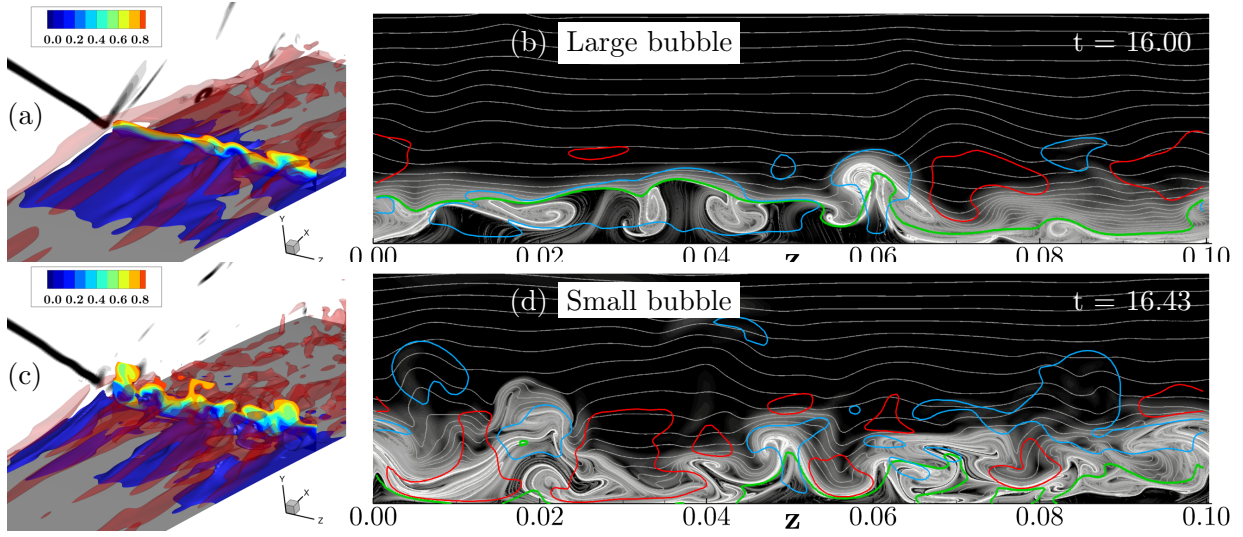
\begin{figure}
\centering	
 \begin{overpic}[trim = 2mm 2mm 2mm 2mm, clip,width=0.99\textwidth]{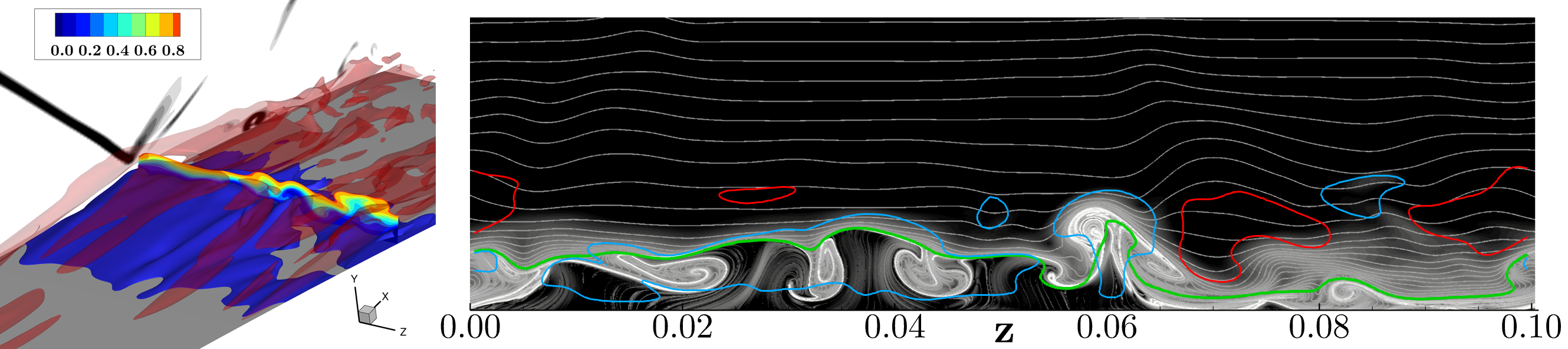}
		\put(0,12){(a)}
            \put(41,18){\white{(b)}}
        \put(45,18){\colorbox{white}{Large bubble}}
         \put(87,18){\white{t = 16.00}}
	\end{overpic} 
  \begin{overpic}[trim = 2mm 2mm 2mm 2mm, clip,width=0.99\textwidth]{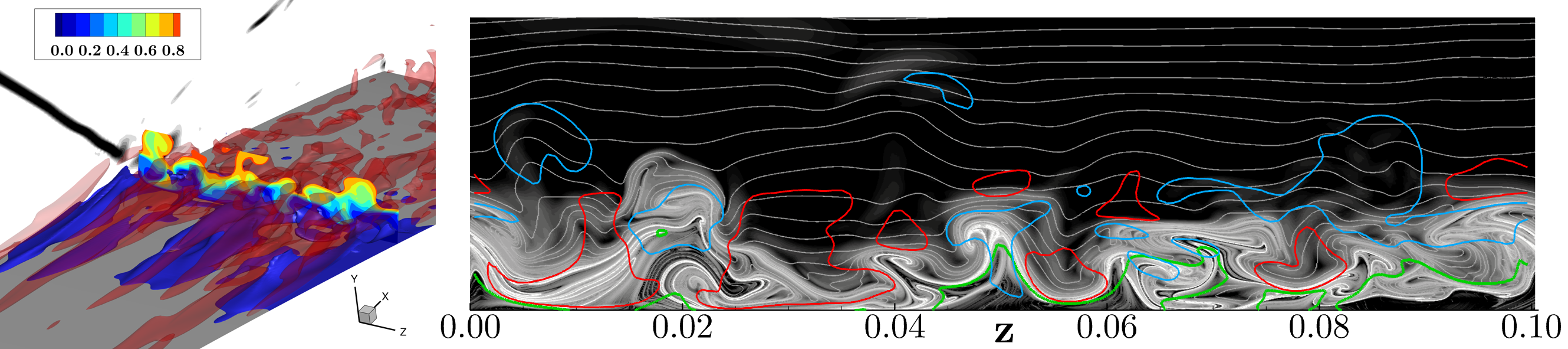}
		\put(0,12){(c)}
        \put(41,18){\white{(d)}}
  \put(45,18){\colorbox{white}{Small bubble}}
   \put(87,18){\white{t = 16.43}}
	\end{overpic} 
	\caption{Interaction between bubble, streaks and vortices downstream of the incident shock for (a,b) large bubble event and (c,d) small bubble event. (a,c) show isosurfaces of instantaneous $u = 0$ velocity (blue) and velocity fluctuations $u' = 0.15$ (red). Shocks are visualized in the background with grayscale contours of $| \nabla \rho |$. (b,d) show instantaneous $z$-$y_n$ planes of the finite-time Lyapunov exponent (white). The green lines delimit the separation regions, and the red and blue lines show streamwise velocity fluctuations of $u' = 0.15$ and $u' = -0.15$, respectively.}
	\label{fig:lcs_TE}
\end{figure}

Figure \ref{fig:lcs_LE} shows the interaction between the bubble, streaks, and vortices upstream of the incident shock, while Fig. \ref{fig:lcs_TE} illustrates the interaction downstream of the shock. The top plots in these figures represent instances when the bubble is large, whereas the bottom plots correspond to small bubble events. The plots in the left column show isosurfaces of instantaneous $u = 0$ velocity (blue) and velocity fluctuations $u' = 0.15$ (red), along with $z-y_n$ slices of instantaneous $u$ contours at the selected position. The right plots show the FTLE fields, where the green lines delimit the regions of separated flow, and the red and blue lines show values of $u' = 0.15$ and $u' = -0.15$, respectively. The velocity fluctuations are normalized by the inlet velocity.
The interaction between the bubble, streaks, and vortices is first analyzed upstream of the incident shock. When the bubble is large, high-speed streaks are advected over the recirculation region, as illustrated in Fig. \ref{fig:lcs_LE}(a), and streamwise vortices are almost absent within the bubble, as shown in Fig. \ref{fig:lcs_LE}(b). In contrast, when the bubble is small, high-speed streaks penetrate the bubble, as seen in Fig. \ref{fig:lcs_LE}(c). These streaks are accompanied by streamwise vortices, as depicted in Fig. \ref{fig:lcs_LE}(d), resulting in a more intense mixing. The small bubble event shows that vortices transport high-speed fluid from the shear layer toward the wall at certain spanwise locations. This high-speed fluid then penetrates the bubble, causing its contraction or even leading to local flow reattachment. 
The large bubble events, on the other hand, display extensive regions of low-speed fluid within the recirculation region, as shown in Fig. \ref{fig:lcs_LE}(b). These regions of low-speed flow are ejected away from the wall by the streamwise vortices in small bubble events. 
A movie corresponding to Fig. \ref{fig:lcs_LE} is provided as supplementary material (Movie 1) to illustrate the dynamics of fluid transport more clearly.

The interaction between bubble, streaks, and vortices is also investigated downstream of the incident shock. When the bubble is large, streamwise vortices become apparent on its trailing edge, mainly concentrated within the recirculation region, as shown in Fig. \ref{fig:lcs_TE}(b). Despite the presence of the vortices, large regions of low-speed fluid are observed within the bubble. Similarly to the results obtained upstream of the shock, when the bubble is small, pronounced fluid mixing occurs in the wall-normal and spanwise directions, with streamwise vortices transporting low- and high-momentum flow away from and toward the wall, respectively. A movie of Fig. \ref{fig:lcs_TE} is provided as supplementary material (Movie 2) to facilitate visualization of the dynamics of fluid transport.

When the separation bubble is small, the streamwise vortices are not fixed at specific spanwise positions but instead meander over time along the entire bubble, from the separation to the reattachment region, as shown in Movies 1 and 2. This effect further reduces the alignment of streamwise structures with the mean flow, as evidenced by the skin-friction contours in Fig. \ref{fig:rms}(d).
This behavior is consistent with the observations of \citet{pasquariello_2017}, although in their study the streamwise vortices developed slightly downstream of the bubble apex. In that case, the vortices originated from a centrifugal instability induced by the streamline curvature along the separation bubble.

Overall, the results indicate that the passage of near-wall high-speed streaks during small bubble events leads to higher values of tangential velocity fluctuations and, hence, $\langle u_t''u_t'' \rangle$, compared to large bubble events. In addition, streamwise vortices, with their meandering motion, induce intense fluid mixing in the wall-normal and spanwise directions, resulting in higher values of $\langle u_n''u_n'' \rangle$, and $\langle w''w'' \rangle$ within the bubble and along its shear layer during small bubble events. Consequently, these quantities affect the turbulent kinetic energy, which is higher in small bubble events, as shown in Fig. \ref{fig:turbulece_statistics_profiles}(d). 

\subsection{Analysis of mass flux along the separation bubble surface}
\label{subsec:mass_imbalance}

\begin{figure}
	\centering	
	\begin{overpic}[trim = 1mm 1mm 1mm 1mm, clip,width=0.89\textwidth]{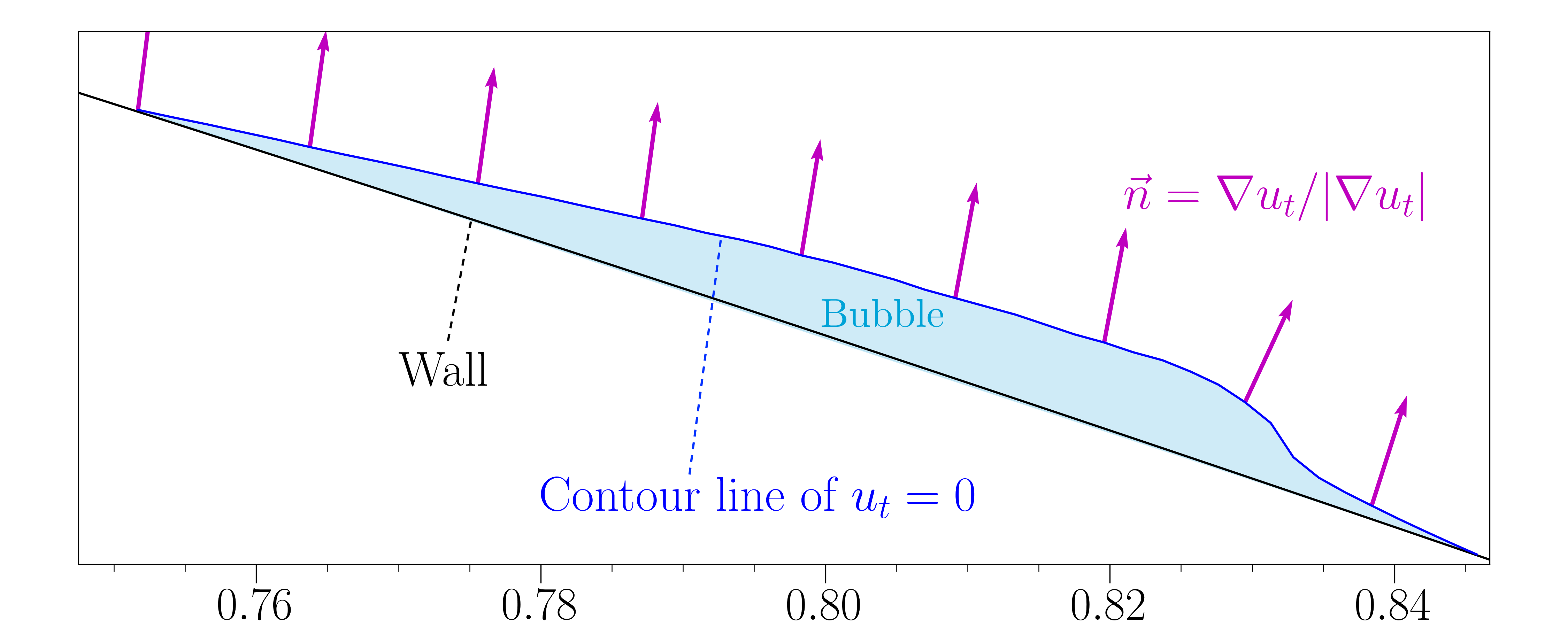}
	\end{overpic} 
	\caption{Schematic of the separation bubble surface, defined by contour lines of $u_t = 0$, along with the unit vector normal $\vec{n}$ to the surface.}
	\label{fig:bubble_drawing}
\end{figure}

Previous flow visualizations have shown that streamwise vortices can influence the mass balance within the separation bubble by transporting low-momentum fluid away from the wall and high-momentum fluid toward the wall. To investigate the mass flux into or out of the separation bubble, the local mass flux, defined as
\begin{equation}
   \dot{m} = \rho(\vec{V_r} \cdot \vec{n}) \mbox{ ,}
\label{eq:mass_flux}
\end{equation}
is computed along its surface. In this equation, $\vec{V}_r = \vec{V} - \vec{V}_{si}$ is the local fluid velocity relative to the bubble surface, where $\vec{V} = [u,v]^T$ denotes the local fluid velocity in a Cartesian system, and $\vec{V}_{si}$ is the local velocity of the bubble surface, both measured in an inertial reference frame. The separation bubble surface is defined by contour lines of $u_t = 0$, as illustrated in Fig. \ref{fig:bubble_drawing}, where $\vec{n} = \nabla u_t/|\nabla u_t|$ is the local outward normal to the surface. The main challenge in computing the mass flux lies in determining $\vec{V}_{r}$. 

To compute $\vec{V}_{r}$, a methodology similar to that used for computing the local entrainment velocity in turbulent/non-turbulent interface studies \citep{POPE1988,holzner2011,Jahanbakhshi_Madnia_2016,Su_Long_Wang_Li_2024} is employed. The time evolution of the contour line $u_t = 0$ is related to the underlying flow field and its surface velocity relative to the fluid. This relation can be written as $\vec{V}_{si} = \vec{V} + \vec{V}_{sf}$, where $\vec{V}_{sf}$ is the local surface velocity relative to the fluid. Mathematically, the local change of $u_t = 0$ on the bubble surface in a frame of reference moving with $\vec{V}_{si}$ is zero, 
\begin{align}
\begin{aligned}
\frac{D^{si}u_{t}}{Dt} &= \frac{\partial u_t}{\partial t} + \vec{V}_{si} \cdot \nabla u_t = 0 \\
&= \frac{\partial u_t}{\partial t} + \left( \vec{V} + \vec{V}_{sf} \right) \cdot \nabla u_t = 0 \mbox{ .}
\end{aligned}
\label{eq:derivada_material}
\end{align}
At each point on the surface, $\vec{V}_{sf} = v_n \vec{n}$, where $v_n$ is the normal velocity component of the bubble surface. Substituting this expression into Eq. \ref{eq:derivada_material},
%
%
\begin{align}
\begin{aligned}
\frac{\partial u_t}{\partial t} + \vec{V} \cdot \nabla u_t &= -v_n 
\underbrace{\frac{\nabla u_t}{|\nabla u_t|} \cdot \nabla u_t}_{|\nabla u_t|} \mbox{ .}
\end{aligned}
\label{eq:derivada_material1}
\end{align}
The normal velocity component of $\vec{V}_{sf}$ can finally be computed by
\begin{equation}
   v_n = -\frac{\frac{\partial u_t}{\partial t} + \vec{V} \cdot \nabla u_t}{|\nabla u_t|} = -\frac{Du_t/Dt}{|\nabla u_t|} \mbox{ .}
\label{eq:v_n}
\end{equation}
In this work, the temporal and the spatial derivatives in Eq. \ref{eq:v_n} are computed using a sixth-order compact finite-difference scheme \citep{Nagarajan2003}. Next, the local fluid velocity relative to the bubble surface $\vec{V}_r$ can be related to $v_n$ by,
\begin{align}
\begin{aligned}
\vec{V}_r &= \vec{V} - \vec{V}_{si} = \vec{V} - \left(\vec{V} + \vec{V}_{sf} \right) = - \vec{V}_{sf} = -v_n\vec{n}  \mbox{ .}
\end{aligned}
\label{eq:V_r}
\end{align}

\begin{figure}[H]
\begin{overpic}[trim = 1mm 1mm 1mm 1mm, clip,width=0.99\textwidth]{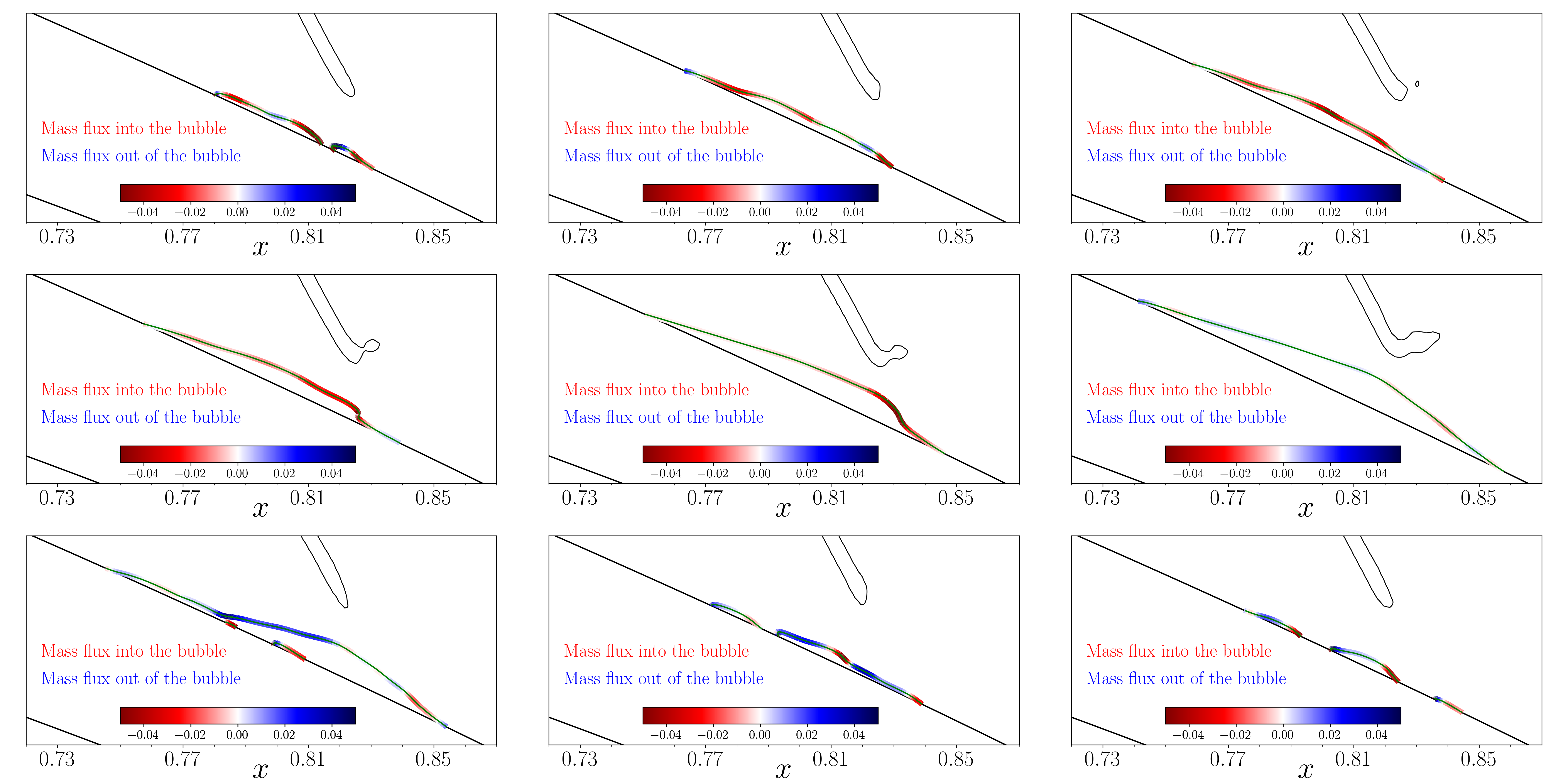}	
\put(2,45){(a)}
\put(35.5,45){(b)}
\put(69,45){(c)}
\put(2,28.5){(d)}
\put(35.5,28.5){(e)}
\put(69,28.5){(f)}
\put(2,12){(g)}
\put(35.5,12){(h)}
\put(69,12){(i)}

\put(22,47){\footnotesize t = 15.79}
\put(55.5,47){\footnotesize t = 15.83}
\put(88.5,47){\footnotesize t = 15.85}

\put(22,30.5){\footnotesize t = 15.87}
\put(55.5,30.5){\footnotesize t = 15.92}
\put(88.5,30.5){\footnotesize t = 16.10}

\put(22,14){\footnotesize t = 16.22}
\put(55.5,14){\footnotesize t = 16.37}
\put(88.5,14){\footnotesize t = 16.39}

	\end{overpic} 
	\caption{Local mass flux $\dot{m}$ along the separation bubble surface (green lines) at different time instants, illustrating the expansion and contraction of the bubble. The black lines display the incident shock through the pressure gradient magnitude.}
	\label{fig:mass_imbalance}
\end{figure}
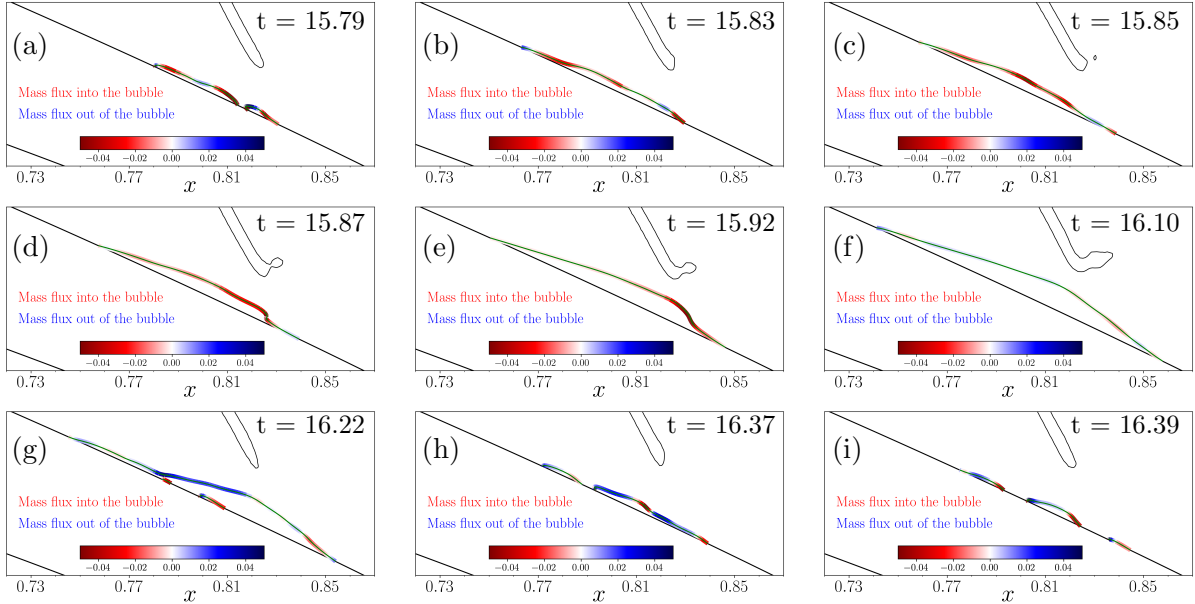

Using Eqs. \ref{eq:v_n} and \ref{eq:V_r}, the local mass flux can be computed with Eq. \ref{eq:mass_flux}. This allows us to identify regions that are particularly susceptible to mass flux entering (negative values) and leaving the bubble (positive values) during its contraction and expansion. Figure \ref{fig:mass_imbalance} shows contours of local mass flux $\dot{m}$ along the surface of the recirculation bubble during the time period highlighted in magenta in Fig. \ref{fig:wavelet}(b), when the bubble undergoes an expansion followed by a contraction. The black lines represent the incident shock, identified by the pressure gradient magnitude, while the green lines define the recirculation bubble. In addition to the figures shown here, a movie covering the entire time period is provided as supplementary material (Movie 3).

Figure \ref{fig:mass_imbalance}(a) shows two small bubbles, with regions of mass flux both into and out of them. The supplementary material (Movie 3) illustrates that, at time instants close to those shown in Fig. \ref{fig:mass_imbalance}(a), the first bubble is expanding while the second one is shrinking until it disappears. These outcomes are likely related to the mass imbalance of the bubbles. Next, the bubble undergoes mild expansion, as shown in Figs. \ref{fig:mass_imbalance}(b) and \ref{fig:mass_imbalance}(c), with predominant mass flux into the bubble, particularly at locations upstream the incident shock. Then, the bubble experiences a strong expansion, as illustrated in Figs. \ref{fig:mass_imbalance}(d) and \ref{fig:mass_imbalance}(e), and also in Movie 3. During this phase, high mass flux entering the bubble is observed at its rear end, especially after the incident shock \citep{piponniau2009, jenquin_2023}.

In Fig. \ref{fig:mass_imbalance}(f), the bubble approaches its maximum size, and there is almost no mass flux into or out of its surface. The bubble size remains nearly constant for a short period, as shown in Movie 3. Once a mass imbalance arises, the bubble starts to shrink, as illustrated in Fig. \ref{fig:mass_imbalance}(g), where mass flux out of the bubble is observed along its frontal edge. As previously discussed, this strong mass imbalance is linked to the passage of high-speed streaks through the bubble and to intense fluid mixing induced by streamwise vortices (see Fig. \ref{fig:lcs_LE}). The interaction between these streaks and vortices disrupts the bubble, causing its contraction, as visualized in Figs. \ref{fig:mass_imbalance}(h) and \ref{fig:mass_imbalance}(i). 

\begin{figure}[H]
\centering
\begin{overpic}[trim = 1mm 1mm 1mm 1mm, clip,width=0.59\textwidth]{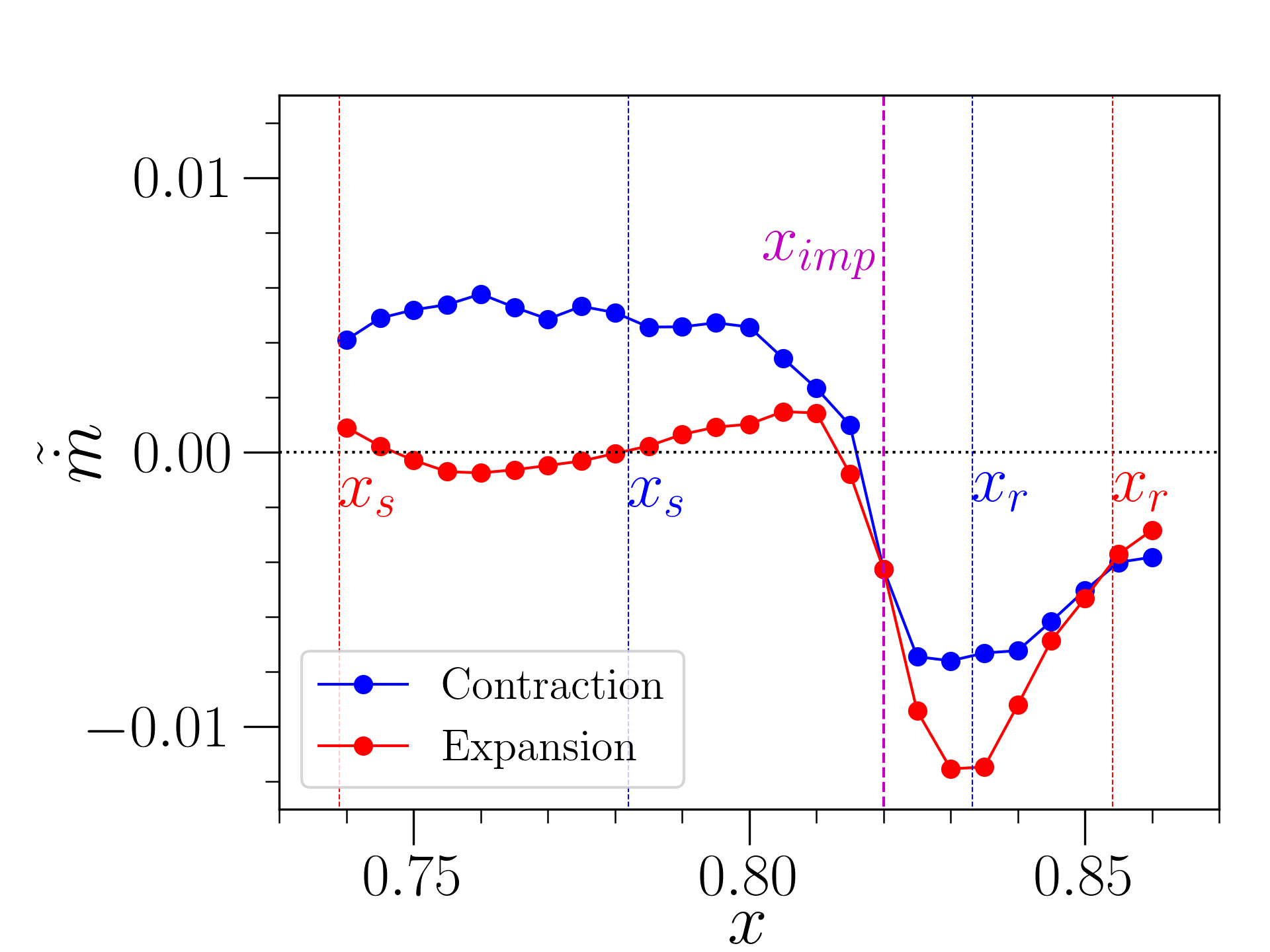}	
	\end{overpic} 
	\caption{Median of mass flux $\tilde{\dot{m}}$ during periods when the bubble is contracting (blue) and expanding (red). Vertical dashed lines indicate the conditionally averaged separation and reattachment  locations for small (blue) and large (red)  bubbles events, respectively. The magenta dashed line indicates the position of the incident shock impingement. Positive values of $\tilde{\dot{m}}$ correspond to mass flux out of the bubble, while negative values indicate mass flux into it.}
	\label{fig:mass_imbalance_median}
\end{figure}

To identify surface regions particularly susceptible to mass flux entering or leaving the bubble, Fig. \ref{fig:mass_imbalance_median} shows the median mass flux $\tilde{\dot{m}}$ computed over periods of bubble contraction (blue) and expansion (red). The vertical magenta dashed line in the plot marks the $x$-position where the incident shock interacts with the boundary layer. For reference, vertical dashed lines indicate the conditionally averaged separation and reattachment locations for small (blue) and large (red) bubbles events, respectively. The results show that during the bubble contraction phase, mass flux leaving the bubble predominantly occurs upstream of the incident shock (positive values), whereas mass flux entering the bubble (negative values) can still occur, taking place downstream of the shock. Since the mass flux leaving the bubble is higher than that entering it, the bubble undergoes a net contraction.

In the bubble expansion phase, both mass flux injection and ejection from the bubble can occur before the incident shock, though at low values. However, significant mass flux into the bubble is observed after the incident shock, as highlighted in Figs. \ref{fig:mass_imbalance}(d) and \ref{fig:mass_imbalance}(e). This result is consistent with the findings of \citet{piponniau2009} and \citet{jenquin_2023}, who proposed that fluid is injected into the bubble in the reattachment region. The supplementary material (Movie 4), which combines Movies 1 to 3, reveals that periods of increased fluid injection ($t = 15.85$ to $t = 16.04$) coincide with the presence of streamwise vortices within the bubble near reattachment, as well as regions of low-speed fluid. This suggests that fluid entrainment by these vortices may play a key role in the governing of the mass flux into the bubble. This interpretation is also supported by previous studies \citep{priebe_2016, pasquariello_2017, hu_2021}, which indicate that a centrifugal instability leads to the formation of streamwise vortices near reattachment.
 
\section{Conclusions}

\label{sec:conc}

The present work extends previous investigations by the authors \citet{lui2022,lui_2024}, which employed wall-resolved LES and post-processing techniques to examine the flow physics and unsteadiness of SBLIs over curved surfaces of a supersonic turbine at Mach 2.0 and Reynolds number of $395\,000$. These studies highlighted the significant role of near-wall streaks in driving bubble breathing motion, the presence of extreme flow bursts, and the formation of streamwise vortices along the suction side bubble. In the present study, these findings are further explored by analyzing the differences between large and small bubble states in terms of mean flow and turbulence quantities, the interplay of streaks and vortices, and the mass flux balance along the bubble surface.


Conditional analysis of the mean flow quantities reveals that, along the recirculation region, the velocity profiles exhibit greater deceleration of the fluid near the wall during large bubble events, while flow reversal is barely noticeable in small bubble events. Downstream of the incident shock, the velocity difference between the low-speed flow near the wall and the high-speed external flow is more pronounced in the large bubble events, indicating a stronger shear between these two regions. Large bubbles are associated with higher wall pressure values upstream of the incident shock as a result of stronger flow compression, while small bubbles exhibit higher wall pressure values downstream of the incident shock as a result of earlier flow reattachment. Moreover, the RMS of wall pressure indicates more intense fluctuations along the separation region and downstream of reattachment for the small bubble events. Similarly, the RMS of the skin-friction coefficient shows that wall shear stress varies more significantly along the separation region for small bubble events, while remaining steadier for large bubble events.

The results of the conditional analysis of turbulence quantities, the conditional PDFs of velocity fluctuations, and the FTLE analysis indicate that the passage of near-wall high-speed streaks through the bubble during small bubble events leads to higher $\langle u_t''u_t'' \rangle$ values upstream of the incident shock compared to large bubble events. These streaks are accompanied by streamwise vortices, which meander over time and induce intense fluid mixing in the wall-normal and spanwise directions, resulting in higher values of $\langle u_n''u_n'' \rangle$ and $\langle w''w'' \rangle$ within the bubble and along its shear layer. Consequently, these quantities contribute to a higher turbulent kinetic energy in small bubble events. In contrast, when the bubble is large, high-speed streaks are advected over it, and streamwise vortices are almost absent of the frontal edge of the bubble; however, they become apparent downstream the incident shock, during flow reattachment. In general, no significant fluid mixing occurs within large bubbles.

The analysis of mass flux along the separation bubble surface reveals that during the bubble contraction phase, mass flux out of the bubble predominantly occurs upstream of the incident shock. This is due to the passage of high-speed streaks through the bubble and intense fluid mixing induced by streamwise vortices. On the other hand, during the bubble expansion phase, significant mass flux into the bubble is observed downstream of the incident shock, near reattachment. This result indicates that periods of increased fluid injection coincide with the presence of streamwise vortices and low-momentum fluid near reattachment, suggesting that fluid entrainment by these vortices affects the mass flux into the bubble.

\section*{Acknowledgments}

W.R.~Wolf and H.F.S.~Lui acknowledge the support from Funda\c{c}\~ao de Amparo \`a Pesquisa do Estado de S\~ao Paulo, FAPESP (grants No.\ 2013/08293-7, 2019/26196-5, 2021/06448-0 and 2025/13174-4). The authors also acknowledge the financial support from the Air Force Office of Scientific Research, AFOSR (grant No.\ FA9550-23-1-0615). We thank the Coaraci Supercomputer for computer time (FAPESP grant No.\ 2019/17874-0) and the Center for Computing in Engineering and Sciences at UNICAMP.

\section{Appendix: Convergence study of conditional analysis}
\label{section:appendix}

\begin{figure}
\centering
    \begin{overpic}[trim = 1mm 1mm 1mm 1mm, clip,width=0.99\textwidth]{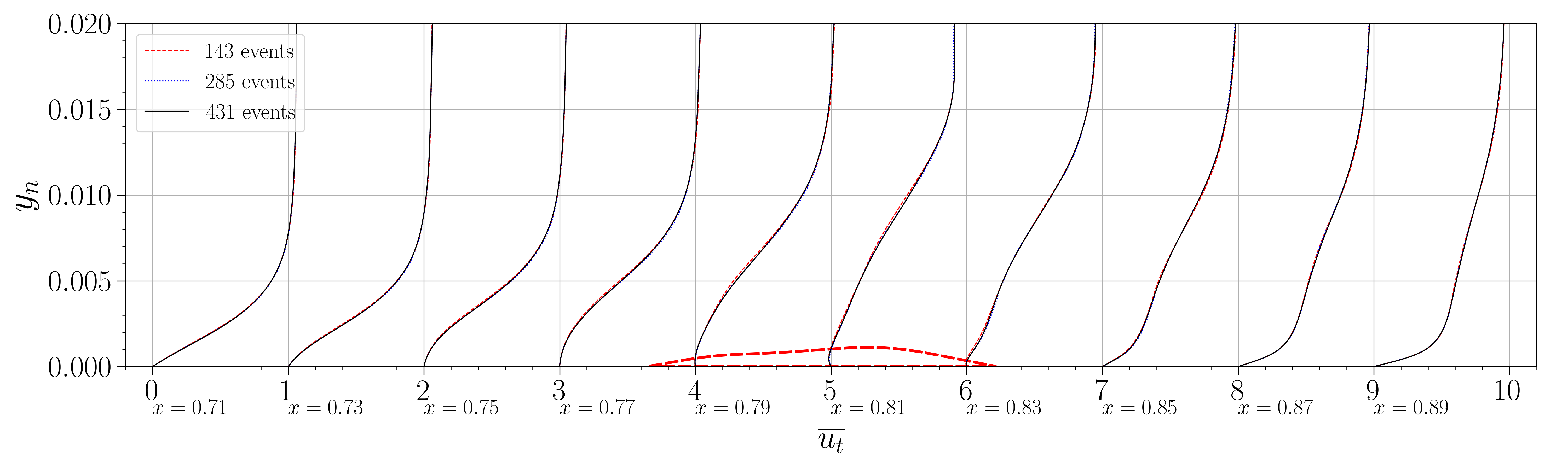}
    \put(0,30){(a)}
	\end{overpic}

    \begin{overpic}[trim = 1mm 1mm 1mm 1mm, clip,width=0.99\textwidth]{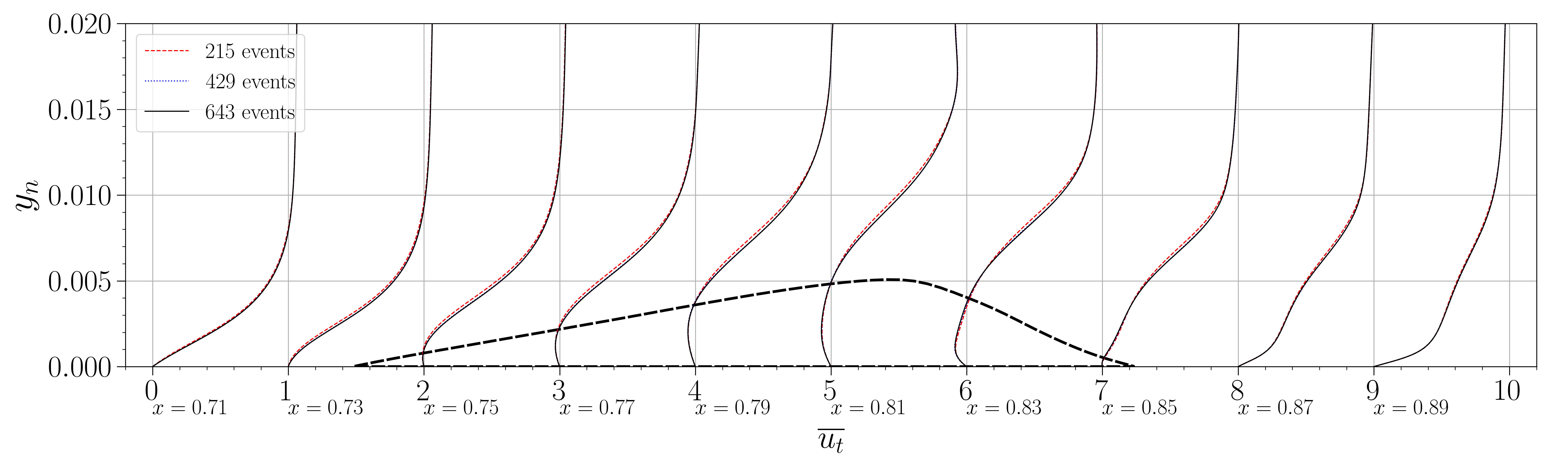}	
    \put(0,30){(b)}
	\end{overpic}
	\caption{Wall-normal profiles of conditionally averaged tangential velocity for the (a) small bubble and (b) large bubble events using different numbers of events. The black and red dashed lines delimit the separation bubble for the large and small bubble events, respectively.}
	\label{fig:ut_profile_conv}
\end{figure}

\begin{figure}
\centering
    \begin{overpic}[trim = 1mm 1mm 1mm 1mm, clip,width=0.99\textwidth]{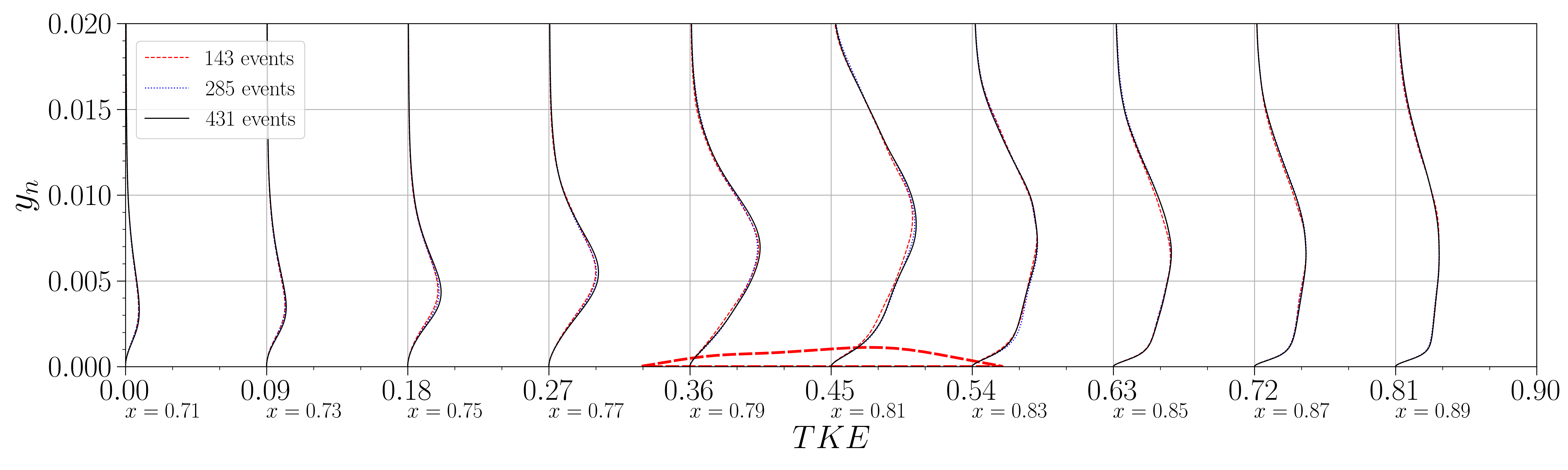}	
    \put(0,30){(a)}
	\end{overpic}

    \begin{overpic}[trim = 1mm 1mm 1mm 1mm, clip,width=0.99\textwidth]{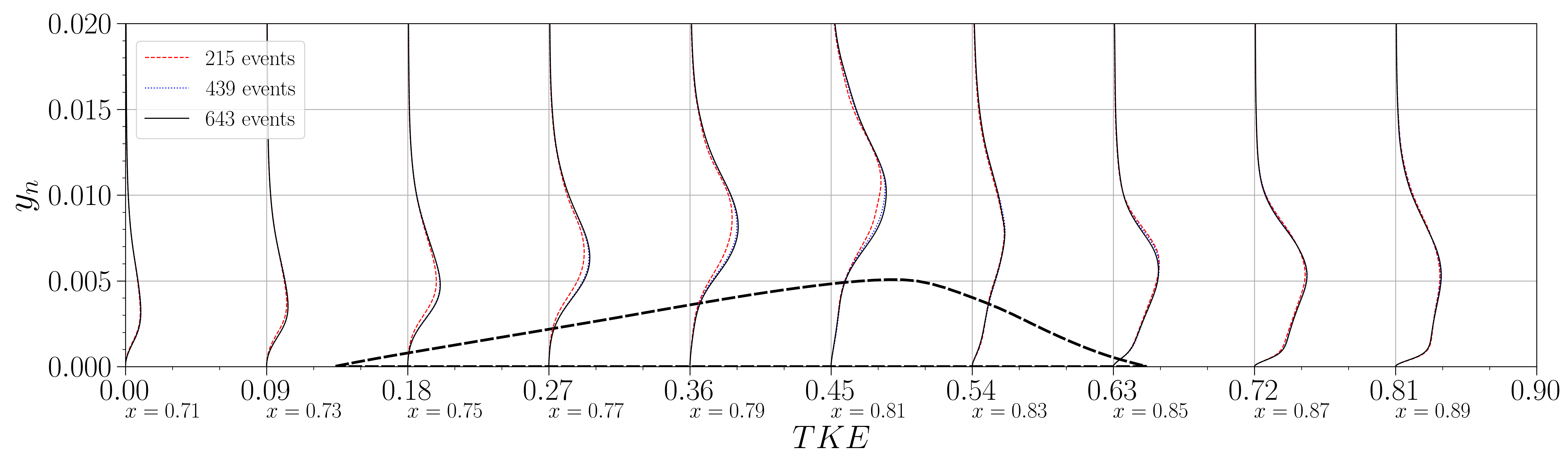}
    \put(0,30){(a)}
	\end{overpic}
	\caption{Wall-normal profiles of conditionally averaged TKE for the (a) small bubble and (b) large bubble events using different numbers of events. The black and red dashed lines delimit the separation bubble for the large and small bubble events, respectively.}
	\label{fig:tke_profile_conv}
\end{figure}

\begin{figure}
\centering
    \begin{overpic}[trim = 2mm 2mm 2mm 2mm, clip,width=0.99\textwidth]{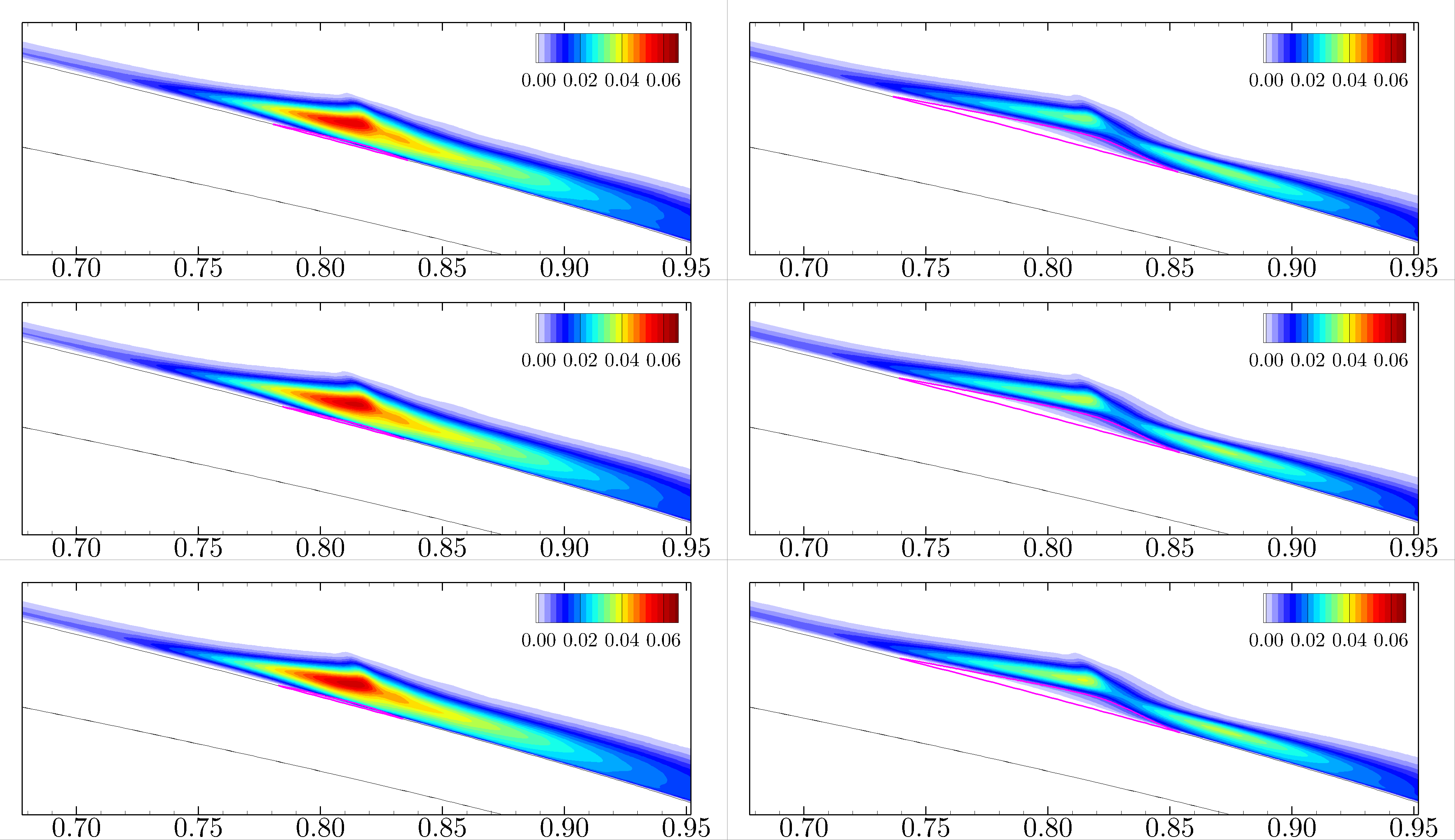}	
    \put(1,57.5){(a)}
    \put(50,57.5){(b)}

    \put(2,3){431 events}
    \put(2,22.5){285 events}
    \put(2,42){143 events}

    \put(52,3){643 events}
    \put(52,22.5){439 events}
    \put(52,42){215 events}

	\end{overpic}
	\caption{Contours of conditionally averaged TKE for the (a) small bubble and (b) large bubble events using different numbers of events. The quantity is normalized by the $u_{\infty}^2$. The purple lines delimit the separation bubble.}
	\label{fig:tke_fields_conv}
\end{figure}
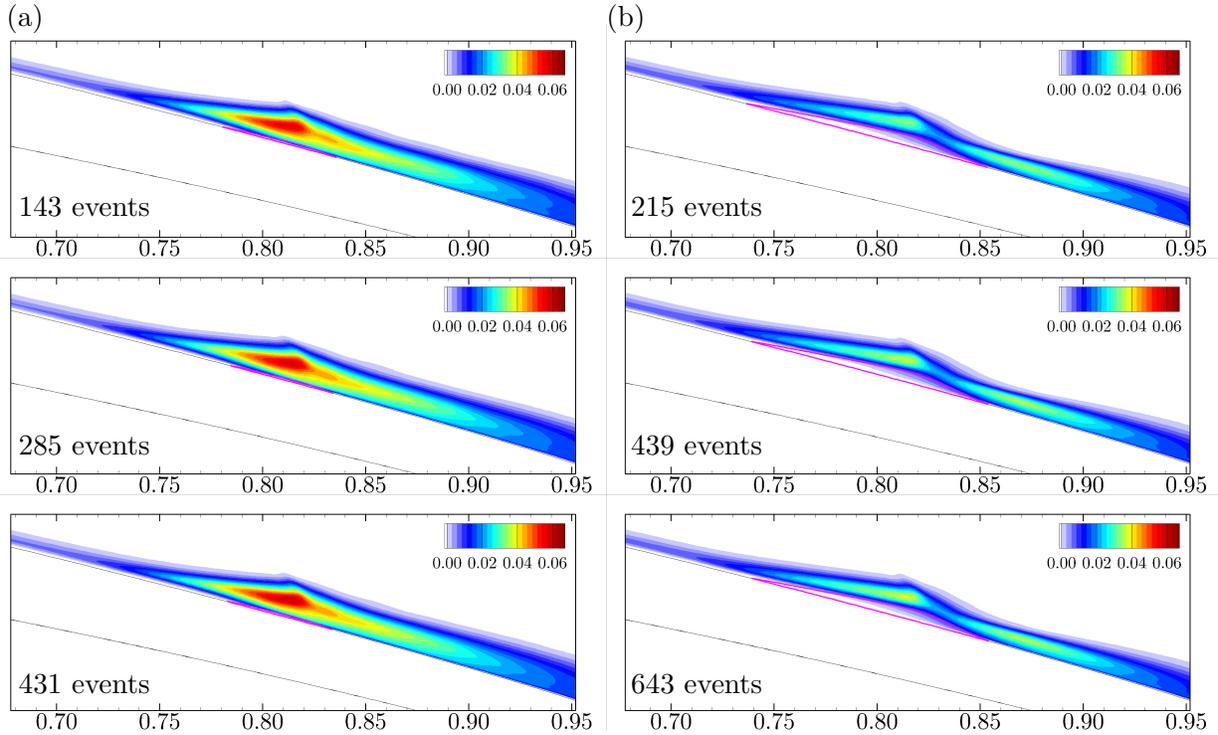

In this appendix, a convergence analysis is performed to demonstrate that the dataset comprising the extreme events is sufficiently large for the conditional analysis. This analysis is conducted by comparing first-order and second-order statistics with different numbers of conditional events, specifically using one third, two thirds, and all conditional events. Figures \ref{fig:ut_profile_conv} and \ref{fig:tke_profile_conv} show the profiles of conditionally averaged tangential velocity and TKE, respectively, for the small bubble and large bubble events using different numbers of events. Additionally, the contours of conditionally averaged TKE are shown in Fig. \ref{fig:tke_fields_conv}. 
 
When the quantities are computed using one third of the conditional events, small variations are observed compared to the other cases. However, the results with two thirds and all conditional events are very similar, indicating that the selected number of extreme events is sufficiently large for the conditional analysis. It is important to note that the flow is 
spanwise-homogeneous; therefore, the conditional averaging operation is performed in both time and spanwise directions. As a result, the total dataset can be interpreted as the number of conditional events multiplied by the number of grid points in the z-direction.

%

\end{document}